\documentclass{jfm}
\usepackage{graphicx}
\usepackage[dvipsnames]{xcolor}
\usepackage{epstopdf, epsfig}
\usepackage{color,soul}
\usepackage{physics}
\usepackage{hyperref}
\usepackage{subcaption}
\usepackage{pdflscape}
\usepackage{tikz}
\usetikzlibrary{calc,tikzmark}

\hypersetup{
    colorlinks=true,
    linkcolor=blue,
    filecolor=magenta,      
    urlcolor=cyan,
}

\newcommand{\R}{\mathit{Re}}

\shorttitle{Linear stability and SPOD of train wake flow}
\shortauthor{X.-B. Li and others}

\title{Linear stability and spectral modal decomposition of three-dimensional turbulent wake flow of a generic high-speed train}

\author{
Xiao-Bai Li\aff{1,2}, 
Simon Demange\aff{2},
Guang Chen\aff{1},
Jia-Bin Wang\aff{1},
Xi-Feng Liang\aff{1},
Oliver T. Schmidt\aff{3}
\and
Kilian Oberleithner\aff{2}
    \corresp{\email{oberleithner@tu-berlin.de}}
}

\affiliation{
  \aff{1}Key Laboratory of Traffic Safety on Track of Ministry of Education, School of Traffic \& Transportation Engineering, Central South University, 410075 Changsha, PR China
  \aff{2}Laboratory for Flow Instability and Dynamics, Technische Universität Berlin, 10623 Berlin, Germany
  \aff{3}Department of Mechanical and Aerospace Engineering, University of California San Diego, La Jolla, CA, USA
}

\begin{document}

\maketitle

\begin{abstract}
{This work investigates the spatio-temporal evolution of coherent structures in the wake of a generic high-speed train, based on a three-dimensional database from large eddy simulation. Spectral proper orthogonal decomposition is used to extract energy spectra and energy ranked empirical modes for both symmetric and antisymmetric components of the fluctuating flow field. The spectrum of the symmetric component shows overall higher energy and more pronounced low-rank behavior compared to the antisymmetric one. The most dominant symmetric mode features periodic vortex shedding in the near wake, and wave-like structures with constant streamwise wavenumber in the far wake. The mode bispectrum further reveals the dominant role of self-interaction of the symmetric component, leading to first harmonic and subharmonic triads of the fundamental frequency, with remarkable deformation of the mean field. Then, the stability of the three dimensional wake flow is analyzed based on two-dimensional local linear stability analysis combined with a non-parallelism approximation approach. Temporal stability analysis is first performed for both the near wake and the far wake regions, showing a more unstable condition in the near wake region. The absolute frequency of the near-wake eigenmode is determined based on spatio-temporal analysis, then tracked along the streamwise direction to find out the global mode growth rate and frequency, which indicate a marginally stable global mode oscillating at a frequency very close to the most dominant SPOD mode. The global mode wavemaker is then located, and the structural sensitivity is calculated based on the direct and adjoint modes derived from a local spatial  analysis, with the maximum value localized within the recirculation region close to the train tail. Finally, the global mode shape is computed by tracking the most spatially unstable eigenmode in the far wake, and the alignment with the SPOD mode is computed as a function of streamwise location. By combining data-driven and theoretical approaches, the mechanisms of coherent structures in complex wake flows are well identified and isolated.}
\end{abstract}

\begin{keywords}
Keywords
\end{keywords}

\section{Introduction}\label{sec:intro}
{In the face of climate change, high-speed rail has gradually developed to become the key to decarbonizing transportation. As a bluff body operating at high Reynolds number ($\R$), the flow around a high-speed train exhibits complex characteristics of a fully developed three-dimensional turbulent flow \citep{schetz_2001}. The unsteady aerodynamics of high-speed trains is then directly characterized by the fluctuating aerodynamic forces and induced slipstreams. Therefore, understanding the dominant dynamics in the complex turbulent flow around the train is crucial to improving and optimizing aerodynamic performance. For this purpose, the search for and identification of physically significant coherent structures, or modes, is a suitable method \citep{taira_2017}. In fact, extracting and understanding the physical mechanisms of instability in complex three-dimensional turbulent flow has been attracting research interest for several decades but remains challenging. In particular, the complexity of the base flow, as well as the high demand for computational resources, causes huge difficulties in solving this problem \citep{theofilis_2011}. However, due to the increasing demand for transportation efficiency, passenger comfort, and operational safety, extracting and understanding the mechanisms of instability in the flow around the train is still of great research interest and significant engineering importance.\par
Despite its aerodynamic design, the high-speed train exhibits bluff-body flow characteristics reminiscent of those of the well-studied Ahmed body \citep{ahmed_1984,lienhart_2002}. Three important structures can be identified in the wake of the body: a recirculation bubble over the slanted surface, a pair of longitudinal C-pillar vortices generated from the two side edges of the slanted surface, and a recirculation zone behind the rear vertical base. Several studies have focused on the interaction and control strategies of these structures. In \citet{zhang_2015,liu_2021}, the unsteady characteristics of Ahmed bodies in the high- and low-drag regimes were investigated using multiple experimental techniques. On the basis of these findings, several steady blow drag reduction strategies have been successfully developed \citet{zhang_2018,li_2022}. Meanwhile, random switching between two reflectional-symmetry breaking states of the wake has been investigated in \citet{grandemange_2013,he_2021}, with appropriate strategies altering the natural bi-stability of the wake proposed in \citet{grandemange_2014,evstafyeva_2017,haffner_2020}.\par
However, the full picture of the three-dimensional coherent structures in the wake, as well as the associated instability information, remains limited. Therefore, further interpretation and understanding of the physical mechanisms that generate disturbances in the flow are limited. In particular, for more aerodynamically shaped high-speed trains, the identification of the flow structures mentioned above is less obvious. To extend understanding and control of wake dynamics in a more complex situation, modal decomposition \citep{taira_2017} of the three-dimensional flow must be taken into consideration, which is the objective of this work.\par
Data-driven analysis has proven to be an effective method to extract coherent structures from flow snapshots as empirical modes. The most classic and widely used data-driven approach, proper orthogonal decomposition (POD), was first introduced to the field of fluid dynamics by \citet{lumley_1967,lumley_1970}. In this approach, the flow is represented as a mean and a superposition of space-time-dependent modes. These resulting modes can then be used for a variety of purposes, from classification to reduced order modeling to control \citep{rowley_2017,taira_2017}. Meanwhile, many other empirical approaches have been proposed. For example, the Balanced POD \citep{rowley_2005} which serves as an expansion for linear input-output relationships, and Dynamic Mode Decomposition (DMD) \citep{schmid_2010} which approximates the dynamics of a higher-order system through a combination of linearly growing or decaying modes.\par
In this paper, we focus on the application of the spectral form of POD, which is called spectral proper orthogonal decomposition (SPOD). This method is derived from the space-time formulation of POD for statistically stationary flow. The resulting modes, which oscillate at a single frequency, are orthogonal under a space-time inner product. In general, SPOD combines the advantages of the classical POD and DMD for statistically stationary flows, meanwhile, it provides an improved robustness over these two methods \cite{towne_schmidt_colonius_2018}. Furthermore, this method has been further extended to achieve more features, such as frequency-time \citep{nekkanti_schmidt_2021} and triadic interaction \cite{Schmidt_2020_BMD} analysis, or better convergence \citep{blanco_2022,schmidt_2022_multitaper} and lower computational cost \citep{SCHMIDT2019}, and therefore has received increasing interest in identifying coherent turbulent structures that are physically meaningful in various flow problems. These applications include jet \citep{schmidt_towne_rigas_2018}, pipe flow \citep{abreu_2020}, flow around airfoils \citep{abreu_2021}, disk wake \citep{nidhan_2022}, and various industrial flows \citep{haffner_2020,he_2021_tip,Li_2021_2,wang_2022}.\par
Considering vehicle wake flow, coherent structures represented by empirical modes are within the research scope of several studies listed above \citep{he_2021,grandemange_2013,haffner_2020,Li_2021_2}. However, these applications are limited to two-dimensional planes in the wake, which do not fully capture the complex three-dimensional space-time coherent structures in the flow. Therefore, this study further extends the previous results shown in \citet{Li_2021_2} where a parameter study using a two-dimensional database from train wake flow is performed, to find coherent structures from a global perspective. However, although SPOD extracts modes related to the dominant fluctuation of the flow, this method is purely data-driven and model-free. As such, it does not reveal the mechanisms driving the coherent structures. However, it includes all non-linear dynamics and may reveal interactions between the structures in a quantitative manner. To further search for the instability mechanisms, one may use either stochastic low order dynamic models \citep{Rigas_2015,sieber_2021} or a mean field stability analysis.\par
Mean field linear stability analysis is considered to provide further insight into the mechanisms driving the flow dynamics. It is known that a self-excited oscillation can be described by an unstable global mode \citep{chomaz_2005} derived from a global stability analysis. Recently, improved feasibility of large-scale linear algebra computations enables tri-global stability analysis of flows with three inhomogeneous directions \citep{theofilis_2011}. Some applications include boundary layer flows with isolated roughness elements \citep{loiseau_2014,kurz_2016,ma_2022}, lid-driven cavity flows \citep{Paredes_2014,gomez_2014}, jets in crossflow \citep{regan_2017}, and wakes of rectangular prisms \citep{zampogna_2023}. When mean flows display homogeneity in the direction normal to convective motion, the Bi-Global stability approach can be utilized. This approach considers two-dimensional modes with a spatial wavenumber in the third dimension, and has been applied to a broad range of canonical flow \citet{THEOFILIS2003249,theofilis_2011}, as well as complex technical flows including swirling flows \citep{Kaiser2018,muller_2020}, reacting flows \citep{Casel_2022,wang_2022_globalLSA}, turbo-machinery flows \citep{Muller_2022} and two-phase flows \citep{schmidt_oberleithner_2023}. For flows that are weakly nonparallel, evolving slowly in streamwise direction, the bi- and tri-global stability can be approximated from a local stability analysis, i.e. local one-dimensional analysis in the lines normal to the streamwise direction to construct bi-global instability. As reviewed by \cite{huerre_1990}, the method is based on a spatio-temporal analysis of the local velocity profile invoking the WKBJ approximation. The relationship between local absolute instability and global modes can be found in \citet{monkewitz_1993,chomaz_2005},  which concluded that a region of local absolute instability is a necessary condition for the existence of global instability. Comparisons between results of local and global stability analysis can be found in \citet{giannetti_2007,juniper_2011,Juniper_2015,Kaiser2018,demange_2022}. In general, local stability analyses require less computational memory than global stability analyzes, since they convert a large matrix eigenvalue problem into several small independent matrix eigenvalue problems \citep{Juniper_2015}. Meanwhile, as the local linearized Navier-Stokes equation is solved at each discrete streamwise position, the eigenvalues can be continuously tracked to provide an accurate spatial description of the mode. Therefore, local stability analysis is still widely used for flows beyond the range of global analysis \citep{pier_2008,oberleithner_2011,rukes_2016}.\par
To the best of the author's knowledge, there are limited studies regarding the linear instability mechanisms of fully developed turbulent wake flows behind vehicles. \citet{zampogna_2023} investigated the global stability of rectangular prisms with rounded front edges, which approximate the geometry of the Ahmed body. However, this study considered a laminar flow without ground effect, such that the problem could be treated with two symmetries. From a more practical perspective, flows behind high-speed trains may consist of a series of vortex structures that are subject only to a spanwise symmetry condition. In addition, the spatial and temporal evolution of these vortex structures differs greatly from free-evolving structures due to the presence of the ground \citep{schetz_2001}. Up to now, the instability mechanisms in typical vehicle wakes of high industrial relevance remain an unanswered question. Is there a linear global mode that drives the instability in the turbulent wake? How does the linear global mode compare to the leading SPOD mode? Which part of the wake serves as the origin of the global instability and is most sensitive to external forcing? These questions need to be answered to serve as a basis for further optimizations, while extending the applications of these two approaches to more complex flow problems and higher $\R$.\par
Since the flow problem considered is only subject to spanwise symmetry, the tri-global stability with three inhomogeneous directions should be considered. The WKBJ approximation then converts the three-dimensional linearized problem into several two-dimensional local problems to account for global instability. This is generally a more complex situation than the research shown in \citet{Juniper_2015,rukes_2016,Kaiser2018}, where local one-dimensional analyzes are used to construct the bi-global mode. Meanwhile, the parallelism or weak non-parallelism in local analysis could be a strong assumption \citep{chomaz_2005,pier_2008,rukes_2016,puckert_2018} in the flow problem considered. How the non-parallelism would affect the results of global instability and how to deal with the non-parallelism in the complex base flow are also important issues to be solved.\par
The outline of the paper is as follows. The large eddy simulation setup used to obtain the three-dimensional flow field around the train is shown in $\S$\ref{sec:methods}, along with the description of the time-averaged wake-flow structures. In $\S$\ref{sec:empiricalmodes}, we use SPOD to extract the dominant empirical modes and provide a first insight into the spatio-temporal characteristic of the coherent structures. Meanwhile, bispectral mode decomposition is considered to compute triadic interactions, which explains the features in the SPOD spectrum. The tri-global stability mode obtained from two-dimensional local stability analysis is presented in $\S$\ref{sec:theoreticalmodes}. In $\S$\ref{sec:comparison}, we compare the SPOD mode with the linear global mode at each streamwise location, and the mechanism of the fundamental instability is interpreted based on the comparison results. The main findings and conclusions are summarized in $\S$\ref{sec:conclusions}.\par
}

\section{Flow problem description}\label{sec:methods}
\subsection{Large Eddy Simulation}\label{sec:LES}
{The database for both the data-driven and the theoretical mode calculations is obtained from a large eddy simulation performed with the commercial code STAR-CCM+ 14.02. A complete description of the numerical setup can be found in \citet{li_2021}, here only the essential information is presented.\par}
\begin{figure}
  \centering
  \includegraphics[width=1\textwidth]{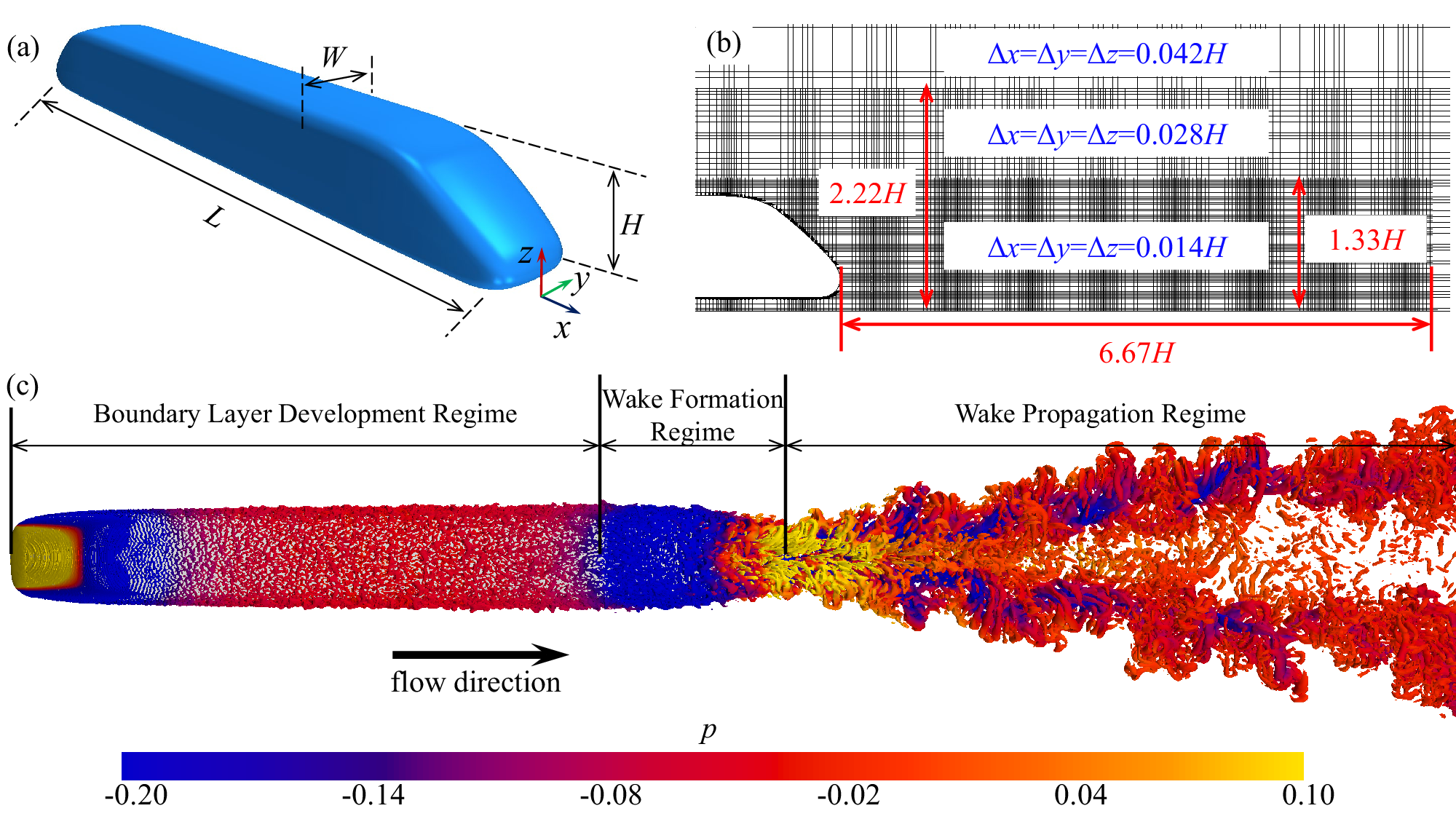}
  \caption{
  Large Eddy Simulation setup. Reproduced from \citet{li_2021} (\textit{a}) Geometric model. (\textit{b}) Distribution of computational grid. (\textit{c}) Instantaneous snapshot of vortex structures; The arrow shows the flow direction.
  }
  \label{fig:LES}
\end{figure}
{A simplified version of the Intercity-Express 2 (ICE2), also known as the aerodynamic train model (ATM), is considered. The simulation is carried out on the scale of 1:10, resulting in the height of the model $H=0.36m$, the width of the model $W=0.30m$ and the length of the model $L=2.5m$, as shown in figure \ref{fig:LES}(\textit{a}). To simulate the relative motion between the train and the surrounding environment, the upstream boundary, located 10$H$ from the train head, is assigned the inflow velocity $U_\infty=4{m/s}$. Meanwhile, the ground boundary, with a distance to the train bottom surface of 0.15$H$, is defined with the same moving velocity. The resulting $\R$ based on the height of the train and $U_\infty$ is $9.5\times10^4$. The downstream boundary is located 30$H$ from the train tail, with a zero static pressure outlet condition. On the side and roof of the computational domain, the symmetry plane boundary condition is assigned, with a distance of 10$H$ from the train model. The coordinate system is shown in figure \ref{fig:LES}(\textit{a}), with the origin located at the ground height of the train tail nose tip. The computational domain is then discretized using unstructured hexahedral volume meshes, with the wall normal and wall parallel distances expressed in viscous units, respectively ${\Delta}y^+=0.16$ and ${\Delta}x^+={\Delta}z^+=28$ for cells attached to the train surface, as shown in figure \ref{fig:LES}(\textit{b}). The total number of volume meshes used in the study is 46.8 million.\par
In the current research, the large eddy simulation based on the wall-adapting local-eddy viscosity (WALE) subgrid-scale model is chosen. The use of a novel form of the velocity gradient tensor in the WALE subgrid scale model allows for much more universal model coefficients compared to other widely used subgrid scale models. Meanwhile, the WALE subgrid scale model does not require any form of near-wall damping but automatically provides accurate scaling at the walls. More details about the WALE subgrid scale model can be found in \citet{nicoud_1999}. An implicit unsteady segregated incompressible finite-volume solver is used, with the convective terms discretized based on a bounded central-differencing scheme, and the diffusion and turbulence terms are discretized with the second-order upstream scheme. The time marching procedure is performed using the implicit second-order accurate three-time level scheme, with the discretized convective time step set to $0.0067t^*$ ($t^*=W/U_\infty$), which leads to the Courant-Friedrichs-Lewy number below unity in most of the computational grids. An instantaneous scene of the flow structures around the generic high-speed train, which briefly illustrates the formation of the turbulent wake, is shown in figure \ref{fig:LES}(\textit{c}). Note that the LES results has been validated in \citet{li_2021}, where the simulated pressure coefficients are compared with experimental results. Here to further enhance the confidence of the results presented in the paper, the LES data are further validated by a wind tunnel experiment, in terms of both the time-averaged flow velocity and turbulent statistics. The detailed comparisons are shown in Appendix \ref{sec:LESvalidation}.\par
\begin{figure}
  \centering
  \begin{subfigure}{0.49\textwidth}
  \subcaption{}
  \includegraphics[width=1\textwidth]{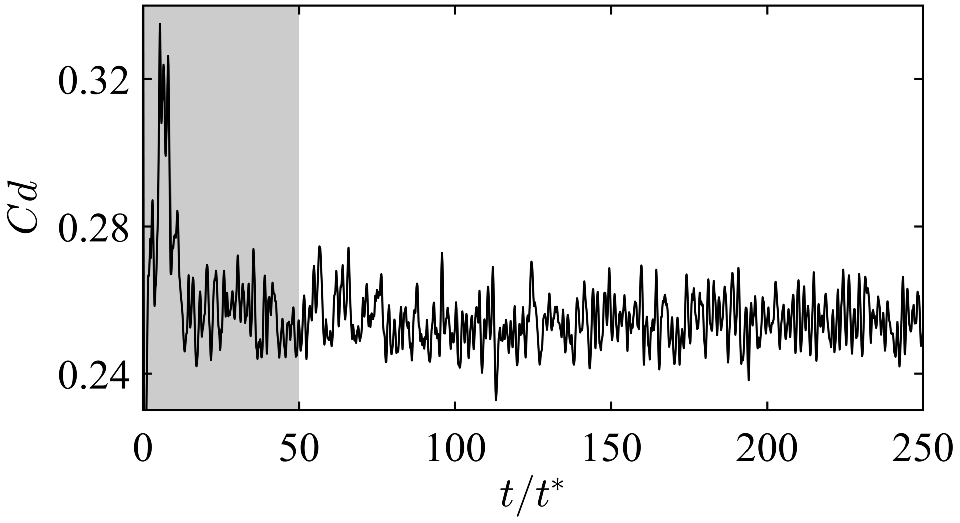}
  \end{subfigure}
  \begin{subfigure}{0.49\textwidth}
  \subcaption{}
  \includegraphics[width=1\textwidth]{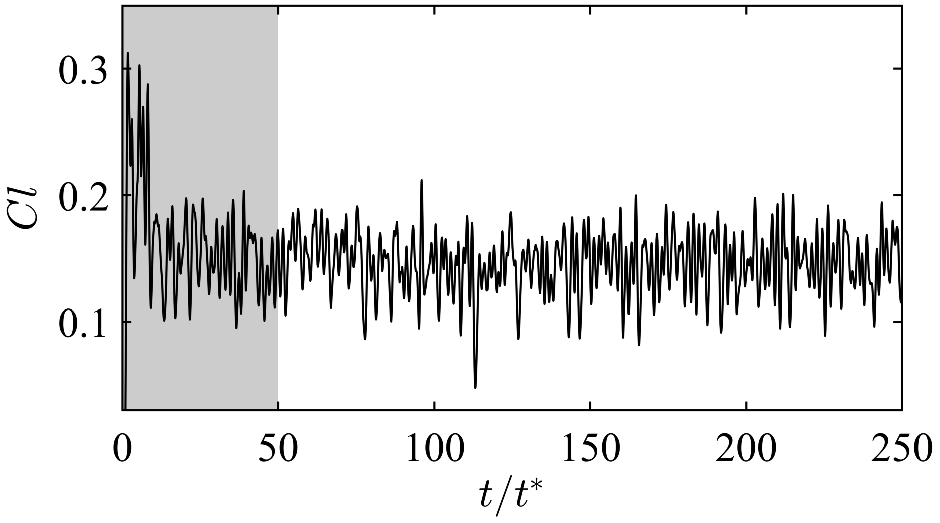}
  \end{subfigure}
  \caption{
  Time histories of the aerodynamic force coefficients. (\textit{a}) Drag force; (\textit{b}) Lift force.
  }
  \label{fig:Force}
\end{figure}
We started to collect the field data after the simulation had been run for $50t^*$. It can be seen from the time-history curves of global quantities shown in figure \ref{fig:Force} that, the flow has already reached the statistically stationary state at this moment. Then the three-dimensional snapshot database was continuously collected from a square box extending from $x/W=-1.33$ upstream of the tail nose tip to $x/W=6.67$ downstream of the tail nose tip, $y/W=1.33$ from the center plane of the train in both spanwise directions, and $z/W=1.33$ from the ground in the vertical direction, for the duration of $800t^*$. The time step between two consecutive snapshots is $0.1t^*$, resulting in a total of 8000 snapshots collected during the simulation. These values have been shown to be sufficient to produce well-converged SPOD results according to the parametric study shown in \citet{Li_2021_2}.\par}

\subsection{Mean flow properties}\label{sec:meanflow}
\begin{figure}
  \centering
  \begin{subfigure}{1\textwidth}
  \subcaption{}
  \includegraphics[width=1\textwidth]{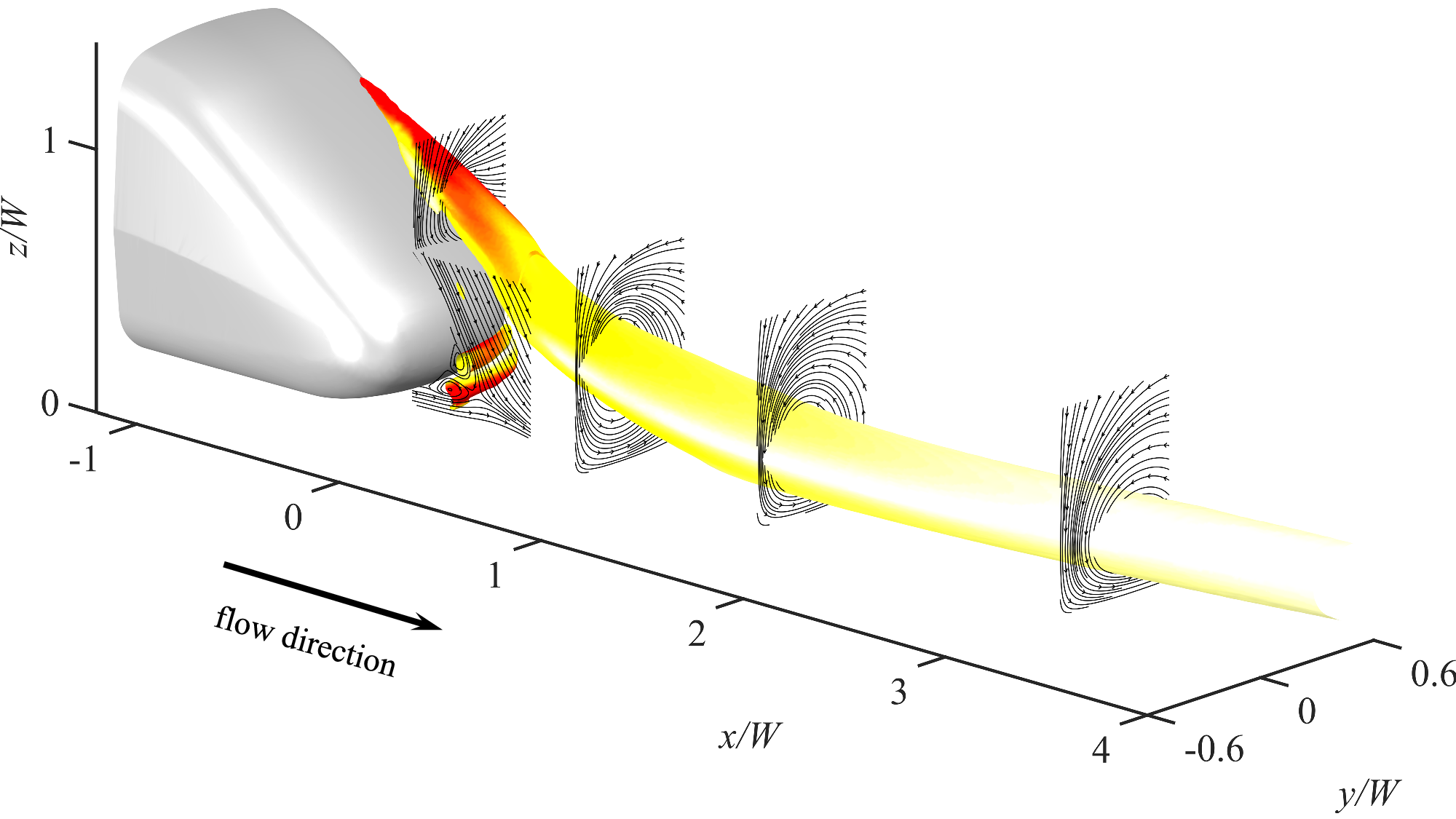}
  \end{subfigure}
  \includegraphics[width=0.55\textwidth]{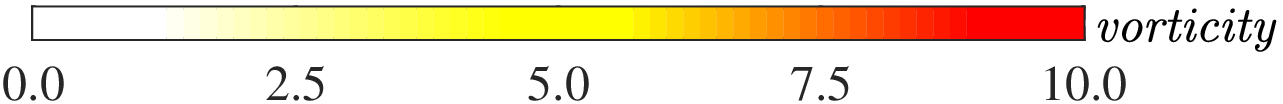}\vspace{0.1cm}
  \begin{subfigure}{1\textwidth}
  \subcaption{}
  \includegraphics[width=1\textwidth]{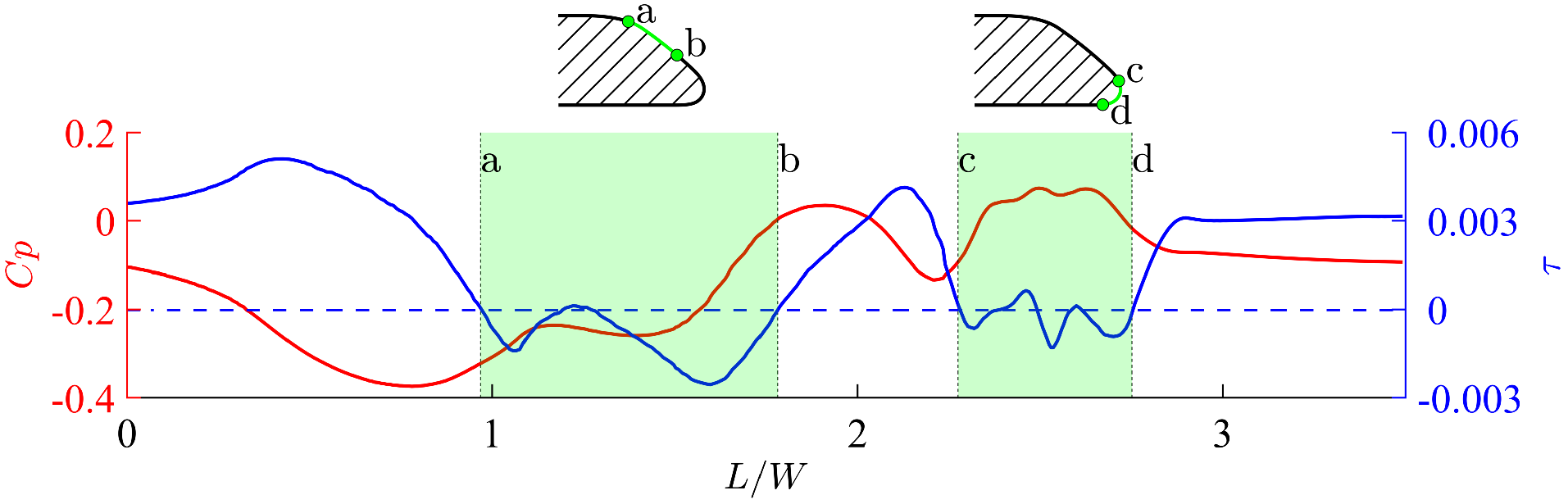}
  \end{subfigure}
  \caption{
  Time-averaged mean flow distributions. (\textit{a}) Flow structures around the train tail visualized by isosurface of vortex identification criterion $\Omega=0.6$ and streamlines; The arrow shows the flow direction. (\textit{b}) Distributions of pressure coefficient and streamwise skin-friction coefficient along the central line of the train tail in clockwise direction; The greed patches highlight the regions with negative skin-friction coefficients.
  }
  \label{fig:Meanfield}
\end{figure}
{We consider the time-averaged field as the base state for the linear stability analysis and introduce it in this section. Furthermore, prior knowledge of the mean field will facilitate understanding of the physics associated with the extracted coherent structures. In figure \ref{fig:Meanfield}(\textit{a}), the time-averaged sectional streamlines in the wake are shown at several representative locations. Meanwhile, vortex regions in the wake are identified by the isosurface of $\Omega$ \citep{liu_2016}, colored by the magnitude of the vorticity. The $\Omega$-method works as a vortex identification criterion similar to the $Q$-criterion and ${\lambda}_2$-criterion, however, it is less sensitive to threshold value to capture both strong and weak vortices in different cases. Note that the wake flow is symmetric about the central plane after long-term time averaging, so only the $y>0$ half is shown here for better visualization. In figure \ref{fig:Meanfield}(\textit{b}), to provide with more quantitative information, the time-averaged pressure coefficient and streamwise skin-friction coefficient along the central line of the train (also shown in this figure) are plotted. The $x$-axis is denoted by the distance from the starting point along the tail central line in clockwise direction. The regions with negative streamwise skin-friction coefficients indicative of flow recirculation are highlighted with green patches.\par
The distribution of the time-averaged flow along the symmetry plane is illustrated first. It can be observed that, as the flow approaches the slanted surface of the tail, the adverse pressure gradient imposed by local flow acceleration forces the attached flow to separate from the surface at point $a$ (see figure \ref{fig:Meanfield}(\textit{b})). However the flow separated from point $a$ is highly deformed and fails to form a strong vortex region, since in figure \ref{fig:Meanfield}(\textit{a}), no obvious structure is identified at this location by the isosurface of $\Omega$. Then in figure \ref{fig:Meanfield}(\textit{b}), the re-attachment can be identified at point $b$, downstream of which the attached flow on the slanted surfaces gradually approaches the rear end. Further downstream, the strong adverse pressure near the tail nose point forces the attached flow on the slanted and bottom surfaces to respectively separate at point $c$ and point $d$, forming the spanwise vortex pair located just behind the tail nose point, as identified in figure \ref{fig:Meanfield}(\textit{a}).\par
In addition to the flow structures related to the separation across the symmetry plane, we can also observe in figure \ref{fig:Meanfield}(\textit{a}) that a pair of longitudinal vortexes is located above the side edges of the tail, which is similar in nature to the C-pillar vortex in the Ahmed body wake flow \cite{zhang_2015,liu_2021,he_2021,li_2022}. This pair of longitudinal vortex is formed by flow separation from the side surface, which exerts a strong pressure-suction effect in this area and continuously rolls up flow from the slanted surface. As the longitudinal vortex propagates downstream, it gradually increases in diameter and lifts away from the tail surface toward the ground. Due to the strong downwash from the slanted surface, the trailing vortex is pushed away from the central symmetry plane as it travels downstream. From $x/W\approx1.5$ onward, the longitudinal vortex structure attaches to the ground and then propagates nearly parallel to the ground in the downstream wake.\par
In general, the mean field is fully three-dimensional in the near wake region, and its complexity gradually decreases downstream of the solid body, developing to be nearly parallel downstream of $x/W\approx2.0$. The two main features, the spanwise recirculation bubble and the streamwise vortex pair, are likely to be related to the mean field instability due to the strong velocity gradient and will therefore be discussed further in the following content.\par}

\section{Data-driven coherent structure identification}\label{sec:empiricalmodes}
\subsection{Spectral proper orthogonal decomposition}\label{sec:SPOD}
{The SPOD approach is the frequency-domain counterpart of the standard POD approach. It seeks modes that optimally represent space-time flow statistics \citep{schmidt_towne_rigas_2018,towne_schmidt_colonius_2018}. A brief overview of this method is provided in this part.\par
We denote the mean subtracted snapshots as
\begin{equation} \label{eqn:q'}
\vb{q}'_i=\vb{q}'(t_i)=\begin{bmatrix} u'(x,y,z,t_i) \\ v'(x,y,z,t_i) \\ w'(x,y,z,t_i) \\ p'(x,y,z,t_i) \end{bmatrix}
\end{equation}
where $i$ represents the number of snapshots. To estimate the spectral contents, the snapshot database is first segmented into $n_{\rm{blk}}$ overlapping blocks, with $n_{\rm{DFT}}$ snapshots in each individual block. Here $n_{\rm{DFT}}=256$, along with the overlap of $50\%$ is used in our study, and the resulting angular frequency resolution is $\Delta\omega=0.245$ ($\omega$ is normalized by the factor of $U_{\infty}/W$). With the above parameters, $n_{\rm{blk}}$ can be determined with the value of 61, which is sufficient for well-converged SPOD energy spectrum and modes \citep{schmidt_2020}. The blocks are then Fourier transformed, and all Fourier realizations at the $k$-th frequency are collected into a new data matrix,
\begin{equation} \label{eqn:Qfft}
\vb{\hat{Q}}_k=\begin{bmatrix} \vb{\hat{q}}^{(1)}_k & \ \vb{\hat{q}}^{(2)}_k & ... & \vb{\hat{q}}^{(n_{\rm{blk}})}_k \end{bmatrix}
\end{equation}\par
At this point, the orthogonal basis can be obtained by solving the eigenvalue problem defined by 
\begin{equation} \label{eqn:Seig}
\left({\vb{\hat{Q}}_k}^H\vb{W}{\vb{\hat{Q}}_k}\right) / n_{\rm{blk}}=\vb{U}_k\vb{\Lambda}_k\vb{U}_k^H
\end{equation}
where $\vb{W}$ is the diagonal matrix containing weight information of each flow quantity at each sampling node. Therefore it defines the energy norm used in SPOD, thus determining physical process to be highlighted \citep{Colonius_2002}. Here, the weight matrix is given as
\begin{equation} \label{eqn:Weight}
\vb{W}=\int_{\Omega}\begin{bmatrix} 1 & & & \\ & 1 & & \\ & & 1 & \\ & & & 0  \end{bmatrix}{\rm{d}}V
\end{equation}
so that the turbulent kinetic energy (TKE) norm can be defined. The matrices $\vb{\Lambda}_k={\rm{diag}}\left(\lambda_{k_1},\lambda_{k_2},...,\lambda_{k_{n_{blk}}},\right)$ contains the SPOD energies which are therefore based only on the turbulent kinetic energy. Then both the velocity modes, as well as the associated pressure modes, can be recovered by
\begin{equation} \label{eqn:SPODmode}
\vb{\Phi}_k=\frac{1}{\sqrt{n_{\rm{blk}}}}\vb{\hat{Q}}_k\vb{U}_k\vb{\Lambda}_k^{-1/2}
\end{equation}\par}
{In addition, since the described flow problem is subjected to the spanwise symmetry, fluctuations of the sampled flow field can be divided into symmetric and antisymmetric contributions \citep{hack_schmidt_2021}. To properly isolate coherent structures defined by the two different types of fluctuations, an additive decomposition was applied to the collected snapshot data before extracting empirical modes. The symmetric contribution of the $i$th snapshot is denined as
\begin{equation} \label{eqn:q'S}
\vb{q}'_S(t_i)=\frac{1}{2}\begin{bmatrix} u'(x,y,z,t_i)+u'(x,-y,z,t_i) \\ v'(x,y,z,t_i)-v'(x,-y,z,t_i) \\ w'(x,y,z,t_i)+w'(x,-y,z,t_i) \\ p'(x,y,z,t_i)+p'(x,-y,z,t_i) \end{bmatrix}
\end{equation}
and the antisymmetric contributions is defined as
\begin{equation} \label{eqn:q'A}
\vb{q}'_A(t_i)=\frac{1}{2}\begin{bmatrix} u'(x,y,z,t_i)-u'(x,-y,z,t_i) \\ v'(x,y,z,t_i)+v'(x,-y,z,t_i) \\ w'(x,y,z,t_i)-w'(x,-y,z,t_i) \\ p'(x,y,z,t_i)-p'(x,-y,z,t_i) \end{bmatrix}
\end{equation}\par}
\begin{figure}
  \centering
  \begin{subfigure}{0.32\textwidth}
  \subcaption{}
  \includegraphics[width=1\textwidth]{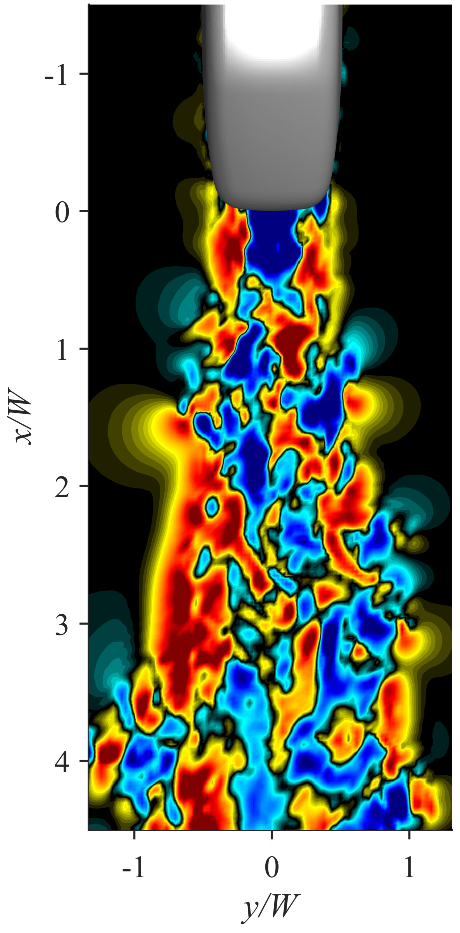}
  \end{subfigure}
  \begin{subfigure}{0.32\textwidth}
  \subcaption{}
  \includegraphics[width=1\textwidth]{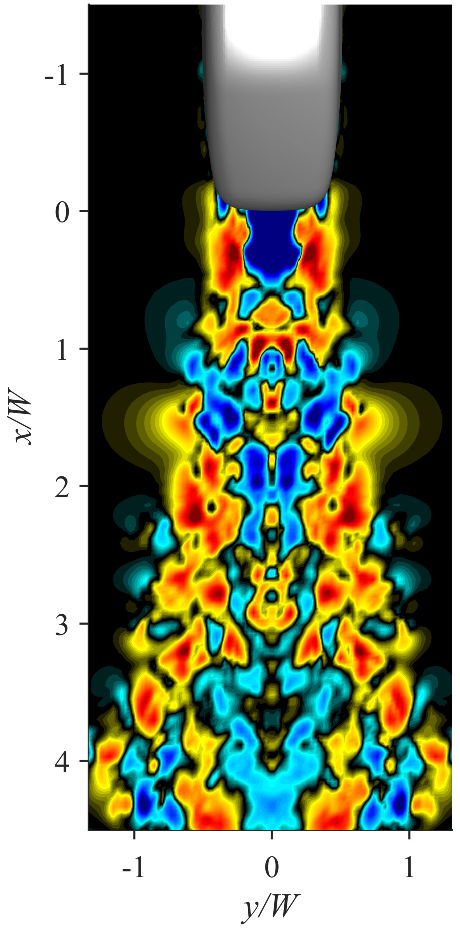}
  \end{subfigure}
  \begin{subfigure}{0.32\textwidth}
  \subcaption{}
  \includegraphics[width=1\textwidth]{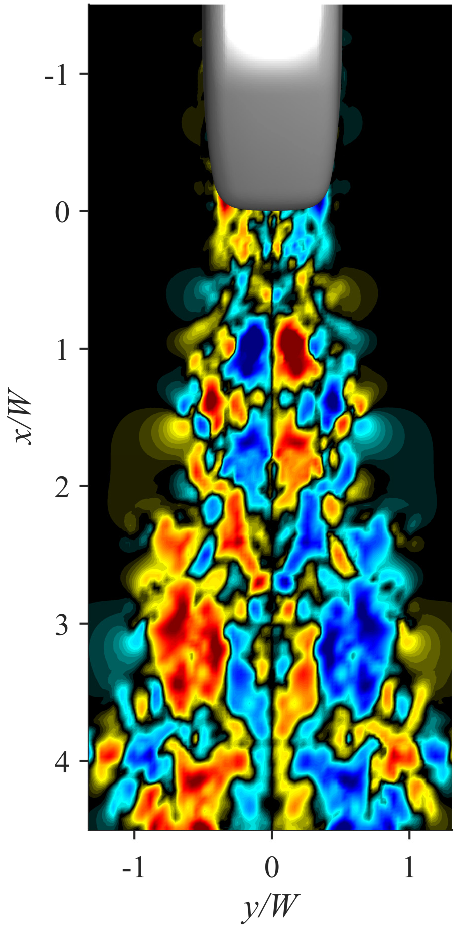}
  \end{subfigure}\vspace{0.1cm}
  \includegraphics[width=0.55\textwidth]{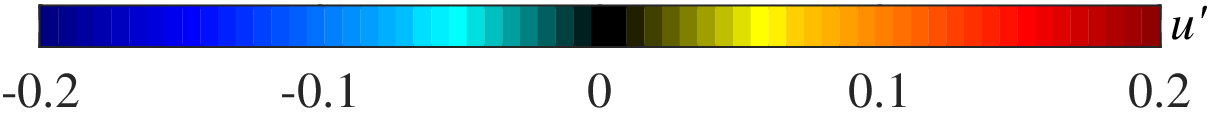}
  \caption{
  Symmetric \& antisymmetric decomposition of a single snapshot. (\textit{a}) Original field; (\textit{b}) Symmetric component; (\textit{c}) Antisymmetric component.
  }
  \label{fig:Snapshot}
\end{figure}
{A visualization of one snapshot of the fluctuating streamwise velocity field, together with the contributions from the symmetric and antisymmetric components, are shown in figure \ref{fig:Snapshot}. The spanwise symmetrical and anti-symmetrical contributions of all collected samples are arranged into two separate data matrices. They are then independently solved following the procedures described above, to extract and analyze, respectively, the symmetric and antisymmetric empirical modes. Note that, due to the zero-integral property of an even and an odd function in the sampling domain, the symmetric and antisymmetric modes yield mutual orthogonality \citep{hack_schmidt_2021}.\par}

\subsection{SPOD energy spectra \& modes}\label{sec:SPODresults}
\begin{figure}
  \centering
  \begin{subfigure}{0.49\textwidth}
  \subcaption{}
  \includegraphics[width=1\textwidth]{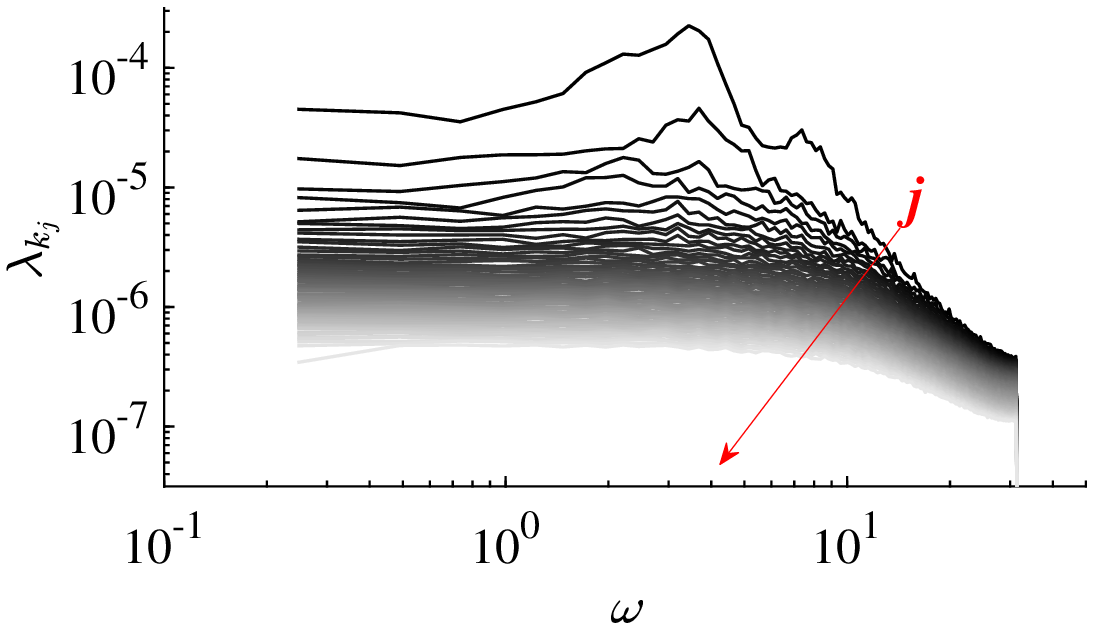}
  \end{subfigure}
  \begin{subfigure}{0.50\textwidth}
  \subcaption{}
  \includegraphics[width=1\textwidth]{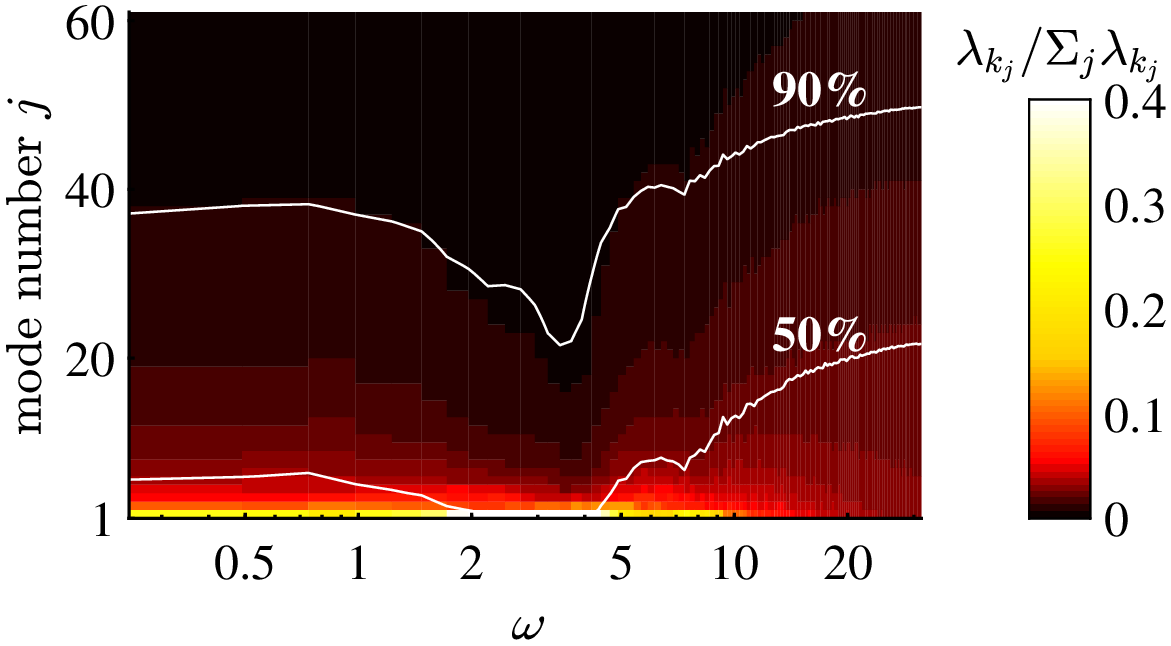}
  \end{subfigure}
  \begin{subfigure}{0.49\textwidth}
  \subcaption{}
  \includegraphics[width=1\textwidth]{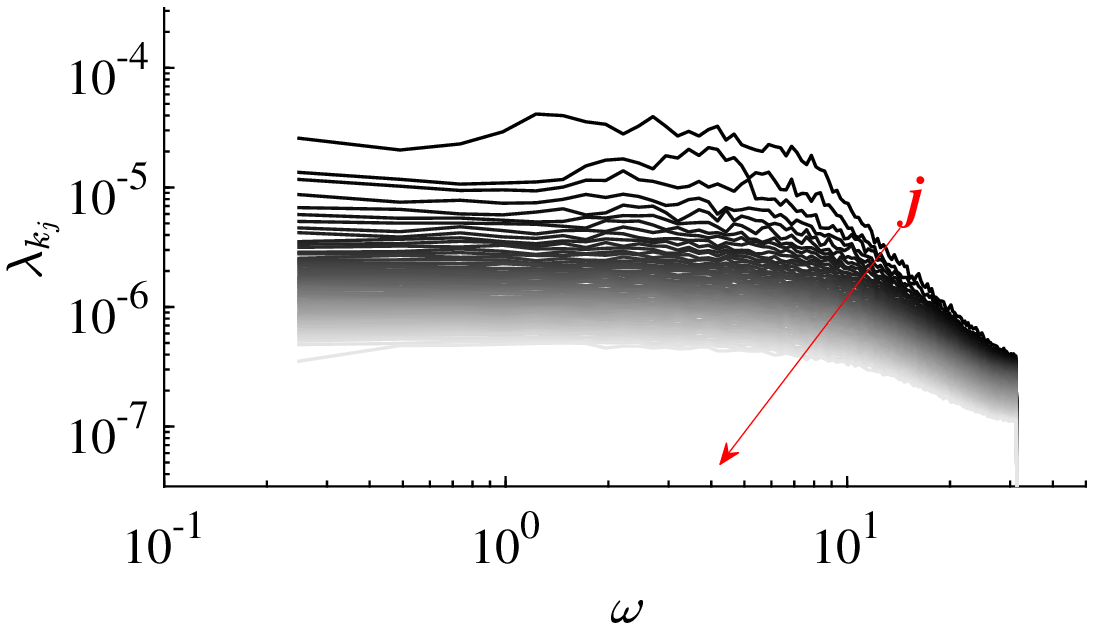}
  \end{subfigure}
  \begin{subfigure}{0.50\textwidth}
  \subcaption{}
  \includegraphics[width=1\textwidth]{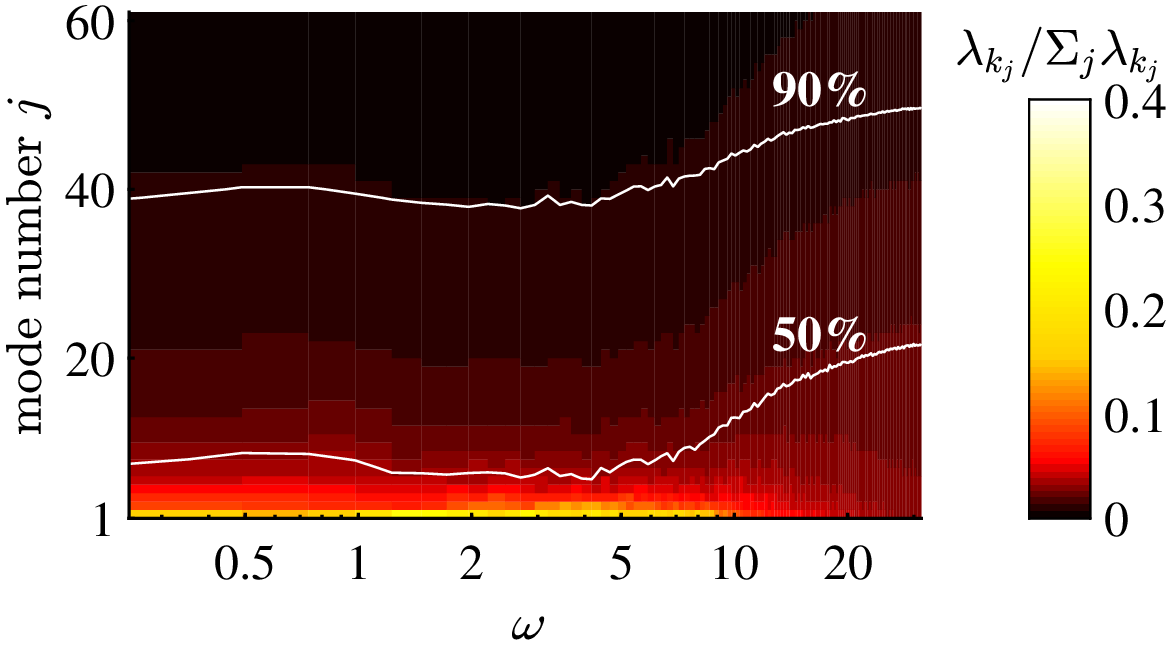}
  \end{subfigure}
  \caption{
  SPOD mode energy spectrum. The left column shows the spectral curves with the shading of the line color representing the increase in the mode number. The right column shows the percentage of energy that each mode accounts for as a function of frequency, with solid lines indicating that the cumulative energy represents $50\%$ and $90\%$ of the total energy at each frequency. (\textit{a,b}) Symmetric component; (\textit{c,d}) Antisymmetric component.
  }
  \label{fig:Spectra}
\end{figure}
{SPOD solves the eigenvalue problems at each discrete frequency independently, and produces energy-ranked eigenvalues. Therefore, the energy distributions of different modes at each frequency can be best visualized in the form of a spectrum \citep{schmidt_towne_rigas_2018}. The left column in figure \ref{fig:Spectra} shows SPOD spectra for both symmetric and antisymmetric components. Meanwhile, the cumulative energy content and the percentage of energy accounted for by each mode as a function of frequency are shown in the right column. The symmetric component displays a higher overall energy level compared to the antisymmetric component, indicating its dominant role in the dynamics of the turbulent wake. The low-rank behavior, characterized by a large separation between the first and second modes, also appears to be more pronounced in the symmetric component. In particular, in the angular frequency ranges of $2<\omega<4$ of the symmetric component, the first modes contribute more than $50\%$ of the fluctuating energy according to figure \ref{fig:Spectra}({\it{b}}). The angular frequency of the dominant symmetric coherent structure is $\omega=3.437$ as shown in figure \ref{fig:Spectra}({\it{a}}). The corresponding Strouhal number is $St=0.547$, based on the free stream velocity $U_{\infty}$ and the train width $W$. Additionally, the spectral energy concentration at $\omega\approx6.9$, and a less pronounced peak around $\omega\approx1.7$ can be identified respectively. These two angular frequencies could be associated with the first harmonic and subharmonic of the fundamental mode. This will be further investigated by checking the mode bi-spectrum in $\S$\ref{sec:triadic}.\par}
\begin{figure}
  \centering
  \begin{subfigure}{0.49\textwidth}
  \centering
  \subcaption{$\omega=3.437$}
  \includegraphics[width=1\textwidth]{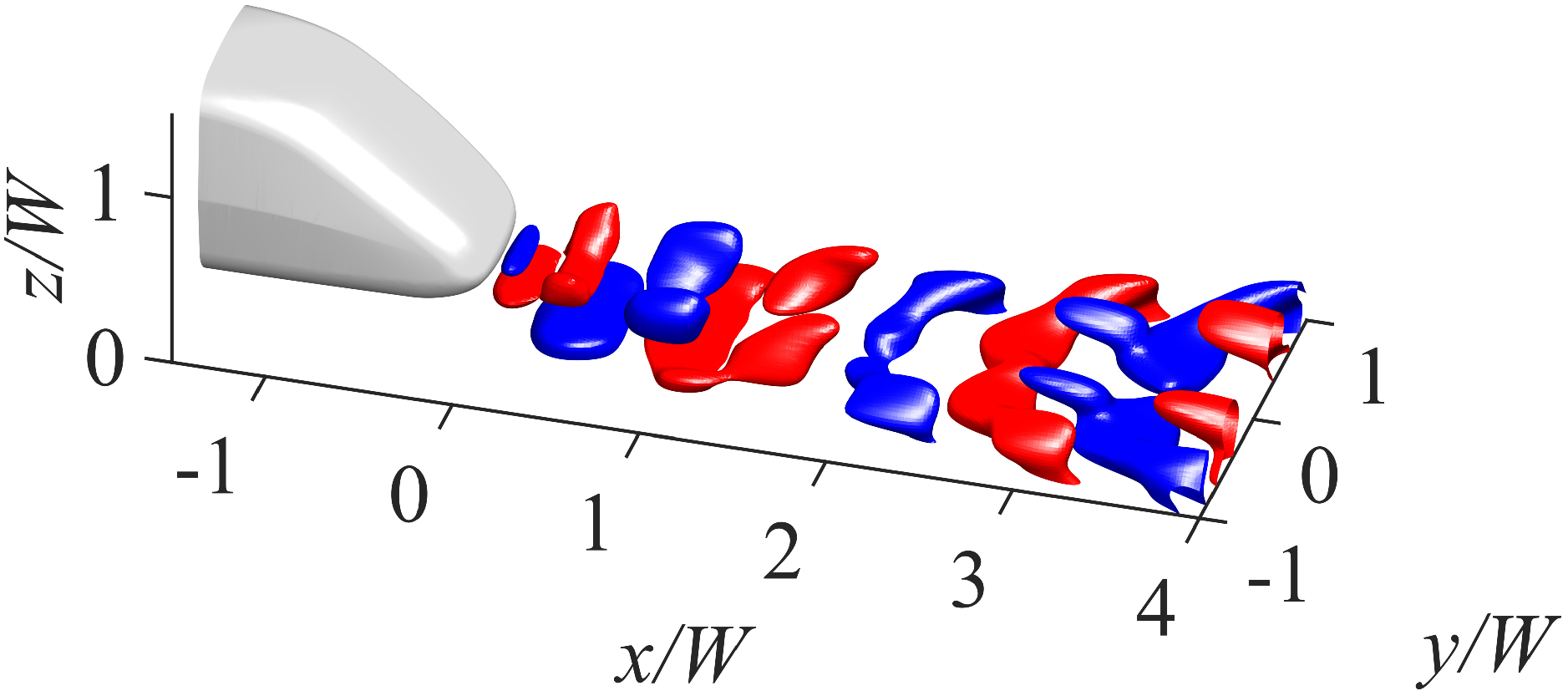}
  \end{subfigure}
  \\
  \begin{subfigure}{0.49\textwidth}
  \subcaption{$\omega=1.718$}
  \includegraphics[width=1\textwidth]{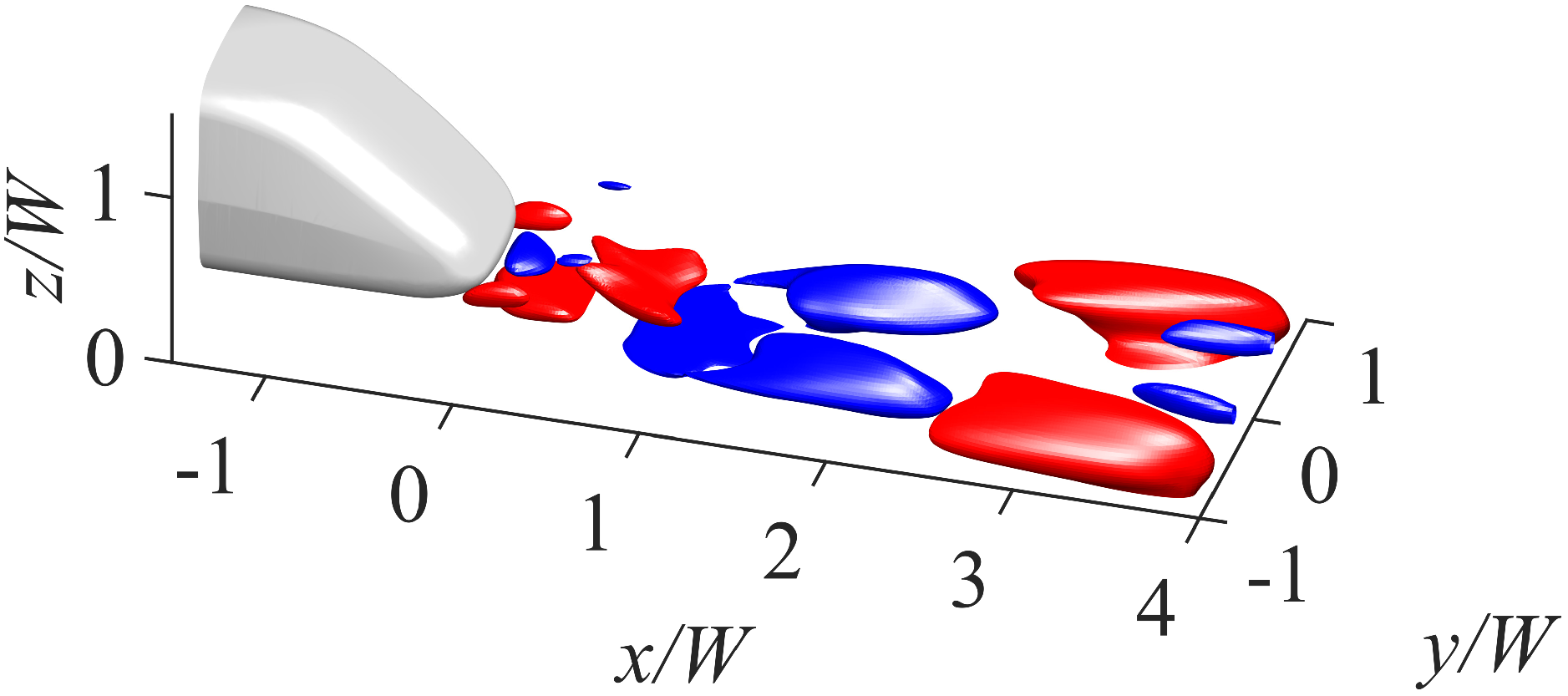}
  \end{subfigure}
  \begin{subfigure}{0.49\textwidth}
  \subcaption{$\omega=6.874$}
  \includegraphics[width=1\textwidth]{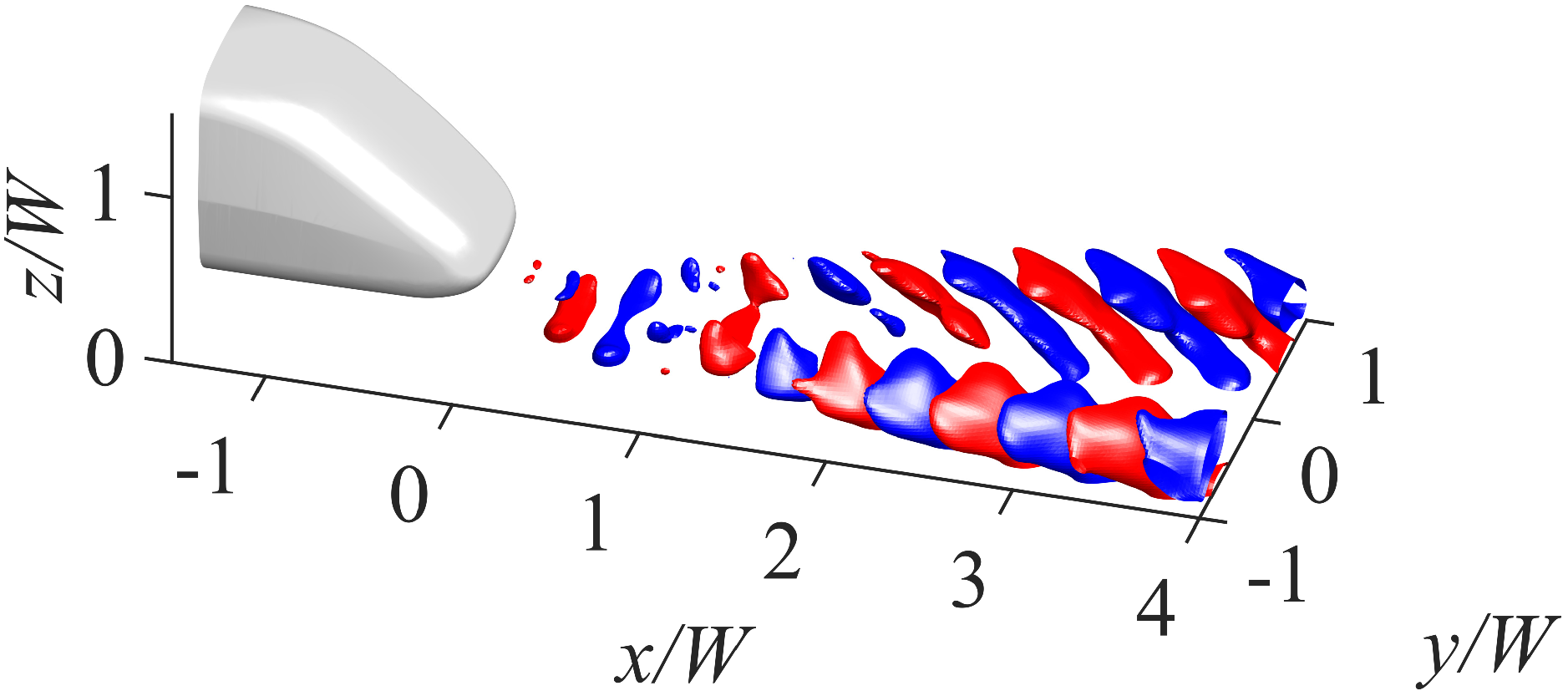}
  \end{subfigure}
  \caption{
  Spatial distributions of the leading symmetric SPOD modes visualized based on isosurface of the streamwise velocity components.
  }
  \label{fig:ModeShape}
\end{figure}
{In figure \ref{fig:ModeShape}, the spatial distributions of the leading symmetric SPOD modes at $\omega=3.437$, $\omega=1.718$, and $\omega=6.874$ are visualized based on the isosurfaces of the streamwise velocity components. A common feature of these modes is that, they all behave very differently between the near and far wake regions, due to the complexity of the mean flow. For the leading SPOD mode at $\omega=3.437$, coherent vortex shedding is observed from the recirculation region just behind the train tail. The generated coherent structures move downstream and gradually approach the bottom boundary due to effect of the moving ground \citep{wang_2023}. Once fully attached, they vanish and separate at the symmetry plane to evolve into far wake coherent structures with nearly constant streamwise wavelength. For the leading mode at $\omega=1.718$, both the alternate shedding of the spanwise vortex and the co-shedding from the side surfaces is observed in the near wake. Similar to the most dominant SPOD mode, these structures slowly evolve into streamwise wavepackets when propagating to the far wake. Then for the leading mode at $\omega=6.874$, no obvious structures can be identified in the near wake, and the far wake is mainly dominated by tilted wavepackets.\par
Another important observation is that, with the increasing of the frequency from $\omega=1.718$ to $\omega=6.874$, the spatial scales of the coherent structures gradually decrease. This phenomenon is more pronounced in the far wake, where the mean flow is nearly parallel and the coherent structures at different frequencies are all characterized by nearly constant streamwise wavenumbers. In particular, the streamwise wavelengths of the far wake coherent structures at $\omega=1.718$ and $\omega=6.874$ can be observed to be approximately twice and half that of the most dominant SPOD mode, respectively. This phenomenon indicates a linear dispersion relation of the wake flow, which will be further discussed and confirmed in the following content.\par}
\begin{figure}
  \centering
  \begin{subfigure}{0.35\textwidth}
  \subcaption{}
  \includegraphics[width=1\textwidth]{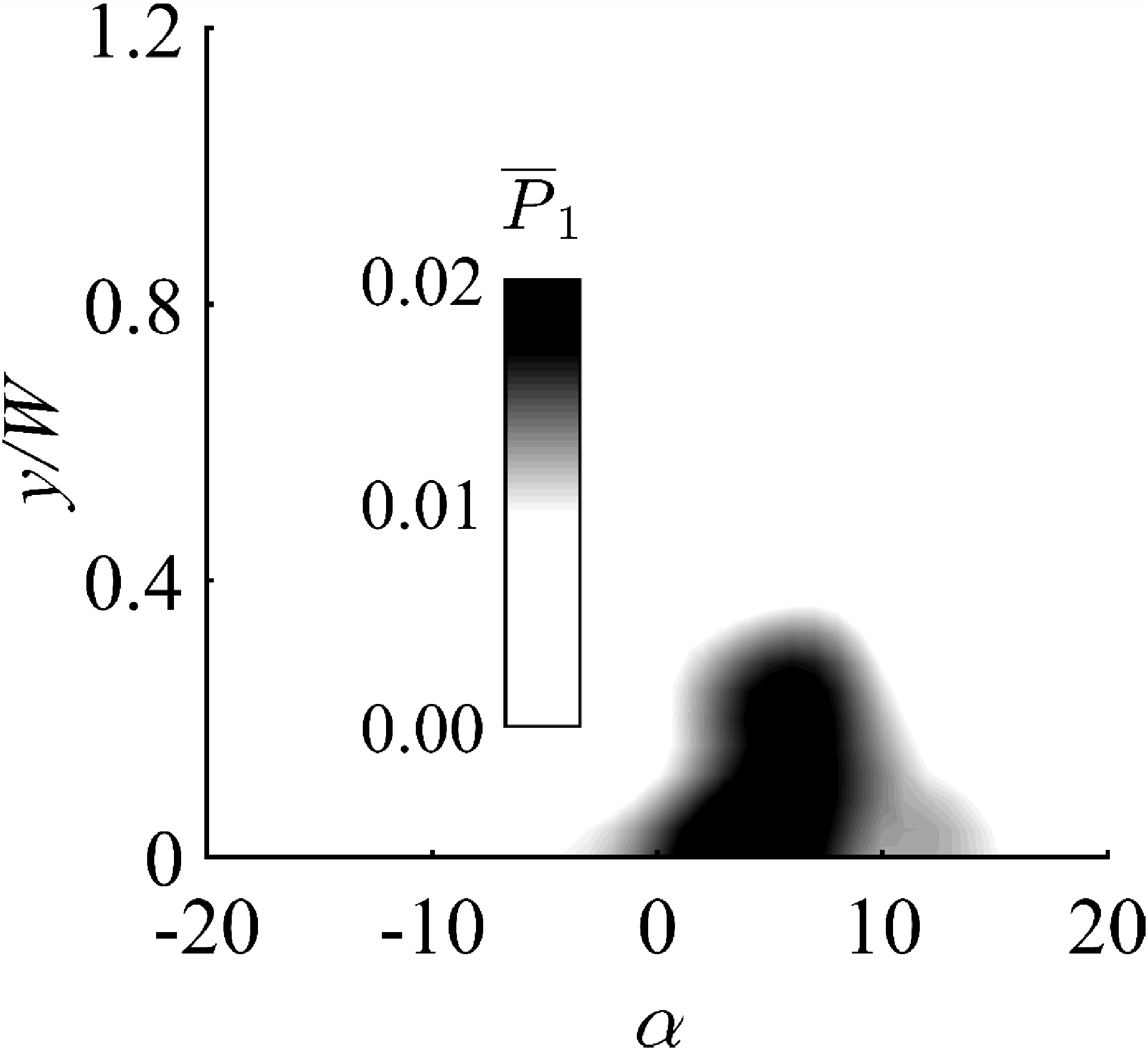}
  \end{subfigure}
  \begin{subfigure}{0.35\textwidth}
  \subcaption{}
  \includegraphics[width=1\textwidth]{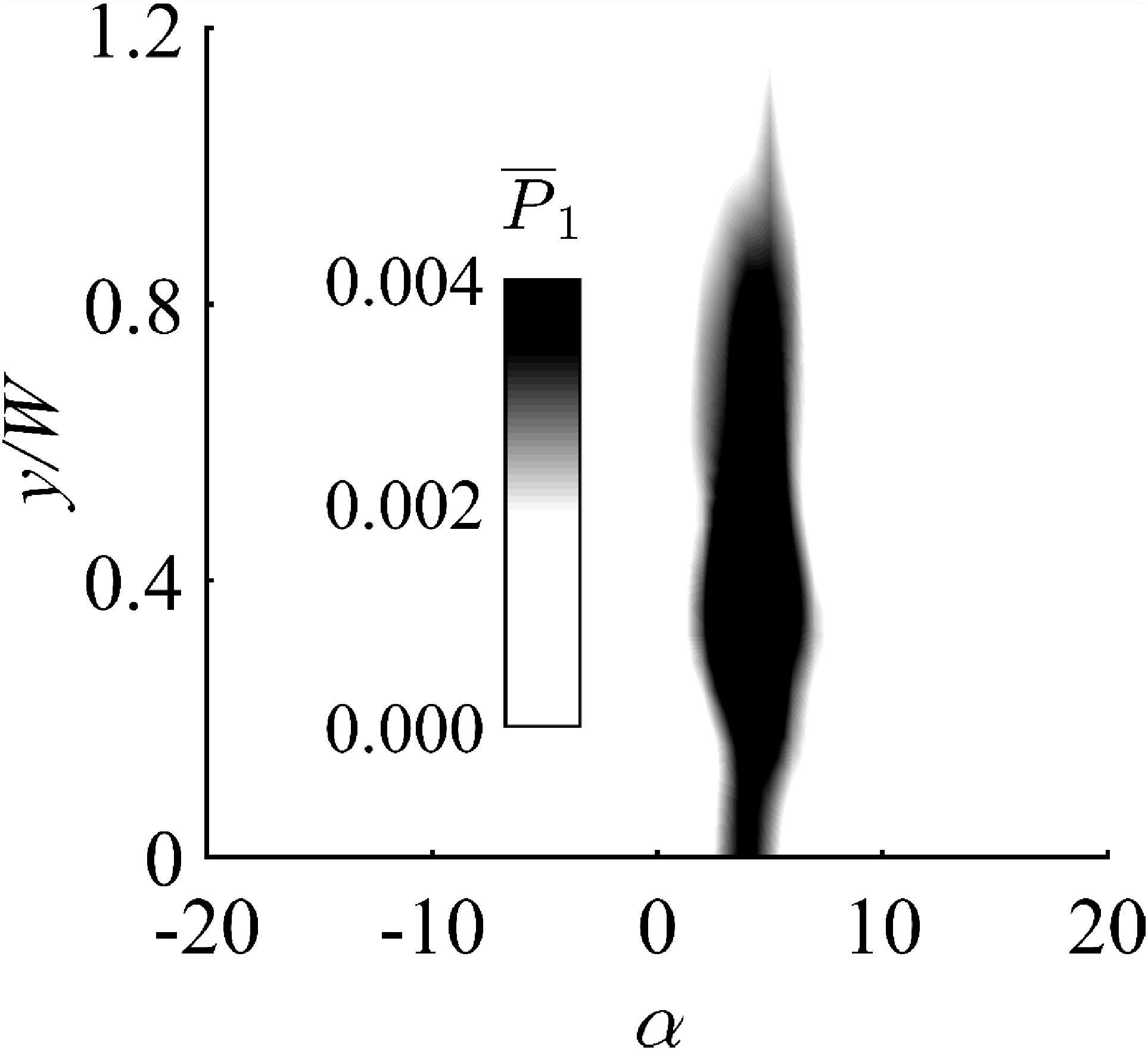}
  \end{subfigure}
  \begin{subfigure}{0.26\textwidth}
  \subcaption{}
  \includegraphics[width=1\textwidth]{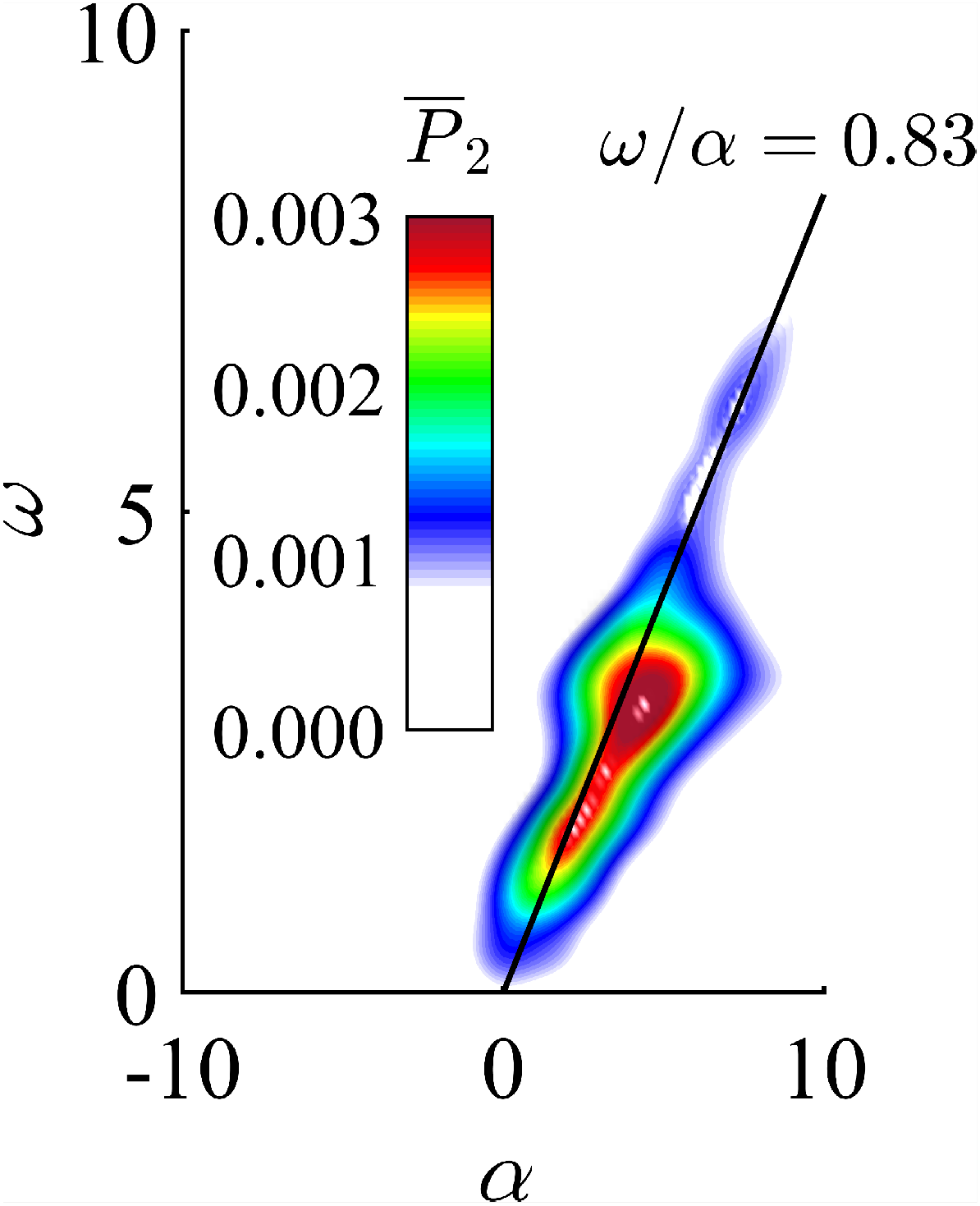}
  \end{subfigure}
  \caption{
  Streamwise wavenumbers of the SPOD modes. (\textit{a}) $0<x/W<1$ at $\omega=3.437$; (\textit{b}) $1<x/W<6.67$ at $\omega=3.437$; (\textit{c}) Angular frequency - streamwise wavenumber diagram.
  }
  \label{fig:Dispersion}
\end{figure}
{To better quantify the frequency-wavenumber characteristics, a streamwise Fourier transform is applied to the streamwise velocity component of the SPOD modes, to convert the signal into the domain of streamwise wavenumber $\alpha$ (normalized by the factor of $1/W$). Due to the dominant role of the symmetric perturbation in the wake, only symmetric modes are considered. First, the power spectral density (PSD) of the streamwise velocity component of the leading mode is averaged along the $z$-axis following the expression
\begin{equation} \label{eqn:P1}
\bar{P}_1(\alpha,\omega,y)=\frac{\int_{z_{\rm{min}}}^{z_{\rm{max}}}\left|\hat{\Phi}_u(\alpha,\omega,y,z)\right|{\rm{d}}z}{z_{\rm{max}}-z_{\rm{min}}}
\end{equation}
where $\hat{\Phi}_u$ denotes the streamwise Fourier transform of the streamwise velocity mode, and $0\leq{z/W}\leq1.3$. Meanwhile, we isolate the mechanism in near wake and far wake regions by using two different spatial window functions \citep{schmidt_towne_rigas_2018} that constrain the signal to $0<x/W<1$ and $1<x/W<6.67$ when the Fourier transform is performed. The results of $\bar{P}_1$ at $\omega=3.437$ using the two different window functions are then shown in figure \ref{fig:Dispersion}(\textit{a}) and figure \ref{fig:Dispersion}(\textit{b}), respectively. In the near-wake region, the maximum PSD is located in the symmetric plane, with $\alpha\approx8$. We can also observe that the high PSD value occurs over a wide range of $\alpha$, which can be attributed to the growth of the wavelength of the spanwise vortex shedding mode. Note that the non-zero values along $\alpha=0$ in figure \ref{fig:Dispersion}(\textit{a}) are due to spectral leakage in the discrete Fourier transform. In the far wake region, the coherent structure has a nearly constant streamwise wavenumber, with the maximum PSD located away from the symmetry plane. This is consistent with what is shown in figure \ref{fig:ModeShape}.\par
Then for the first SPOD modes at all discrete frequency points, a window function including both near-wake and far-wake is used to compute $\bar{P}_1$, followed by the averaging along the $y$-axis, which can be expressed as
\begin{equation} \label{eqn:P2}
\bar{P}_2(\alpha,\omega)=\frac{\int_{y_{\rm{min}}}^{y_{\rm{max}}}\bar{P}_1(\alpha,\omega,y){\rm{d}}y}{y_{\rm{max}}-y_{\rm{min}}}
\end{equation}
with $0\leq{y/W}\leq1.2$, to construct the frequency-wavenumber diagram shown in figure \ref{fig:Dispersion}(\textit{c}). In general, a constant phase velocity of 0.83 can be observed for perturbation waves of all discrete frequencies considered in the research, which confirms the linear dispersion relation of the wake. This value lies between the convective velocity of the streamwise vortex and the outer free stream. Such phase velocity is typical from Kelvin-Helmholtz convective instability found in free-shear flow \citep{schmidt_towne_rigas_2018}. Since the frequency-wavenumber diagram includes both near-wake and far-wake instability waves, the phase velocity tends to decrease at frequencies with a pronounced low rank due to the near-wake mode.\par}

\subsection{Triadic interactions}\label{sec:triadic}
{The dynamic behavior of a nonlinear flow system is significantly influenced by resonance phenomena driven by different mechanisms\citep{tang_2017}. Triadic resonance, resulting from the quadratic nonlinearity of the Navier-Stokes equations, can play an important role in the energy transfer process in turbulent wake flows. Through this mechanism, coherent structures associated with the tonal components of vortex shedding can triadically interact with themselves as well as the background turbulence. As a consequence, the energy of the limit cycle oscillation saturated from a fixed point supercritical bifurcation \citep{sipp_2007} is redistributed over different scales. Therefore turbulent wake flows often exhibit mixed tonal-broadband dynamics, comprising an underlying broadband spectrum and tonal components associated with vortex shedding. The broadband coherent dynamics with several identifiable peaks in the SPOD power spectrum shown in figure \ref{fig:Spectra} indicate such behavior, and is therefore further discussed in this section.\par
Bispectral mode decomposition (BMD, \citep{Schmidt_2020_BMD}) detects triadic interactions by their characteristic phase coupling between two spectral components at frequencies $\omega_1$ and $\omega_2$, and a third frequency at $\omega_3$, obeying the resonance condition $\omega_1\pm\omega_2\pm\omega_3=0$. BMD extends classical bispectral analysis to multidimensional data, thereby simultaneously facilitating the detection of nonlinear triadic interactions from the complex mode bispectrum $\lambda_1(\omega_1,\omega_2)$, as well as the identification of the associated coherent flow structures as bispectral modes. For details on the derivation and implementation, the reader is referred to \citet{Schmidt_2020_BMD}.\par}
\begin{figure}
  \centering
  \begin{subfigure}{0.45\textwidth}
  \subcaption{}
  \includegraphics[width=0.8\textwidth]{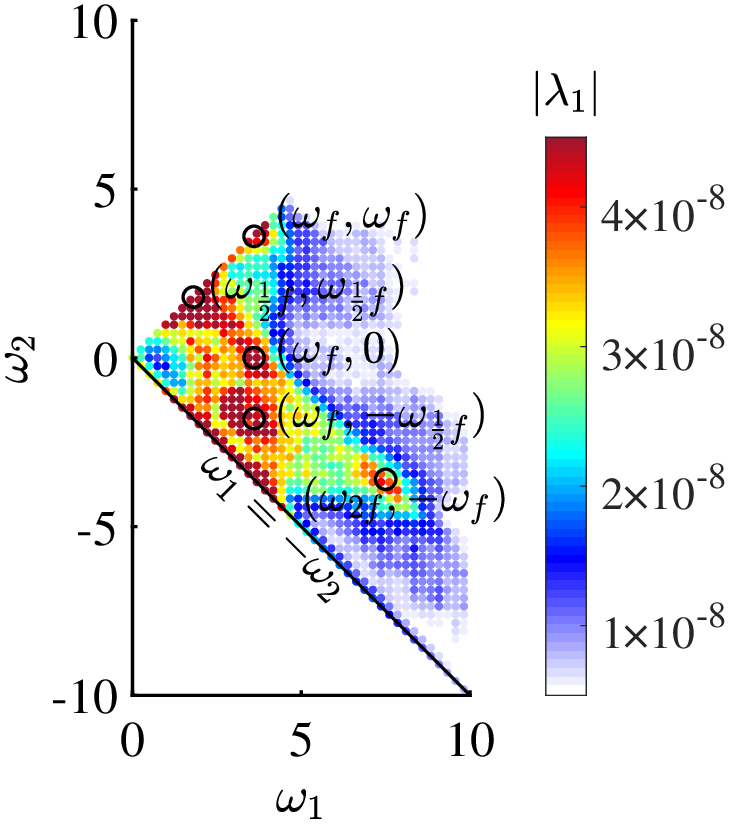}
  \end{subfigure}
  \begin{subfigure}{0.45\textwidth}
  \subcaption{}
  \includegraphics[width=0.8\textwidth]{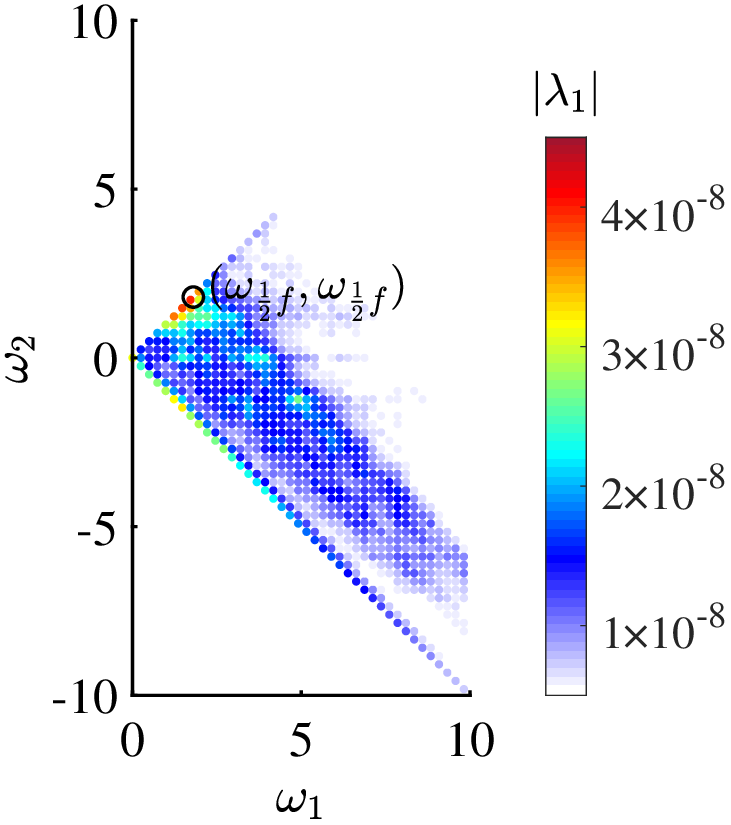}
  \end{subfigure}
  \caption{
  Mode bispectra of triadic interactions in both the sum and difference regions. (\textit{a}) Self-interaction of the symmetric component; (\textit{b}) Self-interaction of the antisymmetric component.
  }
  \label{fig:BMD}
\end{figure}
\begin{figure}
  \centering
  \begin{subfigure}{1\textwidth}
  \subcaption{}
  \includegraphics[width=1\textwidth]{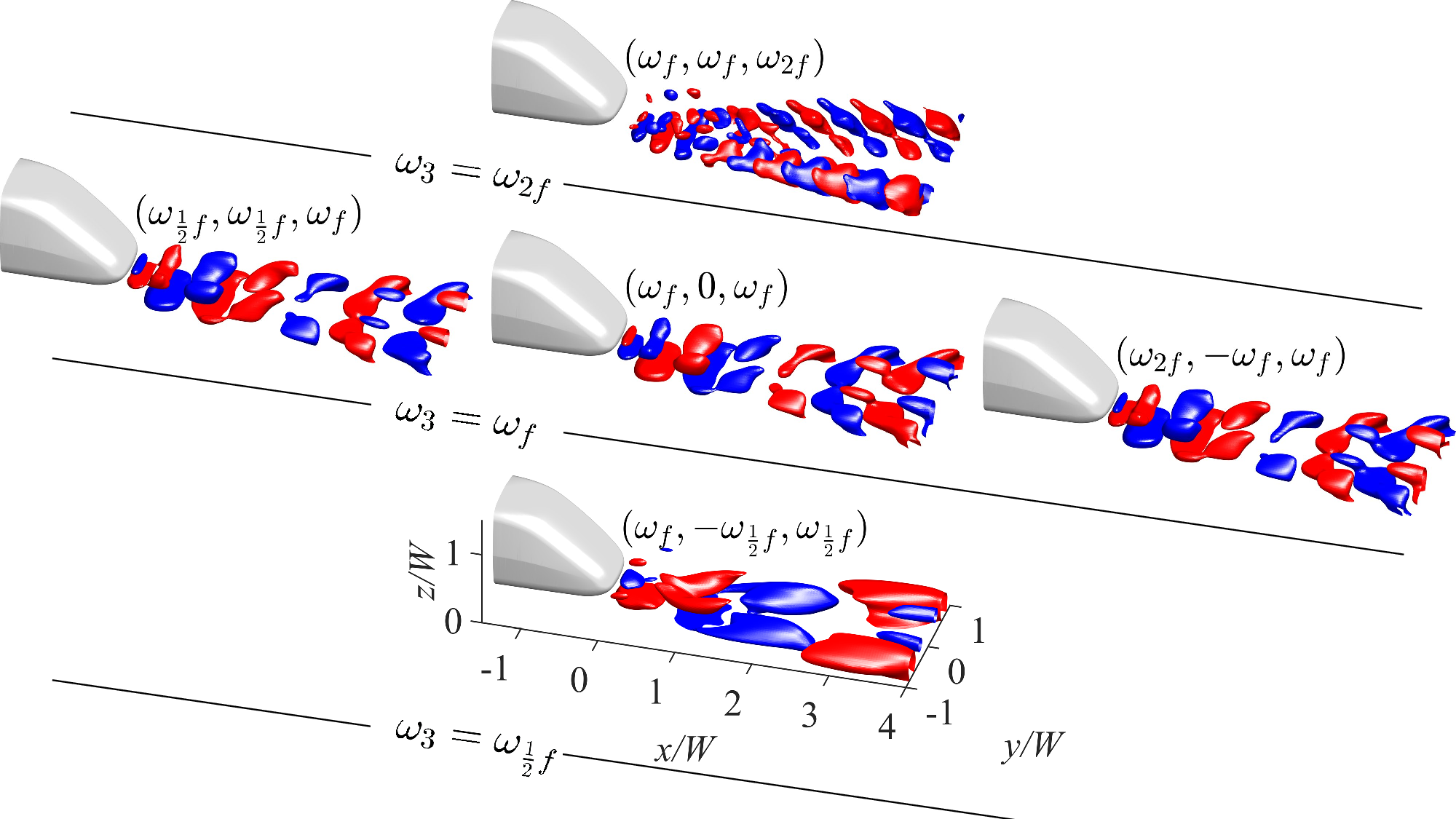}
  \end{subfigure}
  \begin{subfigure}{1\textwidth}
  \subcaption{}
  \includegraphics[width=1\textwidth]{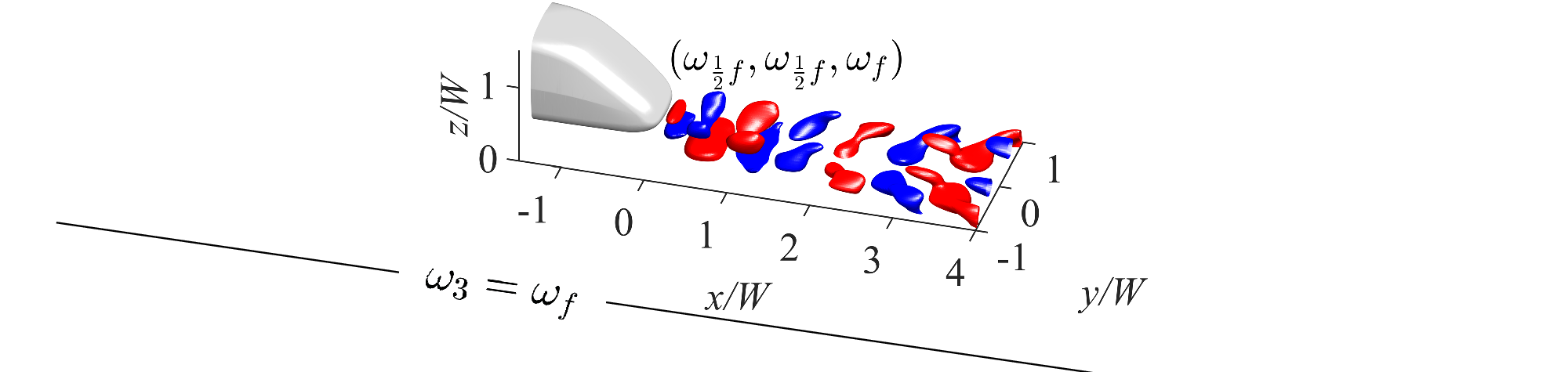}
  \end{subfigure}
  \caption{
  Spatial distribution of the bispectral modes resulted from significant triads, visualized based on isosurface of the streamwise velocity component. (\textit{a}) Self-interaction of the symmetric component; (\textit{b}) Self-interaction of the antisymmetric component.
  }
  \label{fig:bmdModeShape}
\end{figure}
{The mode bispectrum is then computed using the same spectral estimation parameters as in the SPOD of $\S$\ref{sec:SPOD}. Since the flow is subject to the spanwise symmetry, triadic interactions can be further categorized into three different types: self-interaction of the symmetric component, self-interaction of the antisymmetric component, and interaction between the symmetric and the antisymmetric components. Note the third frequency component resulting from the self-interactions of both the symmetric and the antisymmetric components is symmetric, while the third frequency component resulting from the mutual interaction of symmetric and antisymmetric components is antisymmetric. Here we confirm that the mode bispectra of the triadic interaction between the symmetric and antisymmetric components are overall lower in intensity and cannot be identified with distinguishable peaks, therefore will not be further discussed.\par
Hence in figure \ref{fig:BMD}, only the mode bispectra of the symmetric and antisymmetric self-interactions are presented. Consistent with the SPOD spectra, the mode bispectra also appear to be broadband. Here, the self-interaction of the symmetric component shown in figure \ref{fig:BMD}(\textit{a}) is identified with several distinct peaks. We refer to $\omega_f$ as the fundamental frequency detected in SPOD. Conceptually, the triplet $(\omega_f,0,\omega_f)$ on the mode bispectrum can be regarded as the coherent perturbation driven by the mean flow. However, due to the finite sampling frequency, the zero-frequency bin contains unresolved low-frequency components. In this case, the mode bispectrum on the $\omega_1$-axis should not be interpreted \citep{nekkanti_2022}. A local maximum of the distribution can be found that corresponds to the sum-interaction of the fundamental mode with itself, generating the first harmonic at twice the fundamental frequency via the triad $(\omega_f,\omega_f,\omega_{2f})$. At the same time, the doublets $(\omega_{2f},-\omega_f)$ and $(\omega_{\frac{1}{2}f},\omega_{\frac{1}{2}f})$, which include the first harmonic and subharmonic, respectively, interact with $\omega_f$. The triad $(\omega_f,-\omega_{\frac{1}{2}f},\omega_{\frac{1}{2}f})$, analogously, indicates an interaction with the subharmonic frequency. The mean field distortion, indicated by the peak values along line $\omega_1=-\omega_2$ can be identified over a wide frequency range. Meanwhile in figure \ref{fig:BMD}(\textit{b}), a weak triad at $(\omega_{\frac{1}{2}f},\omega_{\frac{1}{2}f})$ can be identified, representing the sum-self-interaction of the antisymmetric subharmonic contributing to the fundamental frequency in the symmetric component.\par
Selected dynamically relevant bispectral modes are shown in figure \ref{fig:bmdModeShape}. Rows in this figure represent the constant third frequencies of $\omega_3=\omega_{\frac{1}{2}f}$, $\omega_3=\omega_{f}$ and $\omega_3=\omega_{2f}$. An important observation is that the modes along the diagonals are similar in shape, and resemble the SPOD modes at corresponding frequencies as shown in figure \ref{fig:ModeShape}. This confirms that the spatial structures involved in or generated by nonlinear triadic interactions with strong phase correlations are also the most energetic coherent structures \citep{nekkanti_2023}. The analysis confirms the expected nonlinear dynamics that are characterized by a fundamental vortex shedding frequency and different interactions of the fundamental mode at $(\omega_f,0,\omega_f)$ and its sub- and super-harmonics, including the mean flow distortion. Also apparent and consistent with the SPOD analysis, the train wake behaves stochastically, with the uncertainty of the fundamental frequency $\omega_f$ leading to broad spectral and bispectral peaks. Nonetheless, the continuous evolution and phase coupling of these structures results in the low-rank behavior observed over a wide range of frequencies, as shown in figure \ref{fig:Spectra}. In addition to shedding light on the different nonlinear interactions that cumulatively lead to peaks in the SPOD energy spectra, the BMD analysis also reveals the self-interaction of the antisymmetric component at the subharmonic frequency in the mode bispectrum in figure \ref{fig:BMD}(\textit{b}) that couples the antisymmetric to the symmetric component.\par}

\section{Physics-based coherent structure modeling}\label{sec:theoreticalmodes}
{Linear global stability analysis is conducted to identify the mechanisms that drive the dominant coherent structure identified in the SPOD. In this work, instead of directly solving the three-dimensional stability equation, we conduct a two-dimensional local analysis to obtain more local information meanwhile reducing computational resource. In this manner, the flow is assumed to be weakly nonparallel in the streamwise direction. Then the full three-dimensional matrix eigenvalue problem is replaced by several local independent matrix problems, each based on the two-dimensional cross-flow planes at different streamwise locations. To determine the global mode, the concept of local convective / absolute instability is applied within the WKBJ approximation \citep{huerre_1990}. To be clear, the approach used in the research does not conceptually correspond to the bi-global analysis \citep{theofilis_2011}, but is similar to the approach used in \citet{huerre_1990,monkewitz_1993,juniper_2014}, in which the bi-global stability is approximated using one-dimensional local analysis.\par}
\subsection{Linearized operator and treatment of the nonparallel flow}\label{sec:LSA}
\begin{figure}
  \centering
  \includegraphics[width=0.75\textwidth]{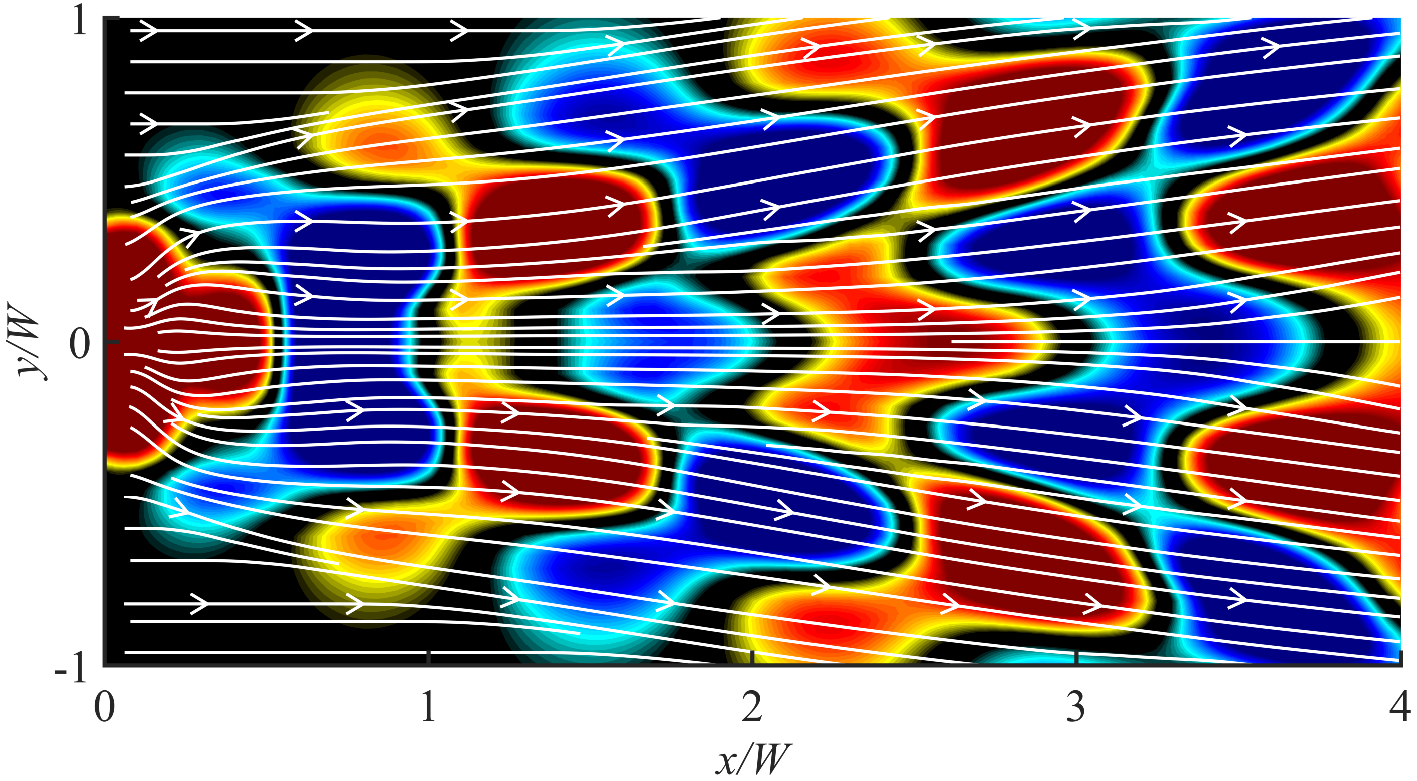}
  \caption{
  Two-dimensional slice of the streamwise component of three-dimensional SPOD mode showing the oblique downstream traveling wavepackets. The streamlines are drawn based on the vector field ($\rm{cos}\theta$, $\rm{sin}\theta$) at each computational node.
  }
  \label{fig:NonParallelTreatment}
\end{figure}

{Coherent structures can be described by the triple decomposition \citep{reynolds_1972}, which leads to a further decomposition of the fluctuating component into coherent and stochastic parts as
\begin{equation} \label{eqn:Triple}
\vb{q}'(\vb{x},t)=\vb{\widetilde{q}}(\vb{x},t)+\vb{q}''(\vb{x},t)
\end{equation}
where $\vb{\widetilde{q}}(\vb{x},t)$ is the coherent fluctuation part and $\vb{q}''(\vb{x},t)$ is the stochastic fluctuation part. This decomposition is
substituted into both the momentum and continuity equations, and both are time-averaged and phase-averaged. Then by subtracting the time-averaged set of
equations from the phase-averaged set of equations, the equations governing the evolution of coherent structures can be written \citep{reynolds_1972}
\begin{equation} \label{eqn:CSNS}
\frac{\partial{\vb{\widetilde{u}}}}{\partial{t}}+(\vb{\widetilde{u}}\cdot\nabla)\vb{\bar{u}}+(\vb{\bar{u}}\cdot\nabla)\vb{\widetilde{u}}=-\nabla{\widetilde{p}}+\nabla\cdot\left(\frac{1}{\R}(\nabla+\nabla^{\rm{T}})\vb{\widetilde{u}}\right)-\nabla\cdot\left(\tau_R+\tau_N\right)
\end{equation}
\begin{equation} \label{eqn:CSC}
\nabla\cdot\vb{\widetilde{u}}=0
\end{equation}
Here $\tau_N$ describes the quadratic interactions of the coherent perturbation. Considering that the energy contribution from this process is higher-order, this term is neglected in the following. The term $\tau_R$ is the fluctuation of the stochastic Reynolds stresses related to the stochastic-coherent interaction, which contributes at leading order, according to the energy considerations of \citet{reynolds_1972}, and is therefore retained in the equation. However, this term is not known a priori and needs to be properly modeled to close the governing equation. In this paper, we use Boussinesq's eddy viscosity model as the closure. The Reynolds stresses are then expressed as
\begin{equation} \label{eqn:RR}
\tau_R=-\nu_t\left(\nabla+\nabla^{\rm{T}}\right)\vb{\widetilde{u}}.
\end{equation}
Here $\nu_t$ is the normalized eddy viscosity, which can be calculated using quantities of the LES mean flow. Since this approach yields an eddy viscosity for each independent Reynolds Stress component, a reasonable compromise is to take a value of $\nu_t$ that minimizes the mean squared error from the six constitutive relations using
\begin{equation} \label{eqn:nut}
\nu_t=\frac{\left<-\overline{\vb{u}'\vb{u}'}+2/3k\vb{I},\bar{\vb{S}}\right>_{\rm{F}}}{2\left<\bar{\vb{S}},\bar{\vb{S}}\right>_{\rm{F}}}
\end{equation}
with $\left<\cdot,\cdot\right>_{\rm{F}}$ denoting the Frobenius inner product, $k$ the kinetic energy, $\vb{I}$ the identity matrix and $\bar{\vb{S}}$ the mean shear strain rate tensor. This approach has been widely used in the linear stability analysis of turbulent flows, as presented in \citet{tammisola_2016,rukes_2016, Kaiser2018,muller_2020,kuhn_2021}.\par
The linearized momentum and continuity equations for the coherent perturbation can be obtained as
\begin{equation} \label{eqn:LNS}
\frac{\partial{\vb{\widetilde{u}}}}{\partial{t}}+(\vb{\widetilde{u}}\cdot\nabla)\vb{\bar{u}}+(\vb{\bar{u}}\cdot\nabla)\vb{\widetilde{u}}=-\nabla{\widetilde{p}}+\nabla\cdot\left(\left(\frac{1}{\R}+\nu_t\right)(\nabla+\nabla^{\rm{T}})\vb{\widetilde{u}}\right)
\end{equation}
\begin{equation} \label{eqn:LC}
\nabla\cdot\vb{\widetilde{u}}=0.
\end{equation}
These equations can be then rewritten as
\begin{equation} \label{eqn:L}
\vb{\mathcal{L}}\vb{\widetilde{q}}=0
\end{equation}
where $\vb{\mathcal{L}}$ is the operator of the linearized equations superimposing the spatial discretization and base state \citep{Paredes_2013}.\par
By assuming that the mean field has much smaller derivatives in the streamwise direction than in the transverse and vertical directions, the system can be  Fourier-transformed in the streamwise direction, following the streamwise weakly nonparallel flow assumption. The coherent perturbation can be then written as
\begin{equation} \label{eqn:qhat}
\widetilde{q}(x,y,z,t)=\hat{q}(y,z){\rm{exp}}\left[{\rm{i}}\left(\alpha{x}-\omega{t}\right)\right]+\rm{c.c}.
\end{equation}
where $\vb{\hat{q}}$ is the complex eigenfunction, $\alpha=\alpha_{\rm{r}}+\rm{i}\alpha_{\rm{i}}$ is the complex streamwise wavenumber, $\omega=\omega_{\rm{r}}+\rm{i}\omega_{\rm{i}}$ is the complex angular frequency, and c.c. is the complex conjugate. Substituting equation \ref{eqn:qhat} into equation \ref{eqn:L} enables stability analysis of the mean field in the cross-flow plane at different streamwise locations. The global stability characteristics can then be recovered based on the concept of absolute/convective instability and the global mode wavemaker \citep{huerre_1990}.\par
However, using the local approach to predict the global mode has been reported to be less accurate than the direct global approach, when the base flow is strongly nonparallel \citep{juniper_2011,Juniper_2015}. In the current case, where a fully developed three-dimensional base flow is considered, the parallel flow assumption is likely to introduce uncertainty into the results. Therefore, in this work, we also make further treatment to approximate the non-parallelism of the three-dimensional base flow.\par
The parallel flow assumption is checked by visualizing the streamwise component of the leading SPOD mode on a horizontal plane as shown in figure \ref{fig:NonParallelTreatment}. Here the coherent perturbations can be observed with clear wavepacket structures; however, they do not travel strictly in the streamwise direction but follow oblique paths in both the near and far wake. This is caused by the downwash flow from the slanted tail surface which gradually separates the wake structures as it propagates downstream, as can be seen in the mean flow structures shown in figure \ref{fig:Meanfield}.\par
We intend to account for the obliqueness of the coherent structure in the linear modeling. To do this, the $x$-dependence of the eigenfunction $\vb{\hat{q}}$ has to be re-considered. Then for a given location $(x_0,y_0,z_0)$, and a streamwise distance $\Delta{x}$ small enough, there exists a set of $(\Delta{y},\Delta{z})$ so that the eigenfunction fulfills
\begin{equation} \label{eqn:flowangle0}
\hat{q}(x_0,y_0,z_0)=\hat{q}(x_0+\Delta{x},y_0+\Delta{y},z_0+\Delta{z})
\end{equation}
This set of $(\Delta{y},\Delta{z})$ is related to the obliqueness of the travelling coherent structure, and equation \ref{eqn:flowangle0} can be replaced by
\begin{equation} \label{eqn:flowangle}
\hat{q}(x_0,y_0,z_0)=\hat{q}(x_0+\Delta{x},y_0+\rm{tan}\theta\Delta{x},z_0+{\rm{tan}}\gamma\Delta{x})
\end{equation}
where $\rm{tan}\theta$ and $\rm{tan}\gamma$ respectively represent the convection direction relative to symmetric plane and horizontal plane. By applying a first-order Taylor expansion to the right-hand side of equation \ref{eqn:flowangle}, the left-hand side can be canceled and the expression can be arranged into
\begin{equation} \label{eqn:xderivative}
\frac{\partial{\vb{\hat{q}}}}{\partial{x}}+{\rm{tan}}\vb{\theta}\frac{\partial{\vb{\hat{q}}}}{\partial{y}}+{\rm{tan}}\vb{\gamma}\frac{\partial{\vb{\hat{q}}}}{\partial{z}}=0
\end{equation}
Therefore, an $x$-derivative applied to the state vector $\vb{\hat{q}}$ is used to account for the oblique traveling direction.\par
To determine $\rm{tan}\theta$ and $\rm{tan}\gamma$, we replace $\vb{\hat{q}}$ with mean field quantities $\vb{\bar{q}}$ in equation \ref{eqn:xderivative}, by assuming that the perturbation waves follow the mean flow convection. Note that, for each node in the computational domain, the four flow variables are used to construct the linear equation system for $\rm{tan}\theta$ and $\rm{tan}\gamma$ and the final results are  obtained using the least-squares solution of the overdetermined linear equation system.\par
Finally, the convection direction calculated from the mean field is validated by comparison with the oblique path of the SPOD mode. In figure \ref{fig:NonParallelTreatment}, streamlines based on the calculated local angle $\theta$ are visualized, by decomposing $\rm{tan}\theta$ at each computational node into streamwise ($\rm{cos}\theta$) and transverse ($\rm{sin}\theta$) components using trigonometric functions. It can be observed that the vector field agrees well with the traveling direction of the SPOD wavepackets. Therefore, this $x$-derivative is assumed to be reasonable to account for the obliqueness of the coherent structure in the linear modeling. Note that the results of the stability analysis without the non-parallelism modeling are also computed and presented in Appendix \ref{sec:stabilityresult} for comparison.\par}

\subsection{Spatio-temporal stability approach}\label{sec:spatialtemporalapproach}
{For the linear stability analysis, the linear operator $\vb{\mathcal{L}}$ is rearranged to construct the generalized eigenvalue problem. In this work, both the temporal and spatial stability formulation is needed. Therefore, either the temporal stability form
\begin{equation} \label{eqn:temporal}
\vb{\mathcal{A}}(x,\alpha)\vb{\hat{q}}=\omega\vb{\mathcal{B}}(x)\vb{\hat{q}}
\end{equation}
or the spatial stability form
\begin{equation} \label{eqn:spatial}
\vb{\mathcal{A}}(x,\omega)\vb{\hat{q}}=\alpha\vb{\mathcal{B}}(x)\vb{\hat{q}}
\end{equation}
are constructed.\par
In the temporal stability form (equation \ref{eqn:temporal}), the streamwise wavenumber $\alpha$ is fixed to a real value, and the eigenvalue problem is solved for a complex $\omega$, the real part of which is the angular frequency and the imaginary part is the temporal growth/decay rate. On the contrary, in the spatial stability form (equation \ref{eqn:spatial}), a real angular frequency $\omega$ is imposed to search for the complex $\alpha$, where the real part corresponds to the streamwise wavenumber, while the imaginary part is the spatial amplification/damping rate. Note that, in the spatial stability equation, quadratic terms appear with respect to $\alpha$. This problem is solved using the companion matrix method \citep{bridges_1984}, which increases the size of the eigenvalue problem compared to the temporal analysis. The complete expression of the operators $\vb{\mathcal{A}}$ and $\vb{\mathcal{B}}$ for both temporal and spatial analysis can be found in Appendix \ref{sec:LSAoperator}.\par
To determine the global stability of the mean flow from a local analysis, a so-called spatio-temporal analysis using equation \ref{eqn:temporal} must be performed to distinguish between convective and absolute instability \citep{huerre_1990}. In this case, both $\alpha$ and $\omega$ are complex-valued. Conceptually, the flow is said to be stable if all perturbations decay in time throughout the entire domain after a localized impulse. Convectively unstable flow gives rise to perturbations that grow in time but convect away from the impulse location, so that the perturbations eventually decay to zero at each spatial location in the long time limit. For an absolutely unstable flow, the perturbations grow both upstream and downstream of the impulse location, contaminating the whole spatial domain in the long time limit.\par
Based on the previous definitions, the convective/absolute nature of a local velocity profile can be determined from the time-asymptotic behavior of the perturbations that remain at the impulse location, that is, for perturbations with zero group velocity: $\partial\omega/\partial\alpha=0$, which is the definition of a saddle point in the complex $\alpha-$plane. Therefore, valid saddle points on the complex $\alpha$-plane that satisfy the Briggs-Bers pinch-point criterion \citep{briggs_1964} must be identified. Therefore, the flow is locally absolutely unstable if the absolute growth rate, given by the imaginary part of $\omega$ at the most unstable valid saddle point, is positive. If the absolute growth rate is negative, the flow is convectively unstable or stable \citep{huerre_1990,rees_2010,juniper_2014,rukes_2016,Kaiser2018,demange_2020}.\par
In a spatially developing flow, a region of absolute instability is a necessary (but not sufficient) condition for global instability \citep{monkewitz_1993,chomaz_2005}. To further link the local absolute instability to global instability, the absolute growth rate needs to be tracked along the streamwise direction to determine the wavemaker, the location where the global mode is selected. This method has been extensively used for one-dimensional local stability analysis in comparison with bi-global stability analysis \citep{giannetti_2007,juniper_2011,Juniper_2015,Kaiser2018}.\par}

\subsection{Solving the eigenvalue problem}\label{sec:eigproblem}
{To solve the eigenvalue problem numerically, cross-flow planes with the dimensions of $0{\leq}y/H{\leq}1.3$ and $0{\leq}z/H{\leq}1.3$ are discretized using Chebyshev spectral collocation methods. This approach has been successively applied to linear stability analysis by \citet{khorrami_1991,parras_2007,oberleithner_2011,demange_2020}. Detailed descriptions or practical guides to spectral collocation methods can be found in \citet{khorrami_1989,trefethen_2000}.\par
To reduce the numerical cost, we further exploit the symmetry of the mean field with respect to the vertical $x$-$z$ plane. This allows us to use only half of the wake plane instead of the full plane, to compute only transversely symmetric or antisymmetric eigenmodes when appropriate boundary conditions are applied \citep{zampogna_2023}. As shown in $\S$\ref{sec:empiricalmodes}, the symmetric perturbations are dominant and potentially related to a global instability, so only symmetric eigenmodes are considered. The corresponding boundary conditions are  given as
\begin{subequations} \label{eqn:directBC}
\begin{align}
&\frac{\partial{\vb{\hat{u}}}}{\partial{y}}=\frac{\partial{\vb{\hat{w}}}}{\partial{y}}=\frac{\partial{\vb{\hat{p}}}}{\partial{y}}=0,\vb{\hat{v}}=0 \ {\rm{on}} \ y/W=0\\
&\frac{\partial{\vb{\hat{u}}}}{\partial{y}}=\frac{\partial{\vb{\hat{v}}}}{\partial{y}}=\frac{\partial{\vb{\hat{w}}}}{\partial{y}}=\frac{\partial{\vb{\hat{p}}}}{\partial{y}}=0 \ {\rm{on}} \ y/W=1.3\\
&\vb{\hat{u}}=\vb{\hat{v}}=\vb{\hat{w}}=0,\frac{\partial{\vb{\hat{p}}}}{\partial{z}}=0 \ {\rm{on}} \ z/W=0\\
&\vb{\hat{u}}=\vb{\hat{v}}=\vb{\hat{w}}=\vb{\hat{p}}=0 \ {\rm{on}} \ z/W=1.3
\end{align}
\end{subequations}
Here, equation \ref{eqn:directBC}$a$ determines eigenmodes to be transversely symmetric. At the wall, homogeneous Dirichlet boundary conditions are imposed for the velocity components. For pressure, by substituting the homogeneous Dirichlet conditions for velocity into equation \ref{eqn:LNS}, a compatibility condition can be obtained \citep{theofilis_2004}. Here, assuming $\partial^2{\vb{\hat{w}}}/\partial{z}^{2}=0$ gives a homogeneous Neumann condition for the pressure. For far-field boundaries, the upper boundary is set to a homogeneous Dirichlet condition, while the side boundary is set to a homogeneous Neumann condition which is necessary to predict far wake eigenmodes.\par
The Krylov-Schur algorithm \citep{stewart_2002}, which serves as an improvement on traditional Krylov subspace methods such as the Arnoldi and Lanczos algorithms, is used to obtain a subset of solutions to the eigenvalue problem. This requires an initial guess of the physical eigenvalue, which can be derived from the dispersion relation shown in $\S$\ref{sec:SPODresults}. To discard spurious eigenmodes caused by the discretization, two criteria are applied: First, all eigenmodes that do not diminish when approaching the upper boundary are discarded. Second, since spurious eigenvalues are very sensitive to discretization, a convergence study of eigenvalues computed using different grid resolutions is used as a criterion for filtering spurious eigenvalues.\par}
\begin{figure}
  \centering
  \begin{subfigure}{0.49\textwidth}
  \subcaption{}
  \includegraphics[ width=0.90\textwidth]{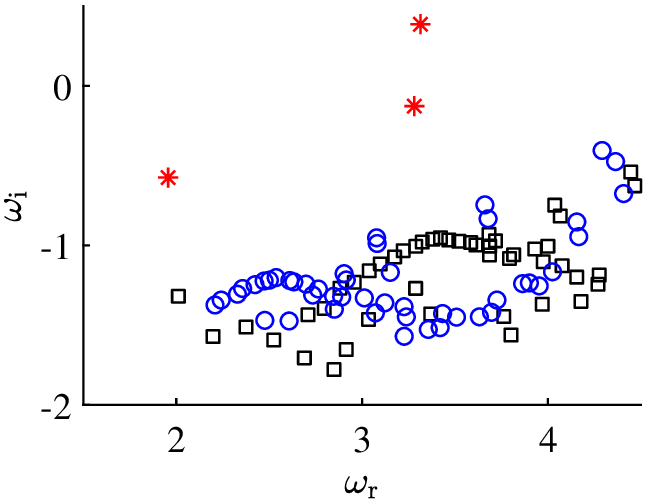}
  \end{subfigure}
  \begin{subfigure}{0.49\textwidth}
  \subcaption{}
  \includegraphics[width=0.90\textwidth]{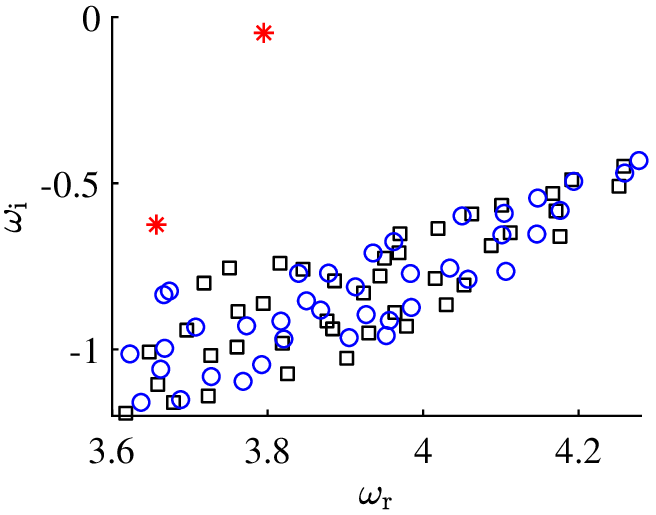}
  \end{subfigure}
  \caption{
  The spectrum of eigenvalues obtained with discretization points of 8,100 (\scalebox{0.6}{$\square$}) and 10,000 ($\color{blue}\circ$), the physical eigenvalues are marked with $\color{red}\ast$. (\textit{a}) Results based on cross-flow plane at $X/W=0.1$ and imposing $\alpha=10$. (\textit{b}) Results based on cross-flow plane at $X/W=3.5$ and imposing $\alpha=5$.
  }
  \label{fig:Eigproblem}
\end{figure}
{The results of the convergence study based on temporal stability analysis are shown in figure \ref{fig:Eigproblem}. Two representative cross-flow planes, located in the near-wake and far-wake, respectively, are considered. Real streamwise wavenumbers of 10 and 5 are, respectively, imposed in the two eigenvalue problems, which are then discretized using two different grid resolutions and solved for the subset of the eigenvalue spectrum. Note that the real streamwise wavenumbers are chosen according to the wavelength distribution of the dominant SPOD mode shown in figure \ref{fig:Dispersion}, so that the SPOD peak frequency can be set as the initial value of the Krylov-Schur iteration procedures. As shown in figure \ref{fig:Eigproblem}, both spectra feature continuous branches and a set of discrete modes. In general, continuous branches are made up of spurious eigenmodes caused by discretization and physical eigenmodes that are highly stable.\par
It can be observed that the locations of these eigenmodes in continuous branches vary significantly when the resolution of the grid is changed. On the contrary, the locations of the discrete modes remain almost stationary with different discretization settings; hence, the discrete modes are considered as physical eigenmodes that can potentially contribute to global instability. In particular, the near-wake cross-flow plane features one unstable mode and two stable modes, while the far-wake plane features two stable modes. Note that these physical eigenvalues do not necessarily correspond between the near-wake and far-wake cross-flow planes. When tracking along the streamwise direction, the physical eigenvalues may become highly stable and fall into the spurious region, and new physical eigenvalues may emerge due to the complexity of the base flow.}

\subsection{Absolute/convective stability analysis}\label{sec:absoluteinstability}
\subsubsection{Streamwise evolution of temporal growth rate}\label{sec:temporalanalysis}
\begin{figure}
  \centering
  \begin{subfigure}{0.49\textwidth}
  \subcaption{}
  \includegraphics[ width=0.90\textwidth]{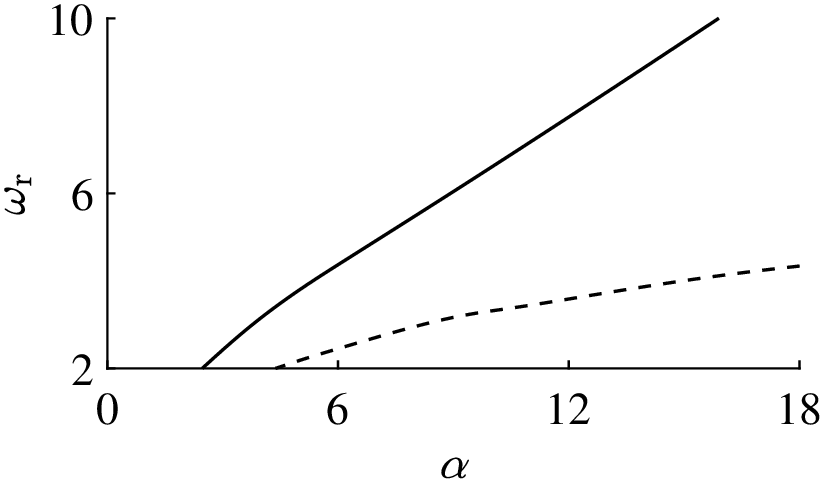}
  \end{subfigure}
  \begin{subfigure}{0.49\textwidth}
  \subcaption{}
  \includegraphics[width=0.90\textwidth]{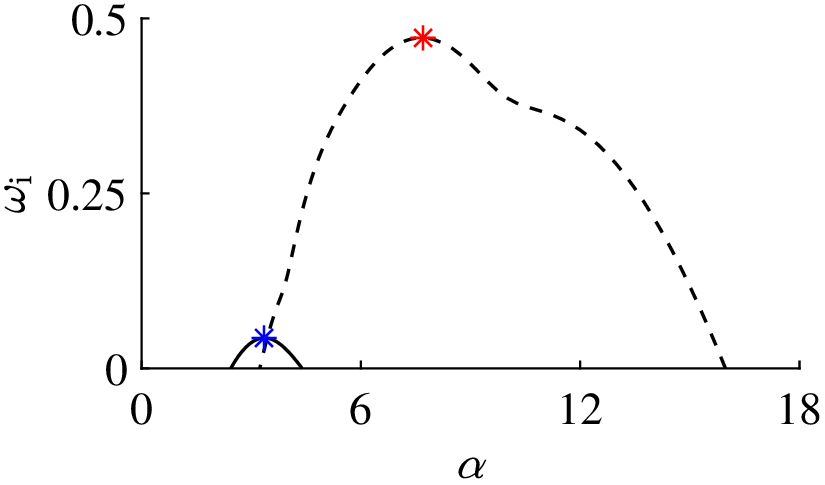}
  \end{subfigure}
  \caption{
  Frequency (\textit{a}) and temporal growth rate (\textit{b}) of the of the temporally most unstable modes at $x/H=0.1$ (dashed line) and $x/H=3.5$ (solid line) as a function of real wavenumber. The maximum temporal growth rates of the two eigenmodes are marked by asterisks.
  }
  \label{fig:OmegaCurve}
\end{figure}

{Before considering the absolute / convective nature of the instability, it is necessary to know which part of the flow is temporally unstable. To this end, the complex frequencies of the previously identified physical modes at $x/H=0.1$ and $x/H=3.5$ are tracked when varying the real streamwise wavenumber. The exact maximum temporal growth rate of each mode is then determined based on a Newton-Raphson iterative method \cite{ypma_1995}. Figure \ref{fig:OmegaCurve} shows the branches of the most unstable modes at the two streamwise locations, with the associated maxima marked by asterisks. These maxima are then tracked along the streamwise direction, with a spacing of $\Delta{x}/W=0.002$, by repeating the iteration procedure. As mentioned above, due to the complexity of the base flow, one physical eigenmode may occur only in a certain range of streamwise location. Therefore, several more cross-flow planes are used to compute the physical eigenvalues, and then the tracking process is repeated to account for the maximum temporal growth rates of all physical eigenmodes in the entire wake.\par}
\begin{figure}
  \centering
  \begin{subfigure}{0.9\textwidth}
  \subcaption{}
  \includegraphics[ width=1\textwidth]{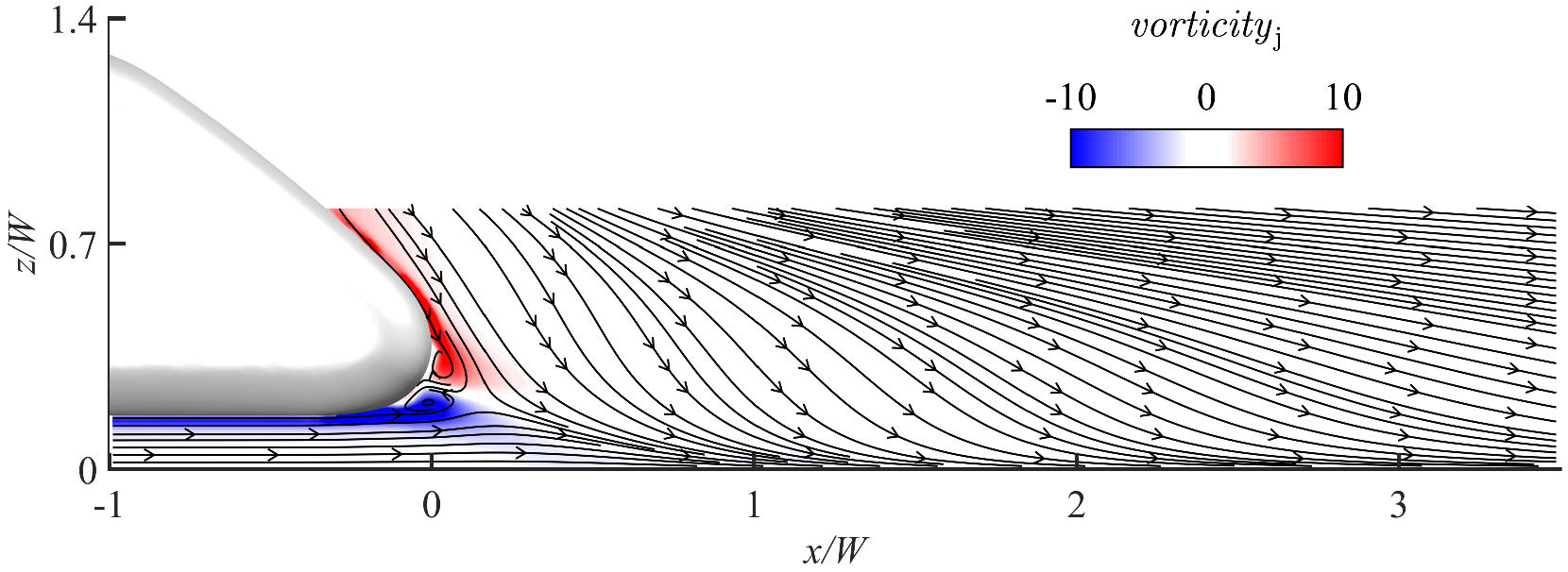}
  \end{subfigure}
  \begin{subfigure}{0.9\textwidth}
  \subcaption{}
  \includegraphics[width=1\textwidth]{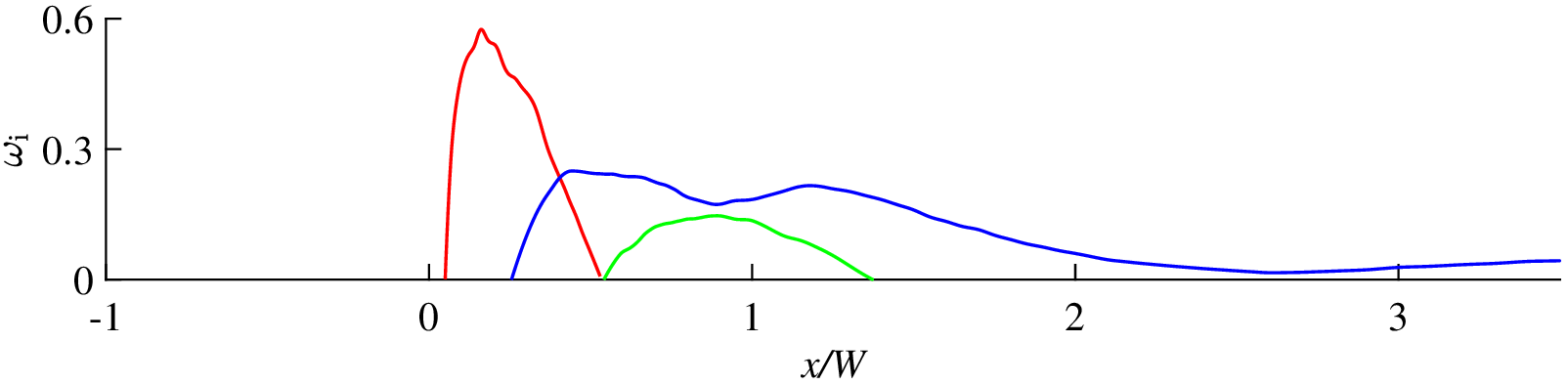}
  \end{subfigure}
  \caption{
  Maximum temporal growth rate as functions of streamwise location. (\textit{a}) Mean flow distribution in the central plane; (\textit{b}) Maximum temporal growth rate of near wake ({\color{red}\rule[0.8mm]{0.5cm}{0.3mm}}), middle wake ({\color{green}\rule[0.8mm]{0.5cm}{0.3mm}}) and far wake ({\color{blue}\rule[0.8mm]{0.5cm}{0.3mm}}) eigenmodes with real $\alpha$ imposed.
  }
  \label{fig:TemporalStability}
\end{figure}
{In figure \ref{fig:TemporalStability}, the maximum temporal growth rates of these physical eigenmodes are displayed as a function of streamwise location. Only the unstable regime ($\omega_{\rm{i}}>0$) is shown here. Three different unstable eigenvalue branches are identified in the entire wake, each dominating in different streamwise regions. Accordingly, the flow becomes temporally unstable very close to the tail of the train and remains unstable in the wake. Based on the spatial locations where these branches become unstable, we conceptually divide the wake into the near, middle, and far wake regimes and name these branches accordingly. It can be found in figure \ref{fig:TemporalStability}(\textit{b}) that the near wake eigenmode is temporally more unstable than all the downstream eigenmodes. According to the mean flow distribution shown in figure \ref{fig:TemporalStability}(\textit{a}), this eigenmode branch is attributed to the transverse recirculation zone located right behind the tail of the train. The far wake eigenmode can be observed across a wide streamwise distance and is located closely to the extension of the streamwise vortex pair. This branch has the largest temporal growth rate at $x/W\sim0.4$, and gradually stabilizes as it extends into the far wake; however, it remains still unstable. The middle wake mode becomes unstable at $0.5<x/W<1.4$. At this location, with respect to the mean field, the flow from the slanted tail surface can be observed attached to the ground according to the figure \ref{fig:TemporalStability}(\textit{a}).\par}

\subsubsection{Spatio-temporal stability analysis in the near wake}\label{sec:spatialtemporalanalysis}
{Spatio-temporal analysis is performed to compute the contour of $\omega$ in the complex $\alpha$-plane, so as to find valid saddle points following the Briggs-Bers criterion. Since the branch in the near-wake region is temporally more unstable than the other branches (figure \ref{fig:TemporalStability}), we start with the near-wake branch.\par}
\begin{figure}
  \centering
  \begin{subfigure}{0.49\textwidth}
  \subcaption{}
  \includegraphics[ width=1\textwidth]{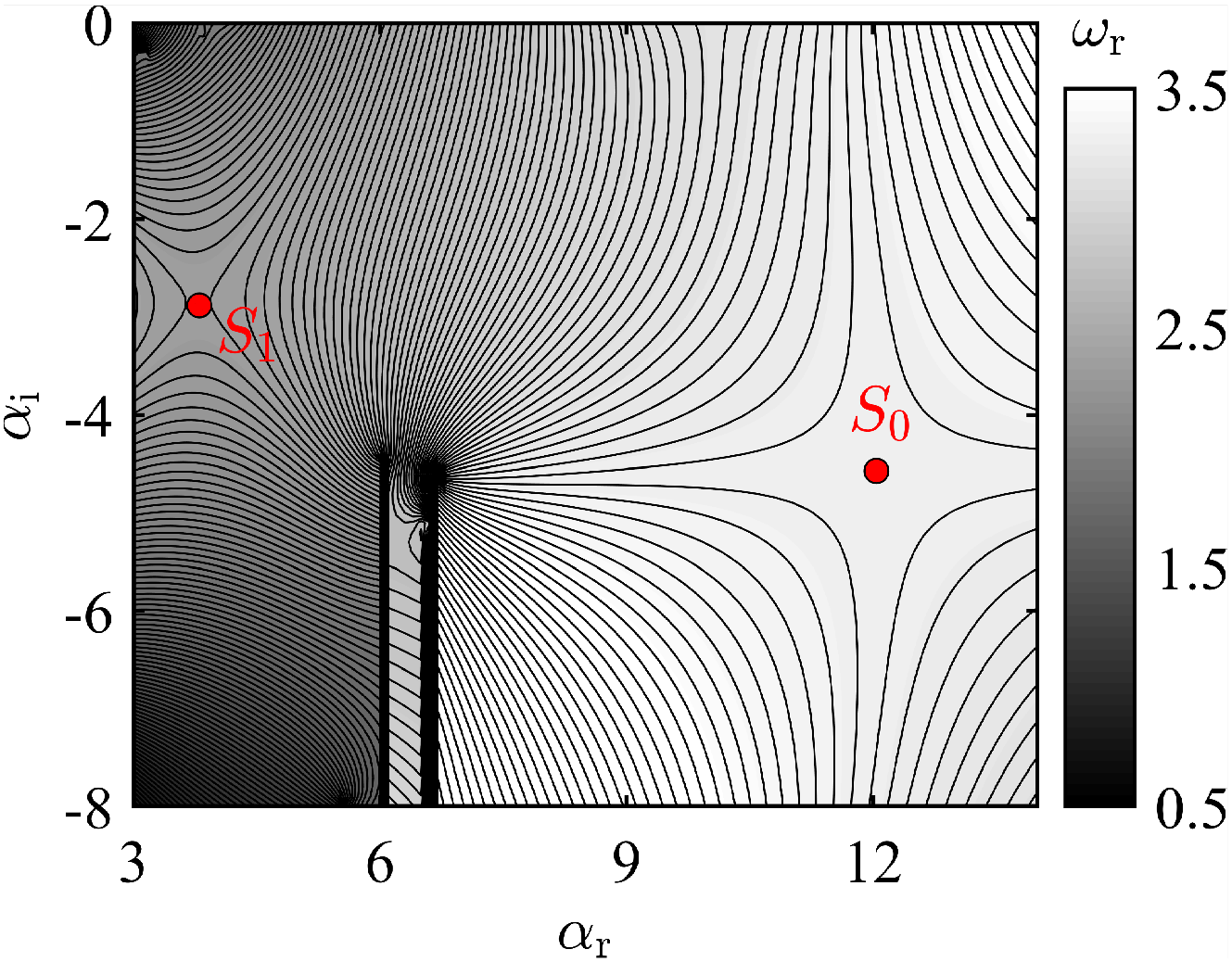}
  \end{subfigure}
  \begin{subfigure}{0.49\textwidth}
  \subcaption{}
  \includegraphics[width=1\textwidth]{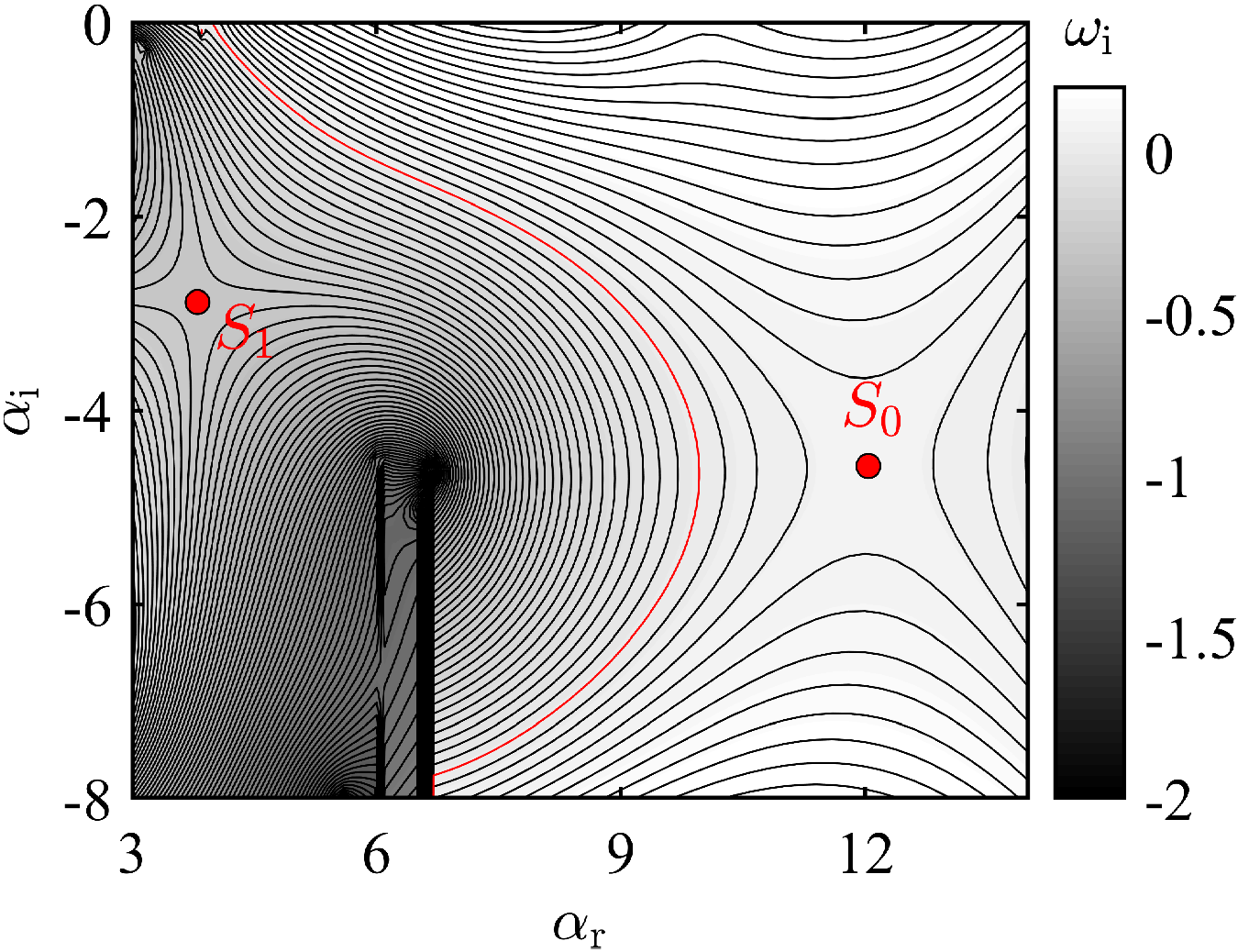}
  \end{subfigure}
  \caption{
  Contour of the complex angular frequency in the complex $\alpha$-plane at $x/W=0.08$. Saddle points on the complex $\alpha$-plane are marked by a red circle. (\textit{a}) The real component; (\textit{b}) The imaginary component, and the $\omega_{\rm{i}}=0$ isocontour is highlighted in red.
  }
  \label{fig:ComplexAlpha}
\end{figure}
{Figure \ref{fig:ComplexAlpha} shows the contour of the complex frequency in the complex $\alpha$-plane, at the streamwise location of $x/W=0.08$. At this location, the near-wake eigenmode reaches its maximum temporal growth rate. In this figure two saddle points are found, $S_0$ and $S_1$. However, valid saddles must be pinched between an $\alpha^+$ and an $\alpha^-$ branch \citep{huerre_1990}. A simple rule based on the Briggs-Bers pinch point criterion \citep{briggs_1964} can be applied by checking the isocontours of the spatio-temporal growth rate (figure \ref{fig:ComplexAlpha}(\textit{b})): Starting from the saddle point, the contours of the growing $\omega_{\rm{i}}$ must reach, respectively, the $\alpha_{\rm{i}}>0$ and $\alpha_{\rm{i}}<0$ half-planes. Here, the validity of both $S_0$ and $S_1$ can be confirmed, with $S_0$ representing the shorter traveling wave at higher frequency, while $S_1$ represents the longer traveling wave at lower frequency. In the long time limit, the nature of the instability is determined by $S_0$, which has a significantly higher absolute growth rate than $S_1$. Then the absolute frequency at this location can be determined as $\omega_0=3.2904+0.0685{\rm{i}}$, which shows that the flow is absolutely unstable at this position.\par
Furthermore, to ensure that all regions with absolute instability have been taken into account, the absolute growth rate has been determined for all unstable branches at several streamwise locations. The near wake branch was the only one to reveal absolute instability.\par}

\subsection{Determining the global mode wavemaker in the near wake}\label{sec:globalwavemaker}
{To determine the global mode, the saddle points are tracked in the streamwise direction to find the global wavemaker. This is done in the local analysis framework based on the frequency selection criterion. Several criteria have been evaluated in \citet{pier_2002}, showing that the criterion for a linear global mode introduced by \citet{chomaz_1991} agrees best with the nonlinear direct numerical simulations. This criterion is obtained from an analytical continuation of the absolute frequency curve into the complex $x$-plane. The wavemaker region is then represented by the saddle point on the complex $x$-plane \citep{huerre_1990} defined as
\begin{equation} \label{eqn:saddleonX}
\frac{\partial\omega_0}{\partial{x}}(x_{\rm{s}})=0
\end{equation}
The global complex angular angular frequency is then given by the value of the absolute frequency at this saddle point
\begin{equation} \label{eqn:globalfreq}
\omega_{\rm{g}}=\omega_0(x_{\rm{s}}).
\end{equation}\par
For this purpose, the saddle point $S_0$ found in figure \ref{fig:TemporalStability} is tracked along the streamwise direction. The three-point Taylor series expansion algorithm \citep{rees_2010_phd} is used to approach $\partial\omega/\partial\alpha=0$ during the tracking process. Then the absolute frequency as a function of streamwise location is analytically continued in the complex $x$-plane using the Pad\'{e} polynomial, which has been shown to be well behaved in the complex plane \citep{cooper_2000}. The Pad\'{e} polynomial takes the form of
\begin{equation} \label{eqn:pade}
f(x)=\frac{P(x)}{Q(x)}=\frac{a_0+a_1x+...+a_nx^n}{1+b_1x+...+b_mx^m}
\end{equation}
To determine the order of the polynomials used in the current work, the procedure described in \citet{juniper_2011} is followed. Polynomials of order 10 have been proven to be sufficient to give a converged approximation of the saddle point.\par}
\begin{figure}
  \centering
  \begin{subfigure}{0.90\textwidth}
  \subcaption{}
  \includegraphics[ width=1\textwidth]{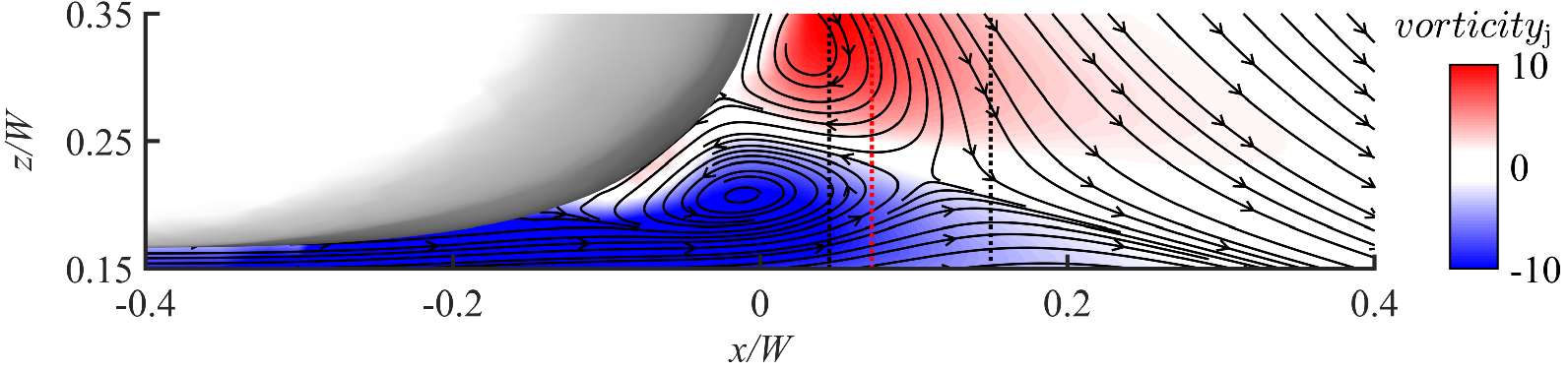}\tikzmark{a0}
  \vspace{0.002cm}
  \end{subfigure}
  \begin{subfigure}{0.90\textwidth}
  \subcaption{}
  \includegraphics[ width=1\textwidth]{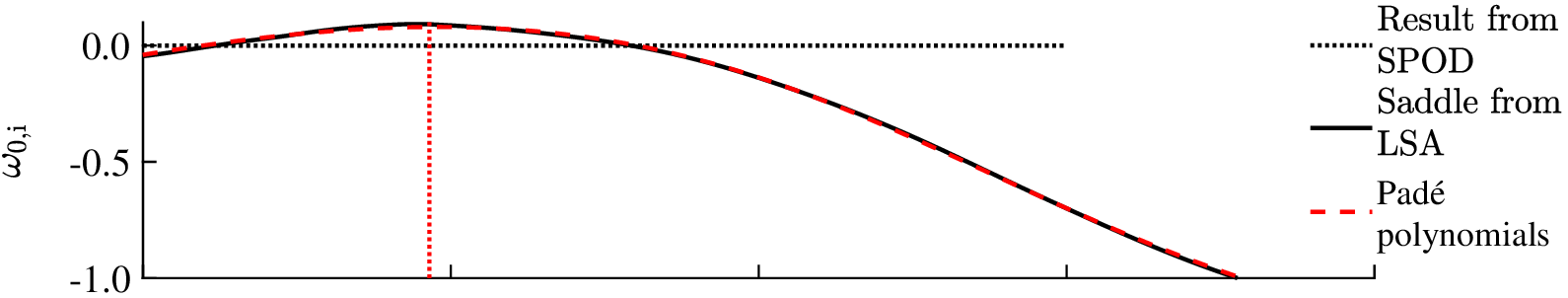}\tikzmark{a1}
  \end{subfigure}
  \begin{subfigure}{0.90\textwidth}
  \subcaption{}
  \includegraphics[ width=1\textwidth]{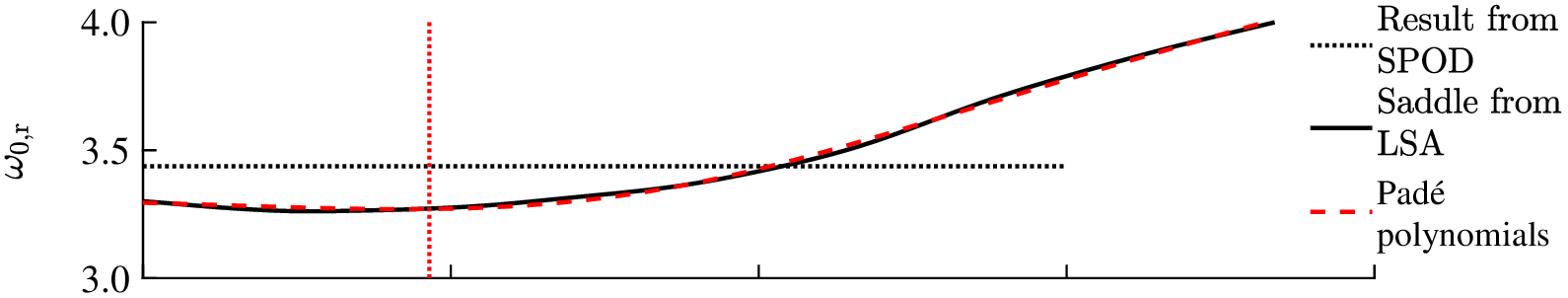}
  \end{subfigure}
  \begin{subfigure}{0.90\textwidth}
  \subcaption{}
  \includegraphics[ width=1\textwidth]{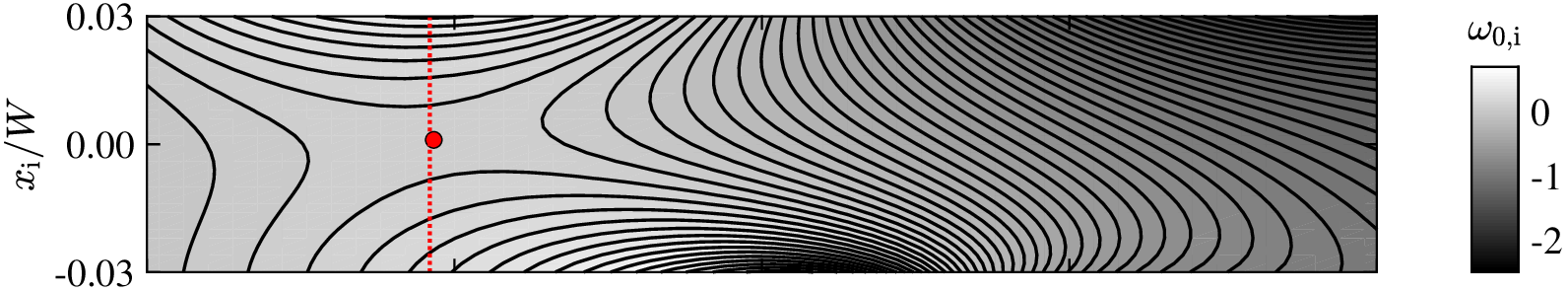}
  \end{subfigure}
  \begin{subfigure}{0.90\textwidth}
  \subcaption{}
  \includegraphics[ width=1\textwidth]{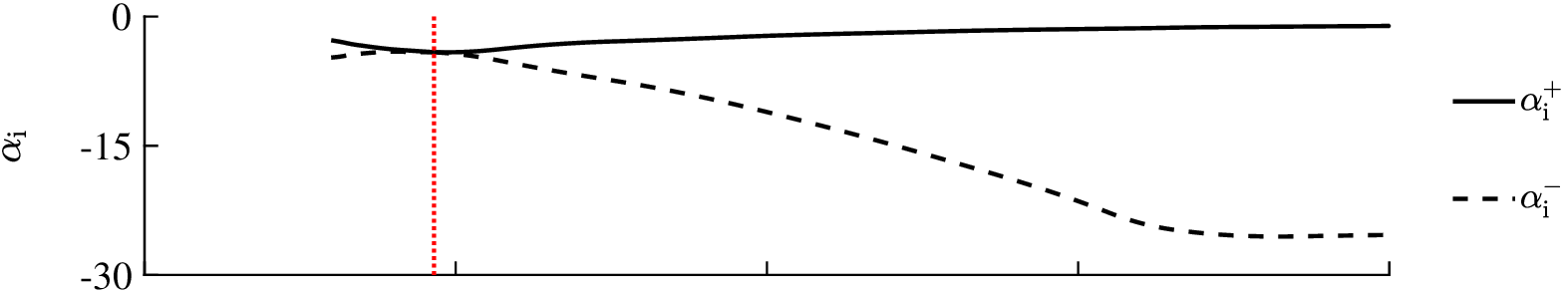}
  \end{subfigure}
  \begin{subfigure}{0.90\textwidth}
  \subcaption{}
  \includegraphics[ width=1\textwidth]{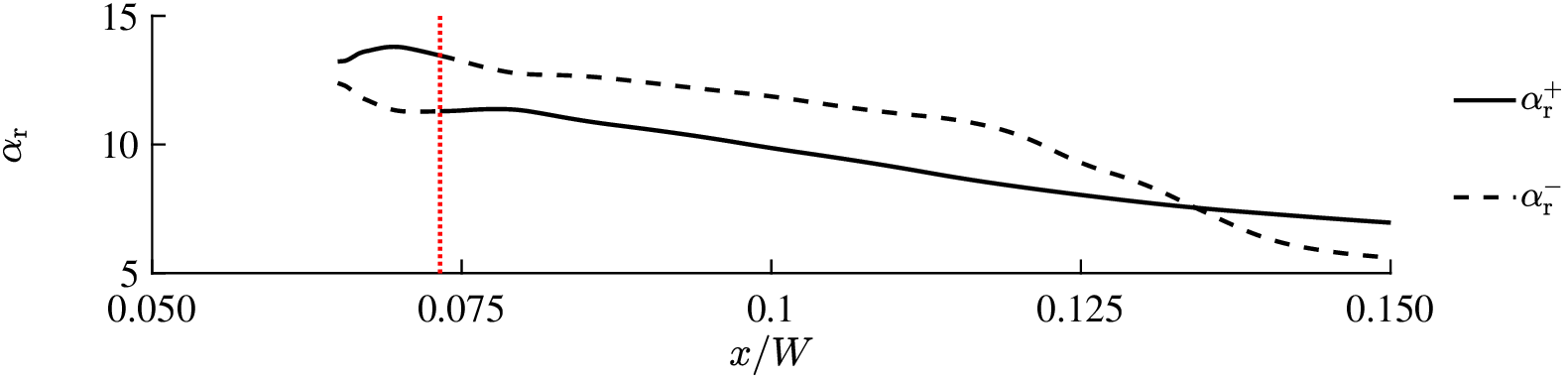}
  \end{subfigure}
  \begin{tikzpicture}[remember picture, overlay, >=latex]
  \draw[arrows=->,black,ultra thin] ($(pic cs:a0)+(-57.5mm,8mm)$) -- ($(pic cs:a1)+(-110mm,24mm)$);
  \draw[arrows=->,black,ultra thin] ($(pic cs:a0)+(-44.5mm,8mm)$) -- ($(pic cs:a1)+(- 15mm,24mm)$);
  \end{tikzpicture}
  \caption{
  (\textit{a}) Streamlines and through-plane vorticity of the mean field in the near wake region. (\textit{b}) Streamwise evolution of the absolute growth rate. (\textit{c}) Streamwise evolution of the absolute angular frequency. (\textit{d}) Contours of $\omega_{\rm{0,i}}$ extruded in the complex $x$-plane, formed by the fitted Pad\'{e} polynomial. The saddle point on the complex $x$-plane is marked by red circle. (\textit{e}) The imaginary component of local complex streamwise wavenumber of $\alpha^+$ and $\alpha^-$ branches calculated at the global frequency $\omega_g$. (\textit{f}) The same as (\textit{e}) with the real component shown. The switching locations between $\alpha^+$ and $\alpha^-$ branches are marked by a red vertical line in each sub-figure.
  }
  \label{fig:Wavemaker}
\end{figure}
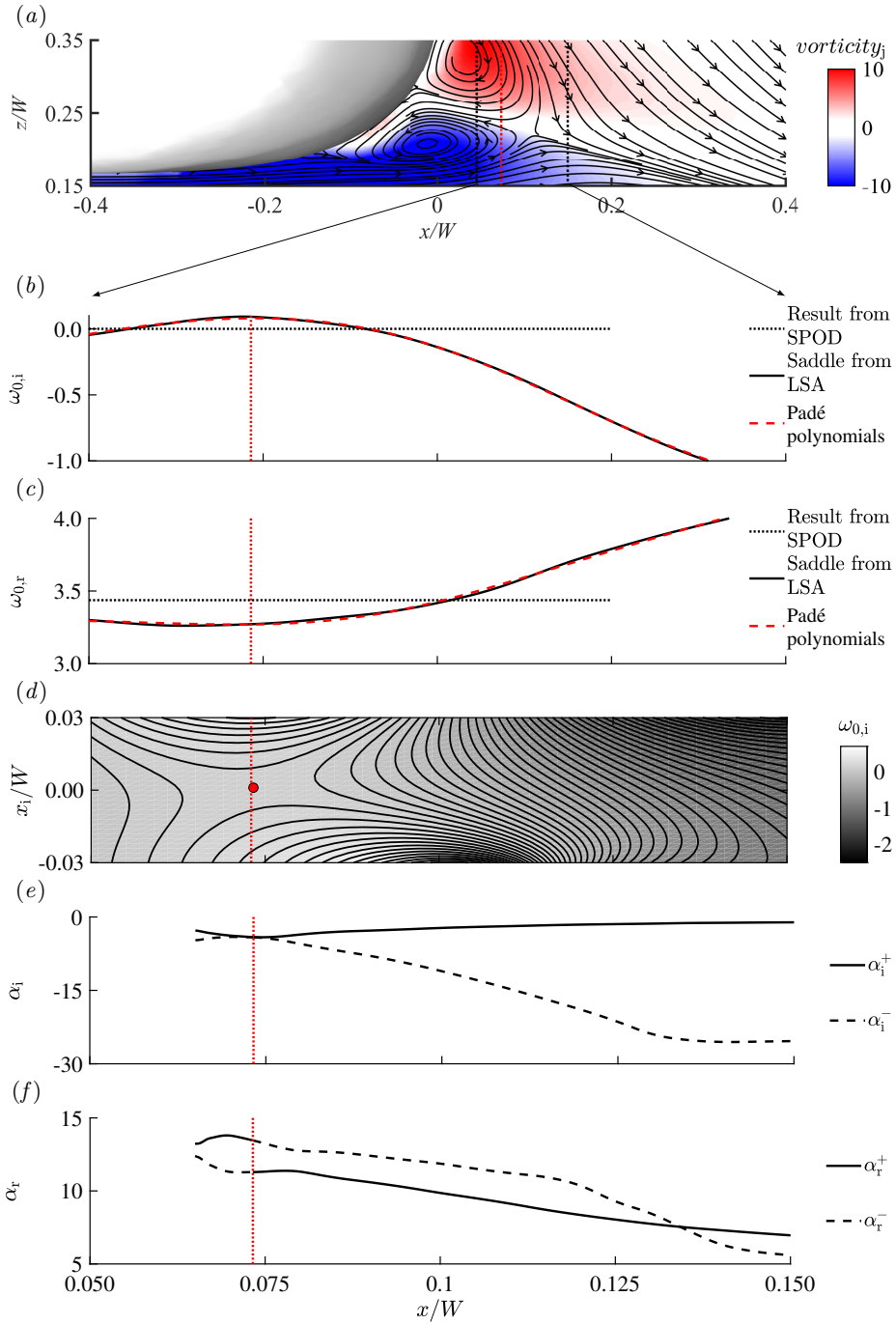
{Figure \ref{fig:Wavemaker} shows the main results of the spatio-temporal stability analysis. For reference, the mean field in the central plane is shown in figure \ref{fig:Wavemaker}(\textit{a}). The following rows include the results of the procedures described above to determine the global instability using local analysis. In figure \ref{fig:Wavemaker}(\textit{b,c}), the imaginary and real parts of the absolute complex angular frequency are shown as a function of streamwise location, respectively. A small region of absolute instability is detected, which is located in the recirculation region. The 10th order Pad\'{e} polynomials show acceptable agreement with the absolute frequency curve of the linear stability analysis. The extrapolated complex $x$-plane is then visualized in figure \ref{fig:Wavemaker}(\textit{d}). The saddle point appears to be very close to the real axis, with $x_{\rm{s}}=0.0733+0.0010{\rm{i}}$, indicating a wavemaker located within the region of absolute instability. The complex global angular frequency associated with this saddle point is $\omega_{\rm{g}}=3.2702+0.0797{\rm{i}}$.\par}
\begin{table}
    \centering
    \begin{tabular}{l c c c}
        & $\omega_{\rm{g,r}}$ & $\omega_{\rm{g,i}}$ & $x_{\rm{s}}$ \\ [1.0ex]
        LSA with non-parallelism approximation & 3.2702 & 0.0797 & 0.0733+0.0010\rm{i} \\
        LSA without non-parallelism approximation & 2.3721 & 0.6082 & 0.0342-0.0046\rm{i} \\
        SPOD & 3.437 & 0 & -- \\
    \end{tabular}
    \caption{Global frequency and wavemaker location predicted by different approaches.}
    \label{tab:Comparison}
\end{table}
{The theoretically predicted global mode frequency can be compared to the frequency of the most energetic SPOD mode, as presented in table \ref{tab:Comparison}. Meanwhile, the global frequency and wavemaker location predicted by the stability analysis without the non-parallelism approximation (see Appendix \ref{sec:stabilityresult}) are also shown for comparison. In general, the global frequency predicted by the stability analysis without the non-parallelism approximation shows a larger deviation from the SPOD result, compared to that with the non-parallelism approximation. By including the spatial variation of the mean field into the stability analysis, the prediction on the global stability has been significantly improved. With the empirical value of $\omega_{\rm{SPOD}}=3.437$, we find that the theoretical prediction deviates by an error of less than $5\%$. This is in very good agreement, given the uncertainties introduced by the non-parallelism and eddy viscosity modeling. Moreover, we expect the global mode to be marginally unstable, representing a limit-cycle oscillation when stability analysis is performed on a temporally averaged mean flow \citep{noack_2003,barkley_2006,oberleithner_2011,rukes_2016,tammisola_2016,Kaiser2018}. However, as reported in \citet{giannetti_2007,juniper_2011,Juniper_2015}, it is a common feature of local analysis to overpredict the growth rate of the linear global mode. In our case, with a growth rate value of 0.0797, the relative error with respect to the real frequency (3.2702) is less than $3\%$, therefore, it is reasonable to say that the mode is marginally unstable. Overall, considering the fact that the flow is not truly parallel and the intrinsic assumption of linearized mean field analysis, it can be concluded that the theoretical mode identifies the dominant SPOD mode as a global mode at limit cycle with remarkable accuracy.\par
The global mode is formed by the switching between the $\alpha^+$ and $\alpha^-$ branches at the global frequency. Therefore, to further truly localize the global mode wavemaker, a spatial stability analysis (equation \ref{eqn:spatial}) imposing $\omega=\omega_{\rm{g}}$ is conducted to compute the $\alpha^+$ and $\alpha^-$ branches \citep{Juniper_2015}. Since the saddle point $x_{\rm{s}}$ on the complex $x$-plane is very close to the real axis, the location of maximum structural sensitivity should also be close to $x_{\rm{s,r}}$. Therefore, in practice, it can be quite straightforward to find the branch pairs by computing the eigenvalue problem at $x/W=x_{\rm{s,r}}$, and following $\alpha_{\rm{i}}^+-\alpha_{\rm{i}}^-\approx0$ in the streamwise direction.\par
In figure \ref{fig:Wavemaker}(\textit{e,f}), the imaginary and real components of the $\alpha^+$ and $\alpha^-$ branches are presented, respectively. The switch between the two branches can be identified to take place at $x/W=0.0731$. The location of the global mode wavemaker is marked by a red vertical line in all sub-figures of figure \ref{fig:Wavemaker}. This location is also within the recirculation region and is nearly identical to the location of the saddle point $x_{\rm{s}}$. The direct global mode then follows the $\alpha^-$ branch upstream of the wavemaker, and the $\alpha^+$ branch downstream. In contrary, the adjoint global mode follows the $\alpha^+$ branch upstream of the wavemaker, and $\alpha^-$ branch downstream \citep{Juniper_2015}. This result also indicates a spatially growing global mode within the recirculation region, with the growth rate gradually decaying to zero as it approaches the downstream boundary of this region.\par
In summary, we identify a global mode with a frequency very close to the dominant SPOD mode and a growth rate close to zero. This suggests that the dominant SPOD mode is a manifestation of a global mode at limit cycle. The wavemaker of this mode is located in the recirculation zone very close to the tail tip. It acts as the origin for the entire coherent wavepacket that propagates far downstream, in a region where the flow is convectively unstable. The spatial shape and mechanisms of the global mode and comparison with the SPOD modes will be shown and discussed in $\S$\ref{sec:comparison}.\par}

\subsection{Structural sensitivity based on adjoint method}\label{sec:adjoint}
{To further analyze where and how intrinsic feedback causes global instability \citep{chomaz_2005,giannetti_2007}, we calculate the structural sensitivity using adjoint linear stability analysis. The structural sensitivity describes the sensitivity of the direct global mode to changes in the linearized Navier-Stokes operator, e.g., base flow modification \citep{marquet_2009}. Therefore, this concept is critical for the development of efficient flow control strategies \citep{giannetti_2007,muller_2020}.\par
In the global framework, structural sensitivity is proportional to the overlap between the direct and adjoint global modes \citep{chomaz_2005}. Following \citet{giannetti_2007}, the $L^2$ norm of the sensitivity tensor is equivalent to the expression
\begin{equation} \label{eqn:Sensi}
\lambda(\vb{x})=\Vert\vb{\hat{m}}(\vb{x})\Vert\Vert\vb{\hat{m}^\dag}(\vb{x})\Vert,
\end{equation}
where $\vb{\hat{m}}(\vb{x})$ and $\vb{\hat{m}^\dag}(\vb{x})$ represent the direct and adjoint momentum vectors, respectively.
In the previous section, the location of the wavemaker was identified by computing the intersection between the $\alpha^+$ and $\alpha^-$ branches (at $x/W=0.0731$). However, the exact location of the maximum structural sensitivity at this cross-flow plane is still unknown. This requires computation of the adjoint mode at this streamwise location. Therefore, the adjoint linear stability analysis is further pursued in this section.\par
Due to the inhomogeneous directions and hence differential dependencies in the linearized operator matrix, taking the Hermitian transpose of the direct operator as the adjoint operator, as has been done in \citet{oberleithner_2014,muller_2020}, does not necessarily apply here. To find the adjoint of the system, the continuous approach is used. First, we define the inner product as
\begin{equation} \label{eqn:innerPd}
\left<\vb{\hat{q}}_1,\vb{\hat{q}}_2\right>\equiv\int_{S}\vb{\hat{q}}_1^H\vb{\hat{q}}_2{\rm{d}}S\equiv\vb{\hat{q}}_1^H\vb{W}\vb{\hat{q}}_2
\end{equation}
The adjoint modes depend on this choice of norm, but when recombined with the direct modes to give the sensitivity measurement of the eigenvalue to changes in $\vb{\mathcal{L}}$, the effect of the norm cancels out \citep{chandler_2012}.\par
By definition, the adjoint operator $\vb{\mathcal{L}^\dag}$ should fulfill
\begin{equation} \label{eqn:Ladj}
\left<\vb{\hat{q}^\dag},\vb{\mathcal{L}}\vb{\hat{q}}\right>\equiv\left<\vb{\mathcal{L}^\dag}\vb{\hat{q}^\dag},\vb{\hat{q}}\right>
\end{equation}
This definition is valid for any pair of vectors, but for convenience, they are expressed in terms of the direct state vector $\vb{\hat{q}}$ and the adjoint state vector $\vb{\hat{q}^\dag}$, as done in \citet{marquet_2009,qadri_2013}. More specifically, we consider the spatial stability form, then the generalized eigenvalue problem is pre-multiplied by the adjoint eigenvector
\begin{equation} \label{eqn:premulti}
\left<\vb{\hat{q}^\dag},\vb{\mathcal{A}}\vb{\hat{q}}\right>-\left<\vb{\hat{q}^\dag},\alpha\vb{\mathcal{B}}\vb{\hat{q}}\right>=0.
\end{equation}
By successively integrating the equation by parts using Green's theorem, equation \ref{eqn:premulti} is rearranged to
\begin{equation} \label{eqn:rearrange}
\left<\vb{\mathcal{A}^\dag}\vb{\hat{q}^\dag},\vb{\hat{q}}\right>-\left<\alpha^\dag\vb{\mathcal{B}^\dag}\vb{\hat{q}^\dag},\vb{\hat{q}}\right>=0.
\end{equation}
Note that the integration-by-parts process would leave boundary terms in the expression. Therefore, appropriate adjoint boundary conditions are applied to cancel out all boundary terms. Then the adjoint of the direct eigenvalue problem is obtained from equation \ref{eqn:rearrange} as
\begin{equation} \label{eqn:Temporaladj}
\vb{\mathcal{A}^\dag}\vb{\hat{q}^\dag}=\alpha^\dag\vb{\mathcal{B}^\dag}\vb{\hat{q}^\dag}
\end{equation}
The detailed derivations as well as the validation of the adjoint operators and boundary conditions used in the current study is presented in Appendix \ref{sec:adjLSAoperator}.\par
It can be shown that, for well-converged adjoint solutions, the adjoint eigenvalues, ordered by the same rule as the direct ones, are the complex conjugates of the direct ones \citep{schmid_2001}, and the bi-orthogonality condition writes
\begin{equation} \label{eqn:BiOrtho}
\left(\left(\alpha_m^\dag\right)^\ast-\alpha_n\right)\left[\left(\vb{\hat{q}^\dag}_m\right)^H\vb{W}\vb{\mathcal{B}}\vb{\hat{q}}_n\right]=0.
\end{equation}
The bi-orthogonality enables the characterization of the receptivity of the open-loop forcing, and the sensitivity to modification of the mean flow \citep{marquet_2009}.\par}
\begin{figure}
  \centering
  \includegraphics[width=0.75\textwidth]{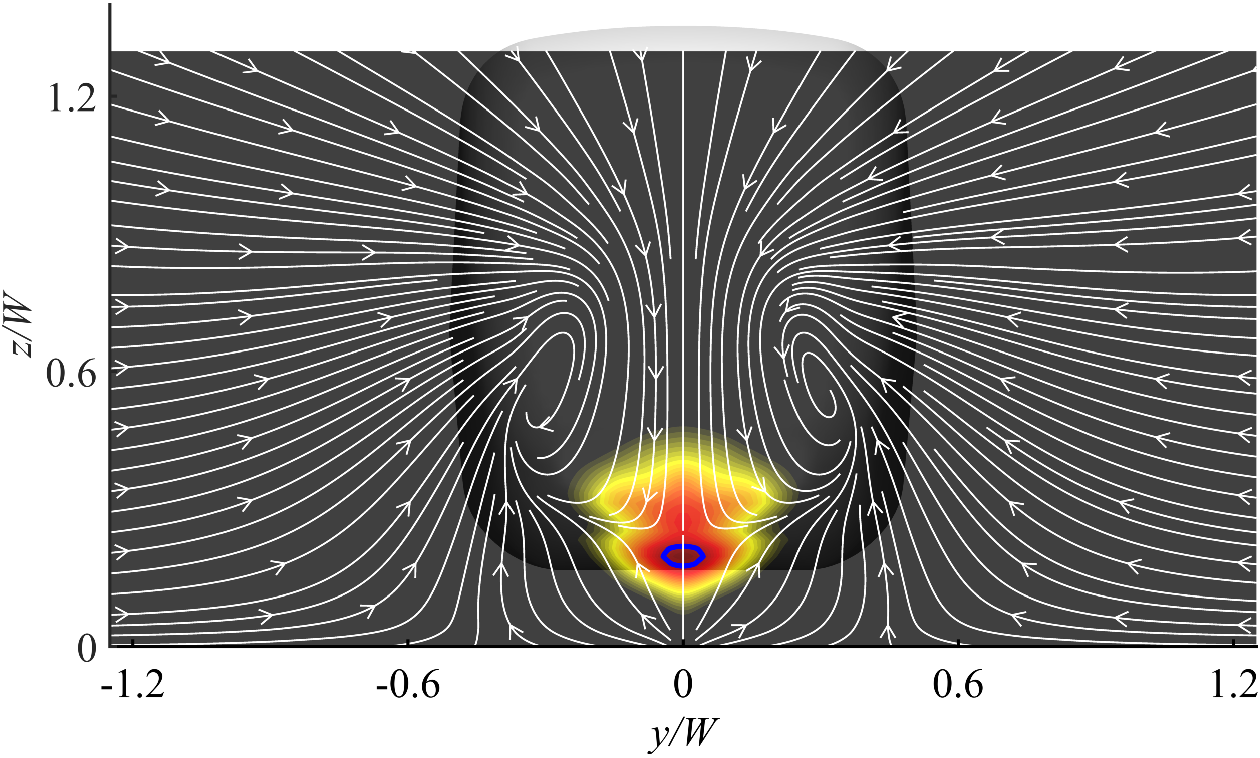}
  \caption{
  Structural sensitivity indicating sensitivity to mean flow modifications, the values are normalized by the maximum value; Streamlines are drawn based on mean flow; The blue line indicates normalized structural sensitivity = 0.90 isocontour.
  }
  \label{fig:Sensitivity}
\end{figure}
{In figure \ref{fig:Sensitivity}, the distribution of the structural sensitivity at the streamwise location where $\alpha^+$ and $\alpha^-$ branches intersect ($x/W=0.0731$) is shown. This field is obtained by first computing the adjoint mode on the considered cross-flow plane based on the methodology presented above and then calculating the $L^2$ norm of the sensitivity tensor following equation \ref{eqn:Sensi}. In addition, mean flow streamlines are also included so that the structural sensitivity can be related to specific flow structures. Based on the results, both the upper and lower recirculation zones are highlighted. In particular, it can be observed that the position of the highest structural sensitivity, enclosed by the blue solid line in figure \ref{fig:Sensitivity}, is located slightly below the stagnation point of the recirculation region. The streamwise vortex pair generated from both sides of the train, on the other hand, seems to have no relevance to the sensitivity of the linear global mode. The highest structural sensitivity at this location, shown in figure \ref{fig:Sensitivity}, suggests that the interaction between the lower and upper shear layers, which drives the self-excitation of the instability, is the most sensitive to a steady external forcing and is therefore of significant importance in terms of controlling of the global mode. Note that in this part, a deeper interpretation of different components of the structural sensitivity tensor, as has been done in \citep{qadri_2013}, is beyond the scope of this work.\par}

\section{Comparison between empirical and theoretical modes}\label{sec:comparison}
{In this section, the near wake eigenmode which serves as the origin of the global mode, is tracked further downstream into the far wake to obtain the full picture of the linear global mode. The linear global mode is then compared with the leading SPOD mode to show the connections between them and to reveal additional physical driving mechanisms.\par}
\begin{figure}
  \centering
  \includegraphics[ width=0.85\textwidth]{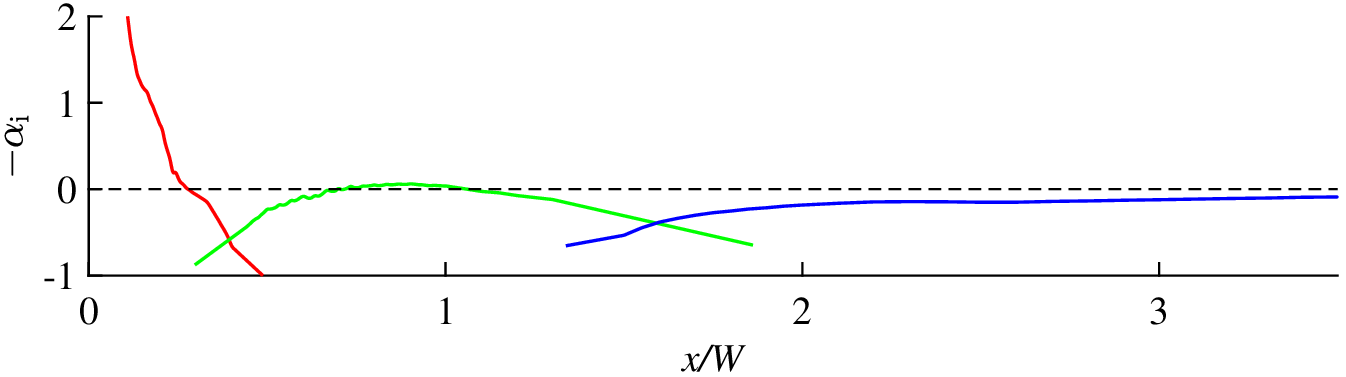}
  \caption{
  Spatial amplification rates of the most unstable near wake ({\color{red}\rule[0.8mm]{0.5cm}{0.3mm}}), middle wake ({\color{green}\rule[0.8mm]{0.5cm}{0.3mm}}) and far wake ({\color{blue}\rule[0.8mm]{0.5cm}{0.3mm}}) branches as functions of streamwise location, computed at the global complex frequency $\omega_g$.
  }
  \label{fig:Xbranch}
\end{figure}
{To reconstruct the shape of the global mode, the $\alpha^+$ branch at $\omega_{\rm{g}}$ must be tracked in downstream direction reaching into the far wake. However, as illustrated in $\S$\ref{sec:eigproblem},  the complexity of the base flow gives rise to the problem that the physical eigenvalue may become highly stable at one location and fall into a spurious region, and then be untraceable at another location. Therefore, we do not only focus on the $\alpha^+$ branch, but also include spatial branches that arise and become unstable during the tracking process.\par
The results of the branch tracking process are shown in figure \ref{fig:Xbranch}. Accordingly, the near wake branch becomes spatially stable when approaching $x/W=0.3$, and cannot be tracked further downstream of $x/W=0.5$. However, at $x/W=0.4$, a new eigenmode, which corresponds to the middle wake branch, becomes the most spatially unstable. The middle wake branch develops to be slightly spatially unstable only within $0.75<x/W<1.05$, and remains spatially stable for most of the streamwise range of its occurrence. However, it is still the most unstable within the streamwise region $0.4<x/W<1.6$. Downstream of $x/W=1.6$, the far wake branch becomes dominant. The spatial amplification rate of the far wake eigenmode is observed to grow slightly upstream of $x/W=2.0$, which is attributed to the growing spatial sizes of the streamwise vortex pair and thereafter the far wake coherent structures. Then the far wake eigenmode remains a nearly constant spatial amplification rate with a marginally stable state after this location, and extends into the farthest streamwise location considered in this study.\par}
\begin{figure}
  \centering
  \begin{subfigure}{0.85\textwidth}
  \centering
  \subcaption{}
  \includegraphics[ width=1\textwidth]{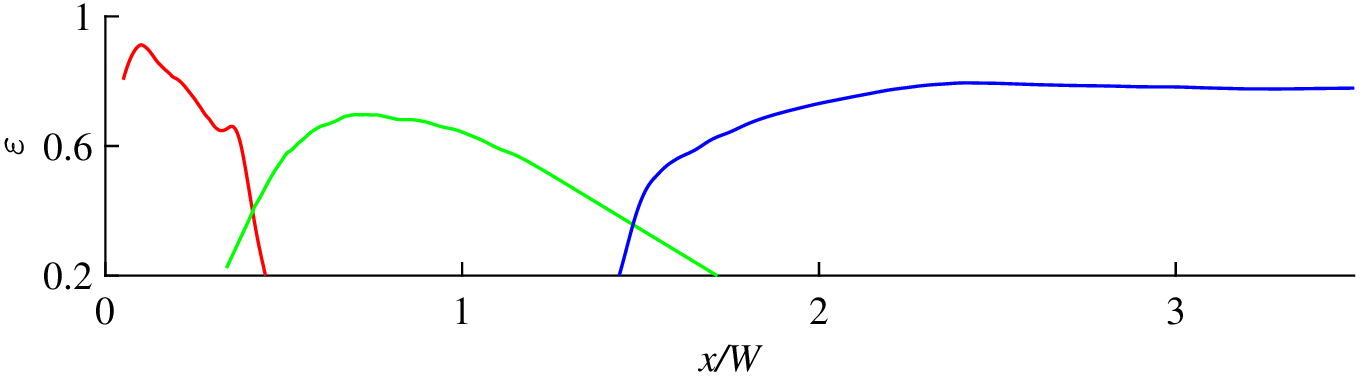}
  \end{subfigure}
  \begin{subfigure}{0.49\textwidth}
  \subcaption{}
  \includegraphics[width=1\textwidth]{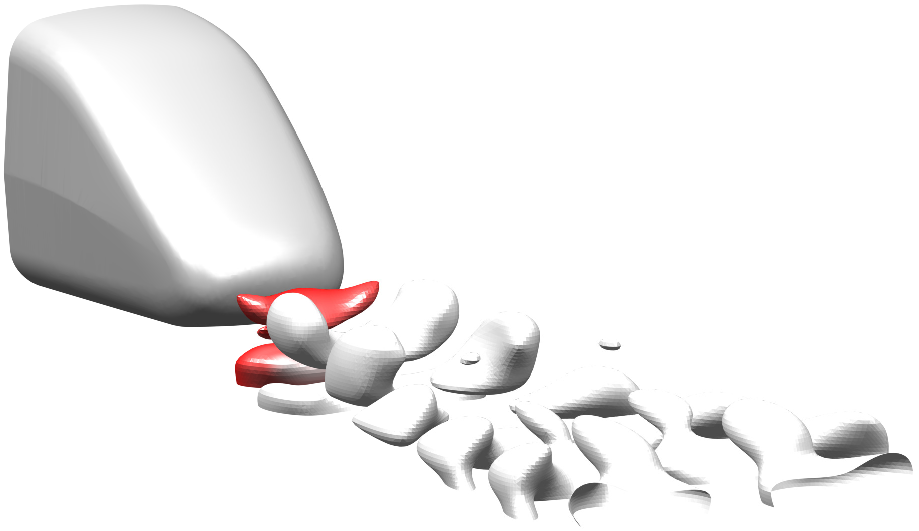}
  \end{subfigure}
  \begin{subfigure}{0.49\textwidth}
  \subcaption{}
  \includegraphics[width=1\textwidth]{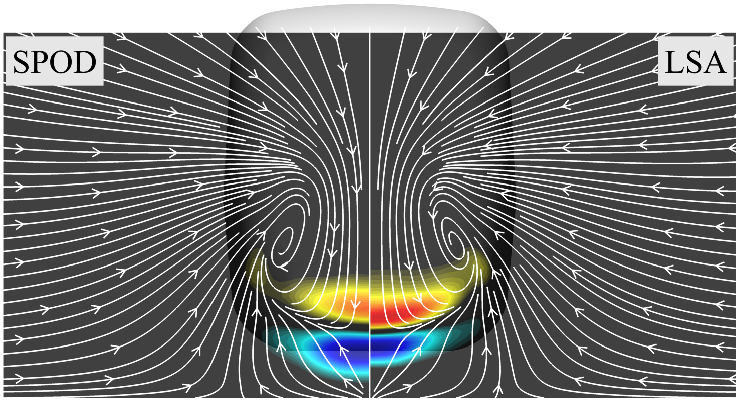}
  \end{subfigure}
  \begin{subfigure}{0.49\textwidth}
  \subcaption{}
  \includegraphics[width=1\textwidth]{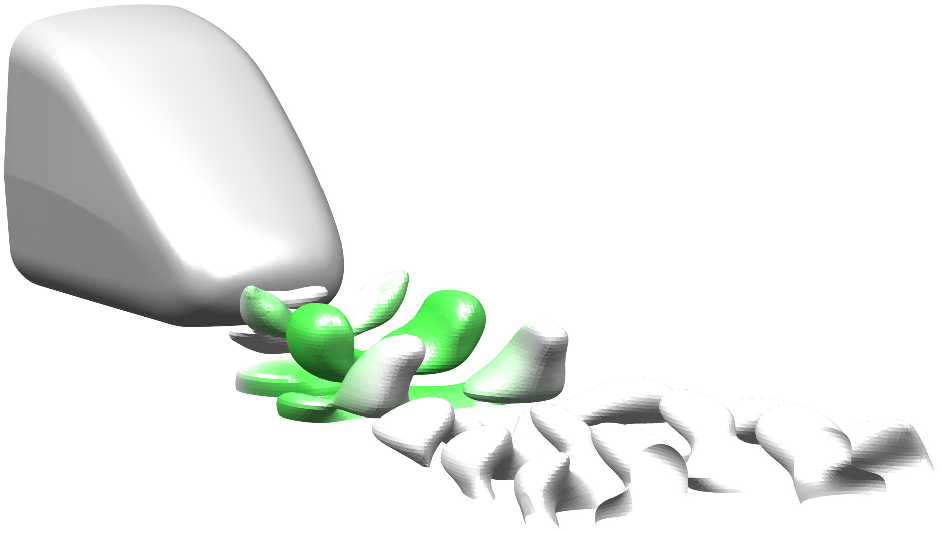}
  \end{subfigure}
  \begin{subfigure}{0.49\textwidth}
  \subcaption{}
  \includegraphics[width=1\textwidth]{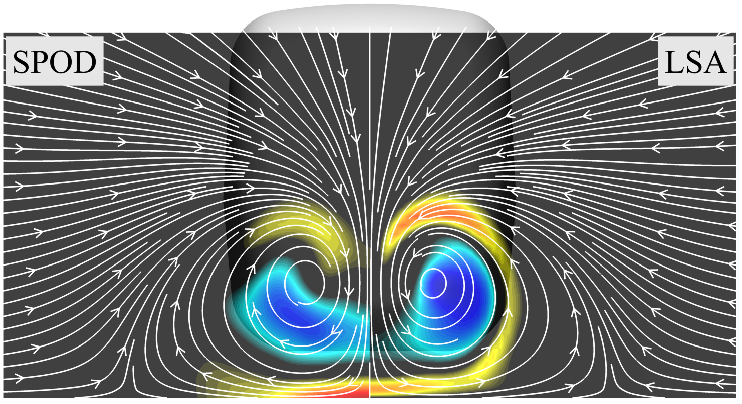}
  \end{subfigure}
  \begin{subfigure}{0.49\textwidth}
  \subcaption{}
  \includegraphics[width=1\textwidth]{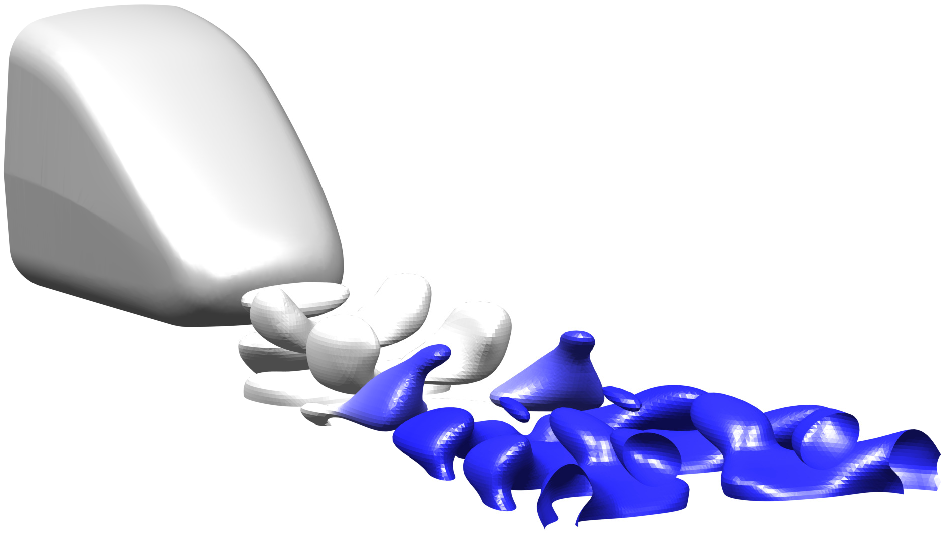}
  \end{subfigure}
  \begin{subfigure}{0.49\textwidth}
  \subcaption{}
  \includegraphics[width=1\textwidth]{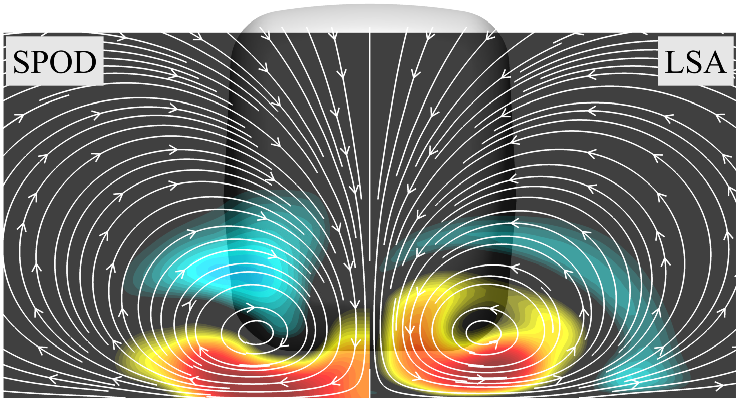}
  \end{subfigure}
  \caption{
  Comparison between the leading SPOD mode and the global linear stability mode from local analysis. (\textit{a}) Alignment between the SPOD mode and the eigenmodes of the three branches (colors of lines correspond to figure \ref{fig:Xbranch}) as functions of streamwise location; (\textit{b,d,f}) The isosurface of the streamwise component of SPOD mode colored respectively by the alignment with near wake, middle wake and far wake branches; (\textit{c,e,g}) The streamwise component of the near wake, middle wake and far wake SPOD modes and eigenmodes; Streamlines are drawn based on mean flow.
  }
  \label{fig:Comparison}
\end{figure}
{To analyze whether these branches actually represent the empirical mode observed in the SPOD, we compute the alignment between the leading SPOD mode and the linear global mode based on the $L^2$ inner product. Since the wake flow is highly nonparallel, with several different spatial branches contributing to the linear global mode, we do not reconstruct the full three-dimensional global mode prior to the alignment measurement. Instead, the alignment between the leading SPOD mode and the three eigenmode branches are computed at all streamwise locations.\par
In figure \ref{fig:Comparison}(\textit{a}), the alignment between the leading SPOD mode and the three eigenmode branches are plotted as a function of $x/W$. It can be observed that the alignment curves generally follow similar trends to the spatial growth rates of the three branches shown in figure \ref{fig:Xbranch}, with the alignment value always being the highest for the most unstable eigenmode branch. In particular, the alignment is, in general, quite high with a value above 0.7, except for locations where the spatial branches switch. Therefore, the three eigenmode branches can be regarded as manifestations of the SPOD mode at different regions of the wake. This result further supports the modeling approach in this research for tracking the linear global mode in a highly complex three-dimensional base flow.\par
For better visualization, in figure \ref{fig:Comparison}(\textit{b,d,f}), the three-dimensional SPOD mode is displayed based on the isosurface of the streamwise component, colored by the alignment of the three eigenmode branches. In addtion, figure \ref{fig:Comparison}(\textit{c,e,g}) shows a  direct comparisons between the cross-flow shapes of the SPOD mode and the linear global mode at different streamwise locations. In general, similar structures can be identified throughout the entire wake region. In the near wake region, the vortex shedding related to the transverse recirculation zone is dominant in both the SPOD and the linear global mode, with slightly different ranges between the two modes. Further downstream in the middle wake and far wake region, the coherent structures related to the streamwise vortex pair become dominant. The SPOD mode generally predicts a higher fluctuation amplitude near the ground and the central line, compared to the linear global mode.\par
In conclusion, the fundamental mechanism of instability in the considered flow problem can be interpreted as follows: Within the recirculation zone, the flow has a small region of absolute instability, which contributes to the global wavemaker located inside. The global frequency is well matched to the SPOD peak frequency and  the linear global mode in this region has very high alignment with the SPOD mode. Further downstream, the flow becomes convectively unstable, amplifying the perturbations received by the global wavemaker, with the oscillation frequency synchronized to the global frequency. In this situation, the most spatially unstable branch becomes dominant and aligns the best with the SPOD mode.\par}

\section{Summary and Conclusions}\label{sec:conclusions}
{Understanding important dynamics in complex technical flow is crucial in engineering practice. In this paper, three-dimensional coherent structures in the turbulent wake flow behind a generic high-speed train are investigated. A large eddy simulation has been performed to simulate and collect a database of the flow problem considered. For the purpose of this research, both the empirical identification approach based on SPOD and the theoretical approach using linear stability analysis are used.\par
The turbulent wake is found to be dominated by spanwise symmetric coherent structures, based on the comparison between the SPOD energy spectrum of symmetric and antisymmetric fluctuations. The most dominant SPOD mode is found to oscillate at the angular frequency of $\omega=3.437$. This dominant mode has an increasing wavelength in the near wake and a nearly constant wavelength in the far wake. The quadratic nonlinear interaction of the velocity perturbation is further checked by computing the mode bispectrum to explain the broadband coherent dynamics. The fundamental mode is identified with strong self-interaction, which generates the first-harmonic and subharmonic triads, meanwhile leading to significant deformation of the mean field. With the continuous evolution of this process, the flow finally appears to be low-rank in a wide frequency range.\par
The global instability is analyzed by employing a two-dimensional local spatio-temporal stability approach following the WKBJ approximation and combined with a non-parallelism treatment. Three spatial branches with positive temporal growth rate are found, with the near wake branch being the most unstable. The absolute frequency of the near wake branch is then found on the basis of a valid saddle point on the complex $\alpha^+$ plane and tracked along the streamwise direction. A confined region of absolute instability is identified in the near wake region. The global frequency is then determined to be $\omega_{\rm{g}}=3.2702+0.0797{\rm{i}}$ based on the frequency selection criterion, indicating a marginally stable global mode. This result is in excellent agreement with the empirical prediction in terms of angular frequency, and the theoretical expectation in terms of growth rate. The near wake recirculation zone is found to be related to the global mode wavemaker, which drives the self-excitation of the instability. The adjoint method is further used to compute the structural sensitivity at the location where the $\alpha^+$ and $\alpha^-$ branches intersect. In the corresponding cross-flow plane, the highest sensitivity is found near the upper and lower shear layers of the recirculation bubble, near the tip of the train nose. Accordingly, the global mode is expected to be most sensitive to mean flow changes in this region.\par
The linear global mode shape is further computed at each streamwise location by imposing the global frequency. Three spatial branches are found to be dominant respectively in the near, middle and far wake regions. The alignment between the linear global and leading SPOD modes is then performed to provide a quantitative comparison between mode shapes. Within the recirculation zone where the global wavemaker is located, the linear global mode has very high alignment with the SPOD mode. Further downstream, the flow becomes convectively unstable and the oscillations within these regions are synchronized with the global frequency by excitation from the wavemaker. Spatial branches with the highest spatial amplification rates become dominant and show highest alignment with the SPOD mode.\par
This work has two main conclusions, one methodological and one physical. Methodologically, a framework to track linear global mode in highly complex 3D base flows is introduced, and this method is justified {\itshape a posteriori} by the very good agreement with the empirical modes. To the authors' knowledge, this is the first research dealing with the linear global stability in such a complex flow problem. Physically, the dominant SPOD mode is identified as being caused by a linear global mode in the wake of the train. This mode exhibits transverse symmetry and may cause significant fluctuations on the train surface, which can cause operational safety problems. The sensitivity analysis further shows that this mode could be quite effectively suppressed by small changes of the base flow near the tail of the train, which is of significant importance in terms of developing efficient flow control strategies.\par
In this work, we have considered a simplified train model as the research object. For a more realistic train model with complex under-body structures and operating under higher Reynolds number, the formation of the spanwise vortex pair is significantly hindered. The global mode driven by the spanwise vortex pair is likely to be suppressed, and the entire wake flow will be convectively unstable. Thereafter, the symmetric global oscillation observed in our current research will develop into asymmetric perturbations with longer wave length attributed to the Crow instability \citep{crow_1970} of the streamwise vortex pair, as have been observed in \citet{bell_2016}. The physical mechanisms of this dominant perturbation in the wake of a realistic train, as well as their influential factors are expected to be further addressed in future works.\par}

\vspace{5mm}
\noindent\textbf{Acknowledgements}
{This work is supported by the National Science Foundation of China (grant no. 52202429), the Project of Scientific and Technological Research and Development of China Railway (P2021J037), the Hunan Provincial Natural Science Foundation (grant no. 2023JJ40747) and the China Scholarship Council (grant no. 202006370204). The authors acknowledge the computational resources provided by the High Performance Computing Centre of Central South University, China.}

\vspace{5mm}
\noindent\textbf{Data availability}
The data and code that support the findings of this study are available from the authors upon reasonable request.

\vspace{5mm}
\noindent{\bf Declaration of Interests} The authors report no conflict of interest.

\appendix
\section{Validation of the LES} \label{sec:LESvalidation}
\begin{figure}
  \centering
  \begin{subfigure}{0.49\textwidth}
  \subcaption{}
  \includegraphics[ width=1\textwidth]{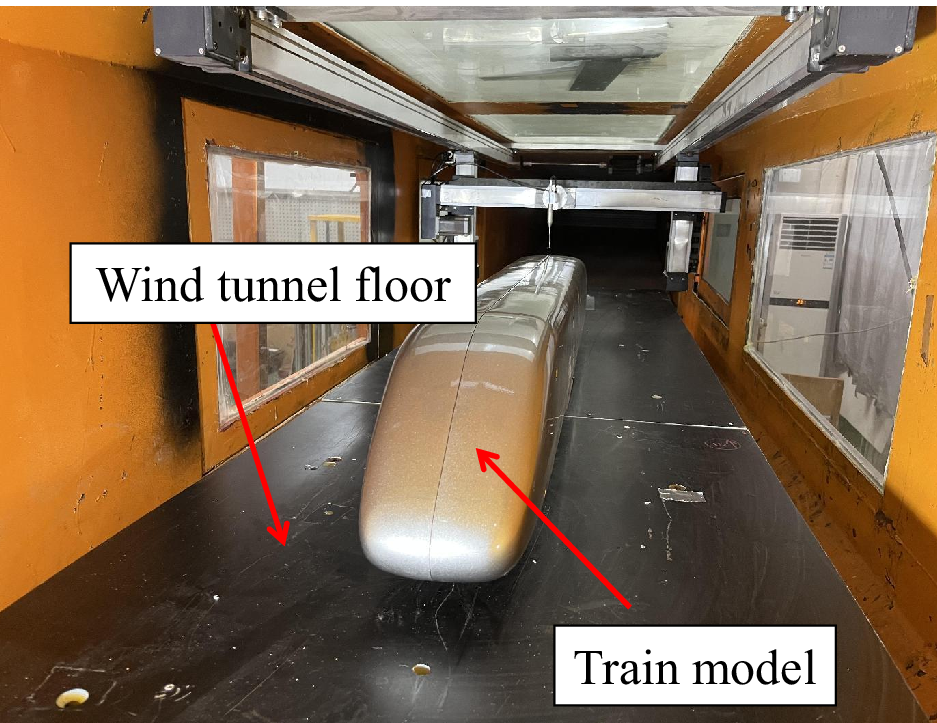}
  \end{subfigure}
  \begin{subfigure}{0.49\textwidth}
  \subcaption{}
  \includegraphics[width=1\textwidth]{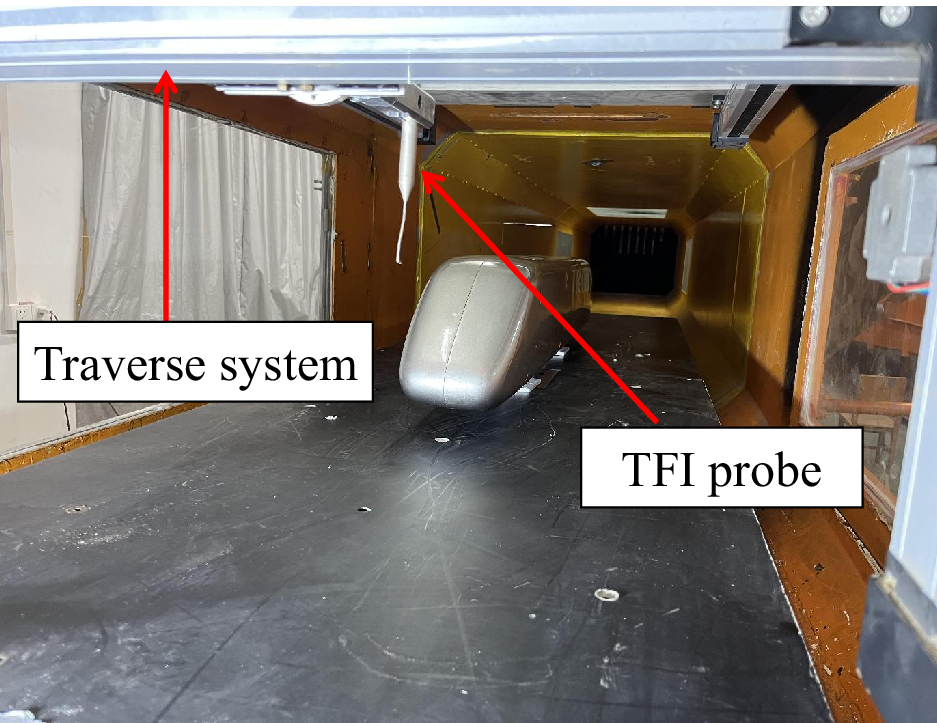}
  \end{subfigure}
  \caption{
  Wind tunnel experiment set-up. (\textit{a}) Windward view; (\textit{b}) Leeward view.
  }
  \label{fig:WindTunnel}
\end{figure}

{In order to  validate the LES results, a wind tunnel experiment is carried out in the high-speed test section of the closed-loop wind tunnel in the Key Laboratory of Traffic Safety on Track, Central South University. The experimental configuration is shown in figure \ref{fig:WindTunnel}. The dimensions of the test section are 3400mm$\times$1000mm$\times$800mm. To eliminate the effect of the wind tunnel boundary layer, the train model is installed on the static wind tunnel floor that is mounted on the bottom surface of the test section. The experimental Reynolds number is $\R=9.5\times10^4$, consistent with the LES.\par
The flow velocity on the vertical symmetry plane $y=0$ is measured using a 4-hole TFI cobra probe installed on a traverse system. The probe is able to measure flow fields within a range of $\pm45^{\circ}$, and can resolve fluctuations up to 2000 Hz. At each measurement point, data are continuously collected for a duration of 20 s, capable of converging the time-averaged velocity and Reynolds stress distributions.\par}
\begin{figure}
  \centering
  \begin{subfigure}{0.49\textwidth}
  \subcaption{}
  \includegraphics[ width=0.9\textwidth]{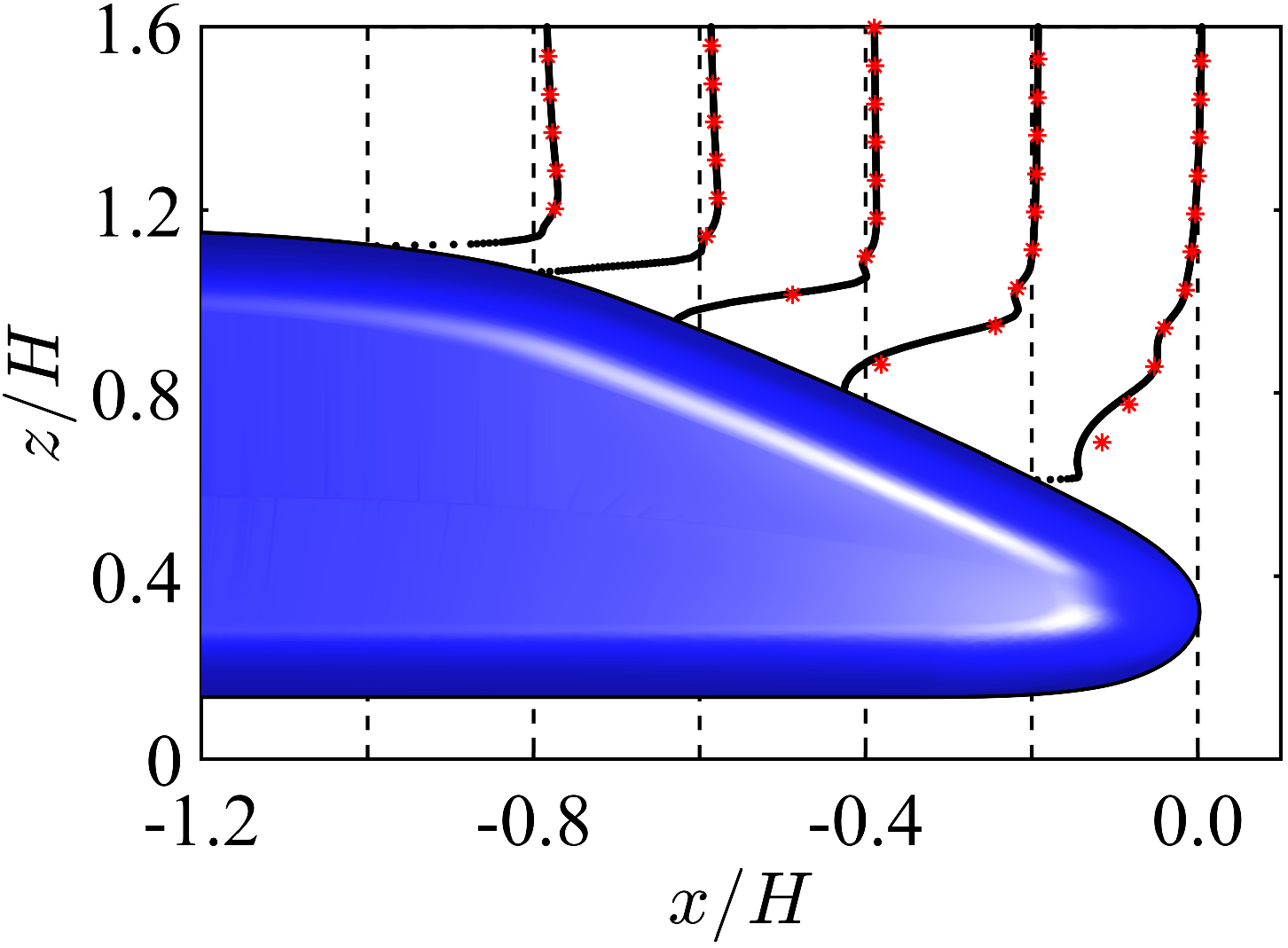}
  \end{subfigure}
  \begin{subfigure}{0.49\textwidth}
  \subcaption{}
  \includegraphics[width=0.9\textwidth]{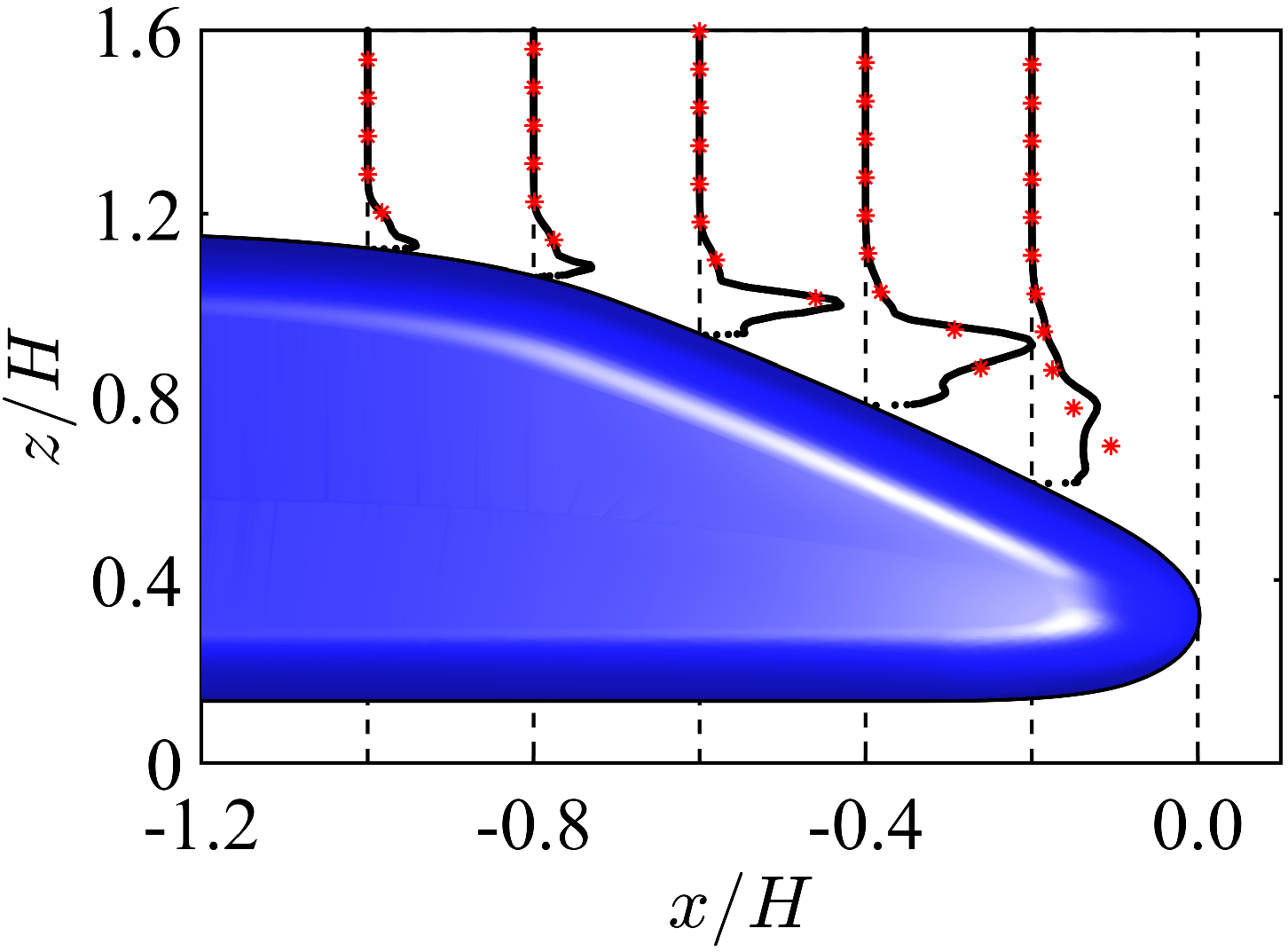}
  \end{subfigure}
  \caption{
  Comparisons between the time-averaged flow quantities from LES modeling ({\color{black}\rule[0.8mm]{0.5cm}{0.3mm}}) and wind tunnel measurement ($\color{red}\ast$). (\textit{a}) Streamwise velocity $\overline{u}$; (\textit{b}) Streamwise Reynolds stress $\overline{u'u'}$. For details see text.
  }
  \label{fig:Validation}
\end{figure}
{Since the ground condition differs between the wind tunnel experiment and the LES, the validation is carried out based on the comparison of flow fields on the upper side of the train tail. In figure \ref{fig:Validation}, the profiles of the time-averaged streamwise velocity $\overline{u}$ and streamwise Reynolds stress $\overline{u'u'}$ are shown. The vertical dashed lines mark the $x$ location of the measured profiles and correspond to $\overline{u}=0$ and $\overline{u'u'}=0$ in figure \ref{fig:Validation}(\textit{a}) and figure \ref{fig:Validation}(\textit{b}), respectively. For the time-averaged streamwise velocity profiles $\overline{u}$, the distance between two adjacent vertical lines corresponds to the range from 0 to $U_{\infty}$. For the streamwise Reynolds stress component $\overline{u'u'}$, the distance corresponds to a range from 0 to the global maxima among all considered data points, which occurs at $x/H=-0.4$.\par
It can be seen from figure \ref{fig:Validation} that, the time-averaged streamwise velocity and Reynolds stress profiles predicted by the large eddy simulation are generally in good agreement with the experimental results. Several discrepancies can be found at $x/H=-0.2$, where the streamwise velocity and Reynolds stresses are overestimated by the LES. Considering the uncertainties introduced by the wind tunnel boundaries, the ground conditions and testing facilities, the discrepancies can be regarded as within reasonable range. In conclusion, the large eddy simulation in this paper predicts the important features and dynamics of the flow at acceptable accuracy.\par}

\section{Stability analysis without the non-parallelism approximation} \label{sec:stabilityresult}
{To justify the non-parallelism approximation method proposed in this paper, we conduct a spatio-temporal analysis using the standard two-dimensional linearized Navier-Stokes equation for comparison.\par}
\begin{figure}
  \centering
  \begin{subfigure}{0.49\textwidth}
  \subcaption{}
  \includegraphics[ width=1\textwidth]{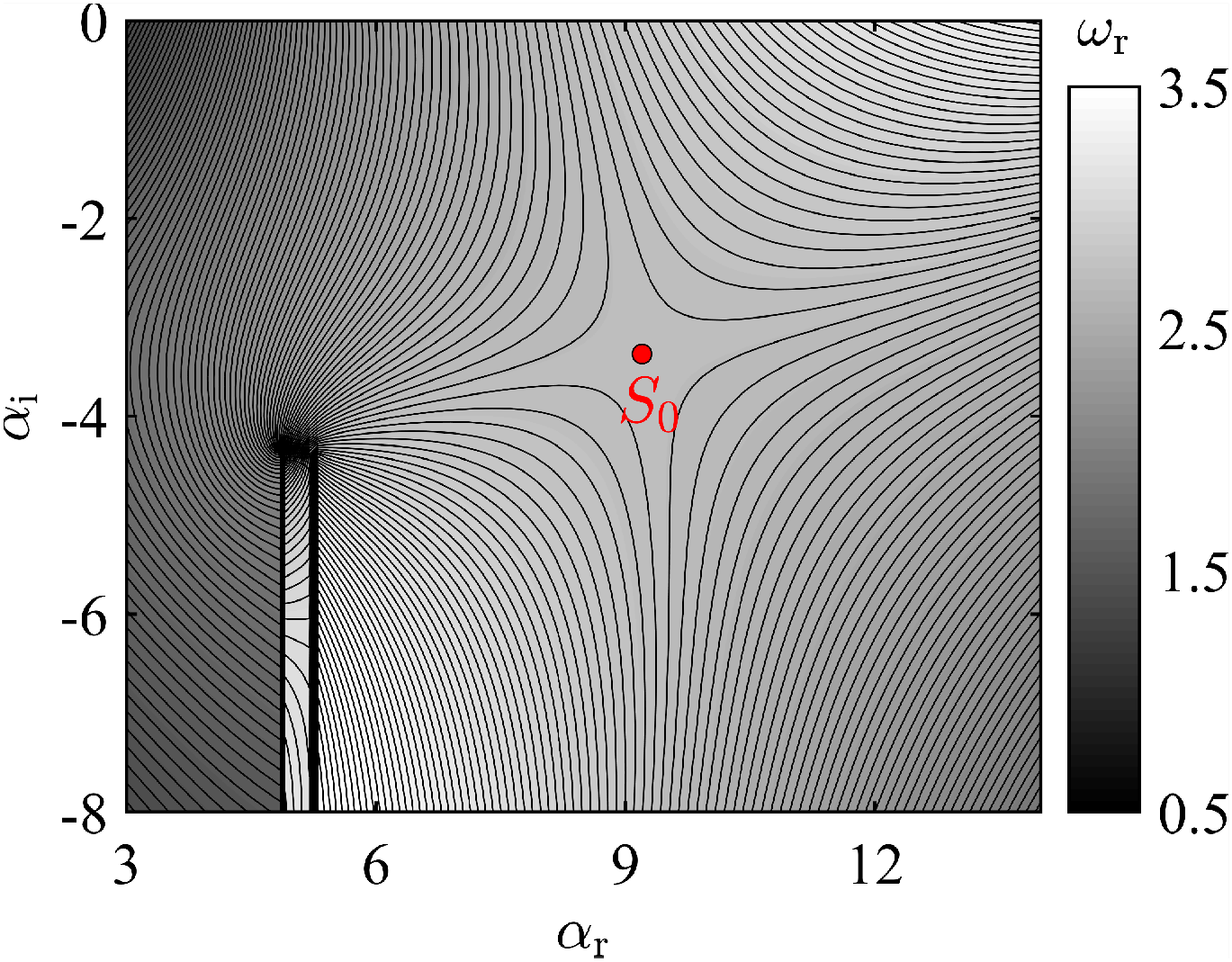}
  \end{subfigure}
  \begin{subfigure}{0.49\textwidth}
  \subcaption{}
  \includegraphics[width=1\textwidth]{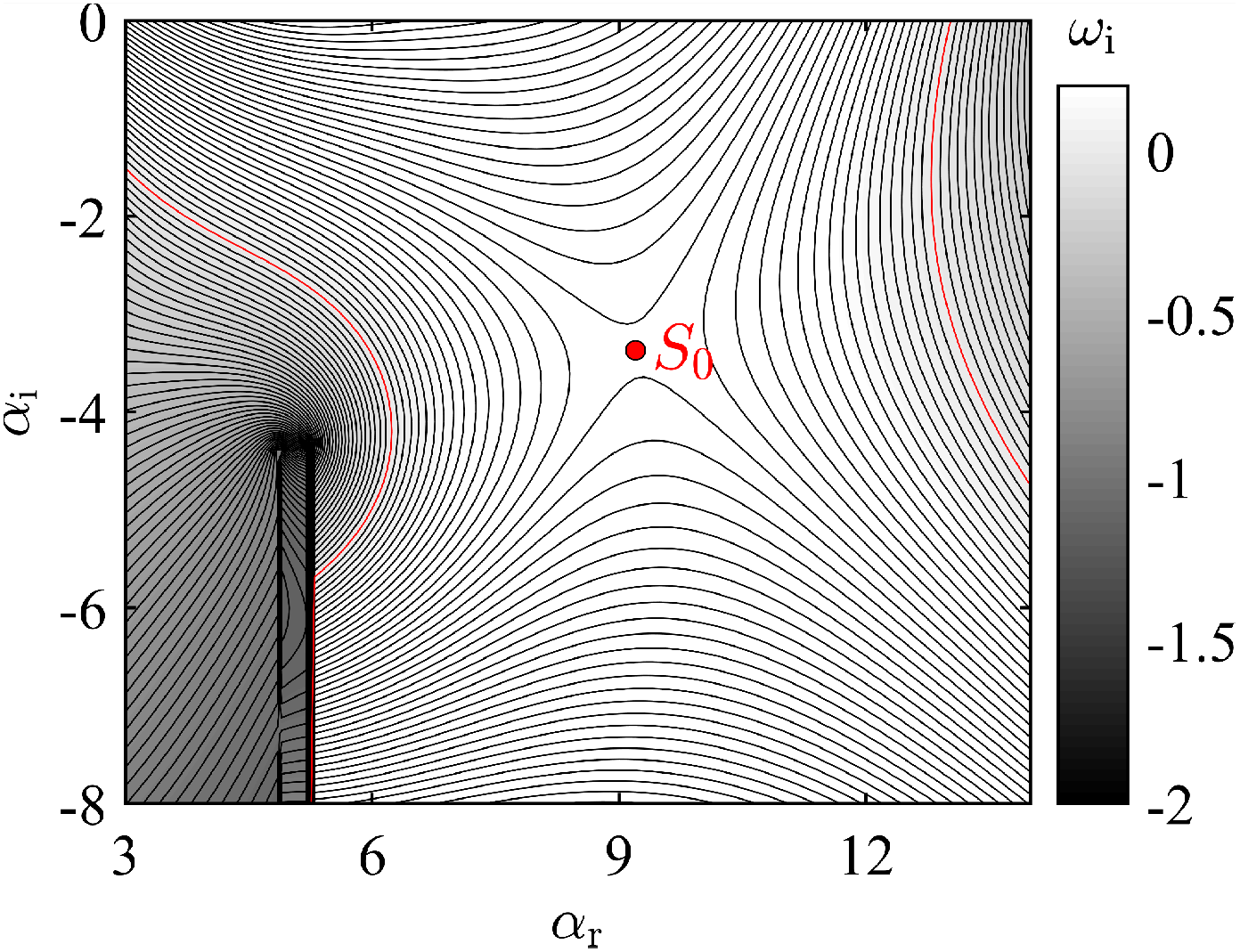}
  \end{subfigure}
  \caption{
  Contour of the complex angular frequency in the complex $\alpha$-plane at $x/W=0.08$. Saddle points on the complex $\alpha$-plane are marked by a red circle. (\textit{a}) The real component; (\textit{b}) The imaginary component, and the $\omega_{\rm{i}}=0$ isocontour is highlighted in red.
  }
  \label{fig:ComplexAlpha_noAngle}
\end{figure}
{In figure \ref{fig:ComplexAlpha_noAngle}, similar to figure \ref{fig:ComplexAlpha}, the contour of the complex angular frequency in the complex $\alpha$-plane, at the streamwise location of $x/W=0.08$ is presented. When the non-parallelism approximation is discarded, the location of the saddle point $S_0$ on the complex $\alpha$-plane moves closer to both the real and imaginary axes, denoting the longer travelling wave with lower spatial amplification rate compared to the one in figure \ref{fig:ComplexAlpha}. Meanwhile, the second saddle $S_1$ in figure \ref{fig:ComplexAlpha} is shifted outside the considered range of the complex $\alpha$-plane, when the non-parallelism approximation approach is not applied. The absolute frequency at this location can be then determined as $\omega_0=2.7114+0.3180{\rm{i}}$, which again indicates the absolutely unstable condition. Compared to the absolute frequency calculated in $\S$\ref{sec:spatialtemporalanalysis}, this value represents the perturbation oscillating with significantly longer time period and higher temporal growth rate.\par}
\begin{figure}
  \centering
  \begin{subfigure}{0.90\textwidth}
  \subcaption{}
  \includegraphics[ width=1\textwidth]{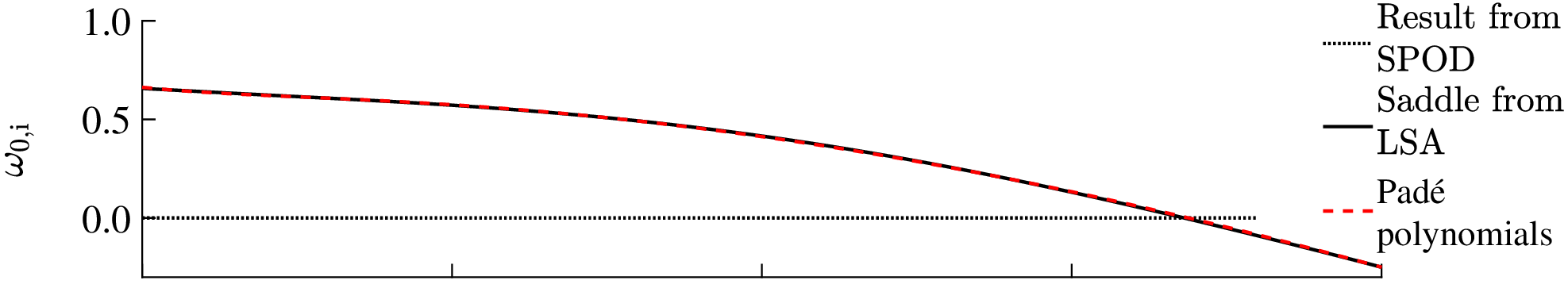}
  \end{subfigure}
  \begin{subfigure}{0.90\textwidth}
  \subcaption{}
  \includegraphics[ width=1\textwidth]{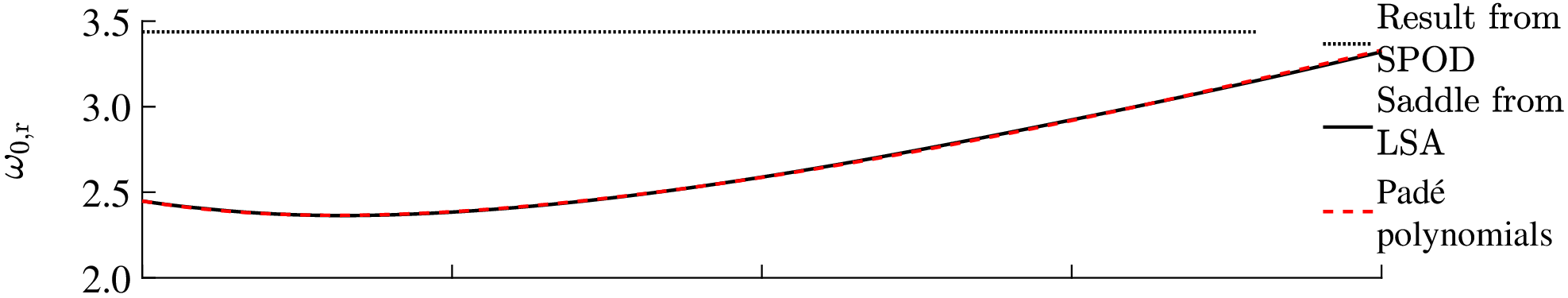}
  \end{subfigure}
  \begin{subfigure}{0.90\textwidth}
  \subcaption{}
  \includegraphics[ width=1\textwidth]{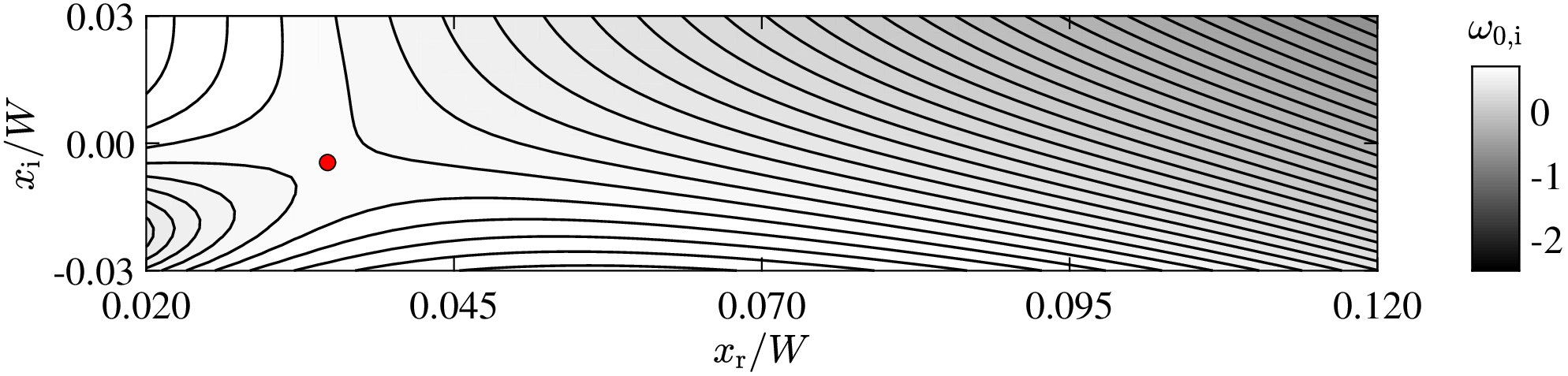}
  \end{subfigure}
  \caption{
  (\textit{a}) Streamwise evolution of the absolute growth rate. (\textit{b}) Streamwise evolution of the absolute angular frequency. (\textit{c}) Contours of $\omega_{\rm{0,i}}$ extruded in the complex $x$-plane, formed by the fitted Pad\'{e} polynomial. The saddle point on the complex $x$-plane is marked by red circle.
  }
  \label{fig:Wavemaker_noAngle}
\end{figure}
{Then the procedures described in $\S$ \ref{sec:globalwavemaker} are followed to account for the global instability, with the main results presented in figure \ref{fig:Wavemaker_noAngle}. In figure \ref{fig:Wavemaker_noAngle}(\textit{a,b}), the imaginary and real parts of the absolute complex angular frequency are respectively shown as a functions of streamwise location. Different from the results shown in figure \ref{fig:Wavemaker}, the flow is predicted to be absolutely unstable within in a much larger streamwise range upstream of $x/W=0.1$. In figure \ref{fig:Wavemaker_noAngle}(\textit{c}), the extrapolated complex $x$-plane based on the 10th order Pad\'{e} polynomials is visualized. The saddle point can be then located at $x_{\rm{s}}=0.0342-0.0046{\rm{i}}$, with the associated global frequency $\omega_{\rm{g}}=2.3721+0.6082{\rm{i}}$. Compared to the global frequency calculated in $\S$\ref{sec:globalwavemaker}, the current value shows larger deviations from both the empirical prediction in terms of angular frequency, and the theoretical expectation in terms of growth rate. This again suggests that, for a flow which is strongly non-parallel, utilizing local approach is likely to give poor prediction on the global instability. By including the approximation approach proposed in this paper, the accuracy of the prediction is much more improved, which justified the ability of the approach to model the non-parallelism of the flow.\par}

\section{Direct LSA operator} \label{sec:LSAoperator}
{The operator matrices $\vb{\mathcal{A}}$ and $\vb{\mathcal{B}}$ used in the temporal stability analysis are shown as follows
\begin{equation} \label{eqn:TemporalA}
\vb{\mathcal{A}}=\begin{bmatrix}
\vb{\mathcal{C}}+\vb{\bar{u}}_x-\vb{\nu}_{tx}(\vb{\mathcal{D}}_x+{\rm{i}}\alpha) & \vb{\bar{u}}_y-\vb{\nu}_{ty}(\vb{\mathcal{D}}_x+{\rm{i}}\alpha) & \vb{\bar{u}}_z-\vb{\nu}_{tz}(\vb{\mathcal{D}}_x+{\rm{i}}\alpha) & \vb{\mathcal{D}}_x+{\rm{i}}\alpha \\
\vb{\bar{v}}_x-\vb{\nu}_{tx}\vb{\mathcal{D}}_y & \vb{\mathcal{C}}+\vb{\bar{v}}_y-\vb{\nu}_{ty}\vb{\mathcal{D}}_y & \vb{\bar{v}}_z-\vb{\nu}_{tz}\vb{\mathcal{D}}_y & \vb{\mathcal{D}}_y \\
\vb{\bar{w}}_x-\vb{\nu}_{tx}\vb{\mathcal{D}}_z & \vb{\bar{w}}_y-\vb{\nu}_{ty}\vb{\mathcal{D}}_z & \vb{\mathcal{C}}+\vb{\bar{w}}_z-\vb{\nu}_{tz}\vb{\mathcal{D}}_z & \vb{\mathcal{D}}_z \\
\vb{\mathcal{D}}_x+{\rm{i}}\alpha & \vb{\mathcal{D}}_y & \vb{\mathcal{D}}_z & {0}
\end{bmatrix}
\end{equation}
\begin{equation} \label{eqn:TemporalB}
\vb{\mathcal{B}}=\begin{bmatrix}
\rm{i} & {0} & {0} & {0} \\
{0} & \rm{i} & {0} & {0} \\
{0} & {0} & \rm{i} & {0} \\
{0} & {0} & {0} & {0}
\end{bmatrix}
\end{equation}
where subscripts $\left(.\right)_x$, $\left(.\right)_y$, and $\left(.\right)_z$ related to flow quantities denoting their first derivatives respect to the three directions, $\vb{\mathcal{D}}_x$, $\vb{\mathcal{D}}_y$ and $\vb{\mathcal{D}}_z$ are the first derivative matrices respect to the three directions. Operator $\vb{\mathcal{C}}$ can be written as
\begin{equation} \label{eqn:TemporalC}
\begin{aligned}
\vb{\mathcal{C}}=&\vb{\bar{u}}(\vb{\mathcal{D}}_x+{\rm{i}}\alpha)+\vb{\bar{v}}\vb{\mathcal{D}}_y+\vb{\bar{w}}\vb{\mathcal{D}}_z-\left(\frac{1}{\R}+\vb{\nu}_{t}\right)(\vb{\mathcal{D}}_{yy}+\vb{\mathcal{D}}_{zz}+\vb{\mathcal{D}}_{xx}+2{\rm{i}}\alpha\vb{\mathcal{D}}_{x}-\alpha^2)\\
&-\vb{\nu}_{tx}(\vb{\mathcal{D}}_x+{\rm{i}}\alpha)-\vb{\nu}_{ty}\vb{\mathcal{D}}_y-\vb{\nu}_{tz}\vb{\mathcal{D}}_z
\end{aligned}
\end{equation}
with $\vb{\mathcal{D}}_{xx}$, $\vb{\mathcal{D}}_{yy}$ and $\vb{\mathcal{D}}_{zz}$ being the second derivative matrices respect to the three directions.\par
In spatial stability analysis, operator matrices $\vb{\mathcal{A}}$ and $\vb{\mathcal{B}}$ take the form
\begin{equation} \label{eqn:SpatialA}
\vb{\mathcal{A}}=\begin{bmatrix}
\vb{\mathcal{C}}+\vb{\bar{u}}_x-\vb{\nu}_{tx}\vb{\mathcal{D}}_x & \vb{\bar{u}}_y-\vb{\nu}_{ty}\vb{\mathcal{D}}_x & \vb{\bar{u}}_z-\vb{\nu}_{tz}\vb{\mathcal{D}}_x & \vb{\mathcal{D}}_x & 0 & 0 & 0 \\
\vb{\bar{v}}_x-\vb{\nu}_{tx}\vb{\mathcal{D}}_y & \vb{\mathcal{C}}+\vb{\bar{v}}_y-\vb{\nu}_{ty}\vb{\mathcal{D}}_y & \vb{\bar{v}}_z-\vb{\nu}_{tz}\vb{\mathcal{D}}_y & \vb{\mathcal{D}}_y & 0 & 0 & 0 \\
\vb{\bar{w}}_x-\vb{\nu}_{tx}\vb{\mathcal{D}}_z & \vb{\bar{w}}_y-\vb{\nu}_{ty}\vb{\mathcal{D}}_z & \vb{\mathcal{C}}+\vb{\bar{w}}_z-\vb{\nu}_{tz}\vb{\mathcal{D}}_z & \vb{\mathcal{D}}_z & 0 & 0 & 0 \\
\vb{\mathcal{D}}_x & \vb{\mathcal{D}}_y & \vb{\mathcal{D}}_z & 0 & 0 & 0 & 0 \\
 0 & 0 & 0 & 0 & 1 & 0 & 0 \\
 0 & 0 & 0 & 0 & 0 & 1 & 0 \\
 0 & 0 & 0 & 0 & 0 & 0 & 1 \\
\end{bmatrix}
\end{equation}
\begin{equation} \label{eqn:SpatialB}
\vb{\mathcal{B}}=\begin{bmatrix}
\vb{\mathcal{E}}+{\rm{i}}\vb{\nu}_{tx} & {\rm{i}}\vb{\nu}_{ty} & {\rm{i}}\vb{\nu}_{tz} & -{\rm{i}} & -1/\R-\vb{\nu}_{t} & 0 & 0 \\
0 & \vb{\mathcal{E}} & 0 & 0 & 0 & -1/\R-\vb{\nu}_{t} & 0 \\
0 & 0 & \vb{\mathcal{E}} & 0 & 0 & 0 & -1/\R-\vb{\nu}_{t} \\
-{\rm{i}} & 0 & 0 & 0 & 0 & 0 & 0 \\
1 & 0 & 0 & 0 & 0 & 0 & 0 \\
0 & 1 & 0 & 0 & 0 & 0 & 0 \\
0 & 0 & 1 & 0 & 0 & 0 & 0 \\
\end{bmatrix}
\end{equation}
with $\vb{\mathcal{C}}$ and $\vb{\mathcal{E}}$ being
\begin{equation} \label{eqn:SpatialC}
\begin{aligned}
\vb{\mathcal{C}}=&-{\rm{i}}\omega+\vb{\bar{u}}\vb{\mathcal{D}}_x+\vb{\bar{v}}\vb{\mathcal{D}}_y+\vb{\bar{w}}\vb{\mathcal{D}}_z-\left(\frac{1}{\R}+\vb{\nu}_{t}\right)(\vb{\mathcal{D}}_{yy}+\vb{\mathcal{D}}_{zz}+\vb{\mathcal{D}}_{xx})\\
&-\vb{\nu}_{tx}\vb{\mathcal{D}}_x-\vb{\nu}_{ty}\vb{\mathcal{D}}_y-\vb{\nu}_{tz}\vb{\mathcal{D}}_z
\end{aligned}
\end{equation}
\begin{equation} \label{eqn:SpatialE}
\vb{\mathcal{E}}=-{\rm{i}}\vb{\bar{u}}+2{\rm{i}}\left(\frac{1}{\R}+\vb{\nu}_{t}\right)\vb{\mathcal{D}}_x+{\rm{i}}\vb{\nu}_{tx}
\end{equation}}

\section{Adjoint LSA operator} \label{sec:adjLSAoperator}
{The $x-$momentum equation is shown as an example of the derivation procedure. By premultiplying the equation by the adjoint eigenvector (equation \ref{eqn:premulti}), the left-hand side can be written as
\begin{equation} \label{eqn:premultiX}
\begin{aligned}
& \iint\left(\vb{\hat{u}^\dag}\right)^H[-{\rm{i}}\omega+\vb{\bar{u}}(\vb{\mathcal{D}}_x+{\rm{i}}\alpha)+\vb{\bar{v}}\vb{\mathcal{D}}_y+\vb{\bar{w}}\vb{\mathcal{D}}_z-(1/\R+\vb{\nu}_{t})(\vb{\mathcal{D}}_{yy}+\vb{\mathcal{D}}_{zz}+\vb{\mathcal{D}}_{xx}+2{\rm{i}}\alpha\vb{\mathcal{D}}_{x}-\alpha^2) \\
& \ \ \ \ \ \ \ \ \ \ \ \ \ \ \ -\vb{\nu}_{tx}(\vb{\mathcal{D}}_x+{\rm{i}}\alpha)-\vb{\nu}_{ty}\vb{\mathcal{D}}_y-\vb{\nu}_{tz}\vb{\mathcal{D}}_z+\vb{\bar{u}}_x-\vb{\nu}_{tx}\vb{\mathcal{D}}_x-{\rm{i}}\alpha\vb{\nu}_{tx}]\left(\vb{\hat{u}}\right){\rm{d}}y{\rm{d}}z+ \\
& \iint\left(\vb{\hat{u}^\dag}\right)^H\left(\vb{\bar{u}}_y-\vb{\nu}_{ty}\vb{\mathcal{D}}_x-{\rm{i}}\alpha\vb{\nu}_{ty}\right)\left(\vb{\hat{v}}\right){\rm{d}}y{\rm{d}}z+\iint\left(\vb{\hat{u}^\dag}\right)^H\left(\vb{\bar{u}}_z-\vb{\nu}_{tz}\vb{\mathcal{D}}_x-{\rm{i}}\alpha\vb{\nu}_{tz}\right)\left(\vb{\hat{w}}\right){\rm{d}}y{\rm{d}}z+ \\
& \iint\left(\vb{\hat{u}^\dag}\right)^H\left(\vb{\mathcal{D}}_x+{\rm{i}}\alpha\right)\left(\vb{\hat{p}}\right){\rm{d}}y{\rm{d}}z
\end{aligned}
\end{equation}
We then present the integration-by-parts procedures for different types of terms in the equation for simplification. From a mathematical point of view, we can categorize terms in the equation based on whether they have a derivative matrix acting on the state vector. For terms that do not have a derivative matrix acting on the state vector, we have
\begin{equation} \label{eqn:iaU}
\iint\left(\vb{\hat{u}^\dag}\right)^H\left[{\rm{i}}\alpha\vb{\bar{u}}\right]\left(\vb{\hat{u}}\right){\rm{d}}y{\rm{d}}z=
\iint\left(-{\rm{i}}\alpha^\ast\vb{\bar{u}}\vb{\hat{u}^\dag}\right)^H\left(\vb{\hat{u}}\right){\rm{d}}y{\rm{d}}z
\end{equation}
If the first derivative matrix $\vb{\mathcal{D}}_y$ act on the state vector
\begin{equation} \label{eqn:Dy}
\iint\left(\vb{\hat{u}^\dag}\right)^H\left[\vb{\bar{v}}\vb{\mathcal{D}}_y\right]\left(\vb{\hat{u}}\right){\rm{d}}y{\rm{d}}z=
-\iint\left(\vb{\mathcal{D}}_y\vb{\bar{v}}\vb{\hat{u}^\dag}\right)^H\left(\vb{\hat{u}}\right){\rm{d}}y{\rm{d}}z+\oint\left(\vb{\bar{v}}\vb{\hat{u}^\dag}\right)^H\left(\vb{\hat{u}}\right){\rm{d}}z
\end{equation}
This procedure will leave boundary term which will have to be canceled then. Also for $\vb{\mathcal{D}}_z$
\begin{equation} \label{eqn:Dz}
\iint\left(\vb{\hat{u}^\dag}\right)^H\left[\vb{\bar{w}}\vb{\mathcal{D}}_z\right]\left(\vb{\hat{u}}\right){\rm{d}}y{\rm{d}}z=
-\iint\left(\vb{\mathcal{D}}_z\vb{\bar{w}}\vb{\hat{u}^\dag}\right)^H\left(\vb{\hat{u}}\right){\rm{d}}y{\rm{d}}z-\oint\left(\vb{\bar{w}}\vb{\hat{u}^\dag}\right)^H\left(\vb{\hat{u}}\right){\rm{d}}y
\end{equation}
For $\vb{\mathcal{D}}_x$, the form $\vb{\mathcal{D}}_x=-{\rm{tan}}\vb{\theta}\vb{\mathcal{D}}_y-{\rm{tan}}\vb{\gamma}\vb{\mathcal{D}}_z$ has to be taken back for the integration-by-parts procedure. Here for simplification, we refer $\vb{a}=-{\rm{tan}}\vb{\theta}$ and $\vb{b}=-{\rm{tan}}\vb{\gamma}$. Then we have
\begin{equation} \label{eqn:Dx}
\begin{aligned}
& \iint\left(\vb{\hat{u}^\dag}\right)^H\left[\vb{\bar{u}}\vb{\mathcal{D}}_x\right]\left(\vb{\hat{u}}\right){\rm{d}}y{\rm{d}}z=
\iint\left(\vb{\hat{u}^\dag}\right)^H\left[\vb{\bar{u}}\vb{a}\vb{\mathcal{D}}_y+\vb{\bar{u}}\vb{b}\vb{\mathcal{D}}_z\right]\left(\vb{\hat{u}}\right){\rm{d}}y{\rm{d}}z \\
= & -\iint\left[\left(\vb{\mathcal{D}}_y\vb{a}\vb{\bar{u}}+\vb{\mathcal{D}}_z\vb{b}\vb{\bar{u}}\right)\vb{\hat{u}^\dag}\right]^H\left(\vb{\hat{u}}\right){\rm{d}}y{\rm{d}}z+\oint\left(\vb{a}\vb{\bar{u}}\vb{\hat{u}^\dag}\right)^H\left(\vb{\hat{u}}\right){\rm{d}}z-\oint\left(\vb{b}\vb{\bar{u}}\vb{\hat{u}^\dag}\right)^H\left(\vb{\hat{u}}\right){\rm{d}}y
\end{aligned}
\end{equation}
Similarly, for all second derivative matrices
\begin{equation} \label{eqn:Dyy}
\iint\left(\vb{\hat{u}^\dag}\right)^H\left[\vb{\nu}_t\vb{\mathcal{D}}_{yy}\right]\left(\vb{\hat{u}}\right){\rm{d}}y{\rm{d}}z=
\iint\left(\vb{\mathcal{D}}_{yy}\vb{\nu}_t\vb{\hat{u}^\dag}\right)^H\left(\vb{\hat{u}}\right){\rm{d}}y{\rm{d}}z+\oint\left[\left(\vb{\nu}_t\vb{\hat{u}^\dag}\right)^H\left(\vb{\mathcal{D}}_y\vb{\hat{u}}\right)-\left(\vb{\mathcal{D}}_y\vb{\nu}_t\vb{\hat{u}^\dag}\right)^H\left(\vb{\hat{u}}\right)\right]{\rm{d}}z
\end{equation}
\begin{equation} \label{eqn:Dzz}
\iint\left(\vb{\hat{u}^\dag}\right)^H\left[\vb{\nu}_t\vb{\mathcal{D}}_{zz}\right]\left(\vb{\hat{u}}\right){\rm{d}}y{\rm{d}}z=
\iint\left(\vb{\mathcal{D}}_{zz}\vb{\nu}_t\vb{\hat{u}^\dag}\right)^H\left(\vb{\hat{u}}\right){\rm{d}}y{\rm{d}}z-\oint\left[\left(\vb{\nu}_t\vb{\hat{u}^\dag}\right)^H\left(\vb{\mathcal{D}}_z\vb{\hat{u}}\right)-\left(\vb{\mathcal{D}}_z\vb{\nu}_t\vb{\hat{u}^\dag}\right)^H\left(\vb{\hat{u}}\right)\right]{\rm{d}}y
\end{equation}
\begin{equation} \label{eqn:Dxx}
\begin{aligned}
& \iint\left(\vb{\hat{u}^\dag}\right)^H\left[\vb{\nu}_t\vb{\mathcal{D}}_{xx}\right]\left(\vb{\hat{u}}\right){\rm{d}}y{\rm{d}}z=\iint\left(\vb{\hat{u}^\dag}\right)^H\left[\vb{\nu}_t\vb{a}^2\vb{\mathcal{D}}_{yy}+2\vb{\nu}_t\vb{ab}\vb{\mathcal{D}}_{y}\vb{\mathcal{D}}_{z}+\vb{\nu}_t\vb{b}^2\vb{\mathcal{D}}_{zz}\right]\left(\vb{\hat{u}}\right){\rm{d}}y{\rm{d}}z \\
= & \iint\left[\left(\vb{\mathcal{D}}_{yy}\vb{a}^2\vb{\nu}_t+2\vb{\mathcal{D}}_{y}\vb{\mathcal{D}}_{z}\vb{ab}\vb{\nu}_t+\vb{\mathcal{D}}_{zz}\vb{b}^2\vb{\nu}_t\right)\vb{\hat{u}^\dag}\right]^H\left(\vb{\hat{u}}\right){\rm{d}}y{\rm{d}}z+ \\ 
& \oint\left[\left(\vb{a}^2\vb{\nu}_t\vb{\hat{u}^\dag}\right)^H\left(\vb{\mathcal{D}}_y\vb{\hat{u}}\right)-\left(\vb{\mathcal{D}}_y\vb{a}^2\vb{\nu}_t\vb{\hat{u}^\dag}\right)^H\left(\vb{\hat{u}}\right)+2\left(\vb{ab}\vb{\nu}_t\vb{\hat{u}^\dag}\right)^H\left(\vb{\mathcal{D}}_z\vb{\hat{u}}\right)\right]{\rm{d}}z- \\
& \oint\left[\left(\vb{b}^2\vb{\nu}_t\vb{\hat{u}^\dag}\right)^H\left(\vb{\mathcal{D}}_z\vb{\hat{u}}\right)-\left(\vb{\mathcal{D}}_z\vb{b}^2\vb{\nu}_t\vb{\hat{u}^\dag}\right)^H\left(\vb{\hat{u}}\right)-2\left(\vb{\mathcal{D}}_y\vb{ab}\vb{\nu}_t\vb{\hat{u}^\dag}\right)^H\left(\vb{\hat{u}}\right)\right]{\rm{d}}y
\end{aligned}
\end{equation}
By applying these procedures to all terms in equation \ref{eqn:premultiX} and rearranging, the adjoint operators $\vb{\mathcal{A}^\dag}$ and $\vb{\mathcal{B}^\dag}$ can be written as
\begin{equation} \label{eqn:SpatialAadj}
\hskip-0.5cm\vb{\mathcal{A}^\dag}=\begin{bmatrix}
\vb{\mathcal{C}^\dag}+\vb{\bar{u}}_x+(\vb{\mathcal{D}}_y\vb{a}+\vb{\mathcal{D}}_z\vb{b})\vb{\nu}_{tx} & 
\vb{\bar{v}}_x+\vb{\mathcal{D}}_y\vb{\nu}_{tx} & \vb{\bar{w}}_x+\vb{\mathcal{D}}_z\vb{\nu}_{tx} & 
-\vb{\mathcal{D}}_y\vb{a}-\vb{\mathcal{D}}_z\vb{b} & 0 & 0 & 0 \\
\vb{\bar{u}}_y+(\vb{\mathcal{D}}_y\vb{a}+\vb{\mathcal{D}}_z\vb{b})\vb{\nu}_{ty} & 
\vb{\mathcal{C}^\dag}+\vb{\bar{v}}_y+\vb{\mathcal{D}}_y\vb{\nu}_{ty} & \vb{\bar{w}}_y+\vb{\mathcal{D}}_z\vb{\nu}_{ty} & 
-\vb{\mathcal{D}}_y & 0 & 0 & 0 \\
\vb{\bar{u}}_z+(\vb{\mathcal{D}}_y\vb{a}+\vb{\mathcal{D}}_z\vb{b})\vb{\nu}_{tz} & 
\vb{\bar{v}}_z+\vb{\mathcal{D}}_y\vb{\nu}_{tz} & \vb{\mathcal{C}^\dag}+\vb{\bar{w}}_z+\vb{\mathcal{D}}_z\vb{\nu}_{tz} & 
-\vb{\mathcal{D}}_z & 0 & 0 & 0 \\
-\vb{\mathcal{D}}_y\vb{a}-\vb{\mathcal{D}}_z\vb{b} & -\vb{\mathcal{D}}_y & -\vb{\mathcal{D}}_z & 0 & 0 & 0 & 0 \\
0 & 0 & 0 & 0 & 1 & 0 & 0 \\
0 & 0 & 0 & 0 & 0 & 1 & 0 \\
0 & 0 & 0 & 0 & 0 & 0 & 1 \\
\end{bmatrix}
\end{equation}
\begin{equation} \label{eqn:SpatialBadj}
\vb{\mathcal{B}^\dag}=\begin{bmatrix}
\vb{\mathcal{E}^\dag}-{\rm{i}}\vb{\nu}_{tx} & 0 & 0 & {\rm{i}} & -1/\R-\vb{\nu}_{t} & 0 & 0 \\
-{\rm{i}}\vb{\nu}_{ty} & \vb{\mathcal{E}^\dag} & 0 & 0 & 0 & -1/\R-\vb{\nu}_{t} & 0 \\
-{\rm{i}}\vb{\nu}_{tz} & 0 & \vb{\mathcal{E}^\dag} & 0 & 0 & 0 & -1/\R-\vb{\nu}_{t} \\
{\rm{i}} & 0 & 0 & 0 & 0 & 0 & 0 \\
1 & 0 & 0 & 0 & 0 & 0 & 0 \\
0 & 1 & 0 & 0 & 0 & 0 & 0 \\
0 & 0 & 1 & 0 & 0 & 0 & 0 \\
\end{bmatrix}
\end{equation}
with $\vb{\mathcal{C}^\dag}$ and $\vb{\mathcal{E}^\dag}$ given the form
\begin{equation} \label{eqn:SpatialCadj}
\begin{aligned}
\vb{\mathcal{C}^\dag}=&{\rm{i}}\omega^\ast-\vb{\mathcal{D}}_y\vb{a}\vb{\bar{u}}-\vb{\mathcal{D}}_z\vb{b}\vb{\bar{u}}-\vb{\mathcal{D}}_y\vb{\bar{v}}-\vb{\mathcal{D}}_z\vb{\bar{w}}-\vb{\mathcal{D}}_{yy}\left(\frac{1}{\R}+\vb{\nu}_t\right)-\vb{\mathcal{D}}_{zz}\left(\frac{1}{\R}+\vb{\nu}_t\right)\\
&-\left(\vb{\mathcal{D}}_{yy}\vb{a}^2+2\vb{\mathcal{D}}_{y}\vb{\mathcal{D}}_{z}\vb{ab}+\vb{\mathcal{D}}_{zz}\vb{b}^2\right)\left(\frac{1}{\R}+\vb{\nu}_t\right)+\vb{\mathcal{D}}_y\vb{a}\vb{\nu}_{tx}+\vb{\mathcal{D}}_z\vb{b}\vb{\nu}_{tx}\\
&+\vb{\mathcal{D}}_y\vb{\nu}_{ty}+\vb{\mathcal{D}}_z\vb{\nu}_{tz}
\end{aligned}
\end{equation}
\begin{equation} \label{eqn:SpatialEadj}
\vb{\mathcal{E}^\dag}={\rm{i}}\vb{\bar{u}}+2{\rm{i}}(\vb{\mathcal{D}}_y\vb{a}+\vb{\mathcal{D}}_z\vb{b})\left(\frac{1}{\R}+\vb{\nu}_t\right)-{\rm{i}}\vb{\nu}_{tx}
\end{equation}
The boundary terms should then be canceled by applying appropriate adjoint boundary conditions. The expression for boundary terms on side boundaries is
\begin{equation} \label{eqn:BCUSide}
\begin{aligned}
&\hskip-0.5cm\oint\left(\vb{a}\vb{\bar{u}}\vb{\hat{u}^\dag}\right)^H\vb{\hat{u}}{\rm{d}}z+\oint\left(\vb{\bar{v}}\vb{\hat{u}^\dag}\right)^H\vb{\hat{u}}{\rm{d}}z-\oint\left[\left(\left(\frac{1}{\R}+\vb{\nu}_t\right)\vb{\hat{u}^\dag}\right)^H\vb{\mathcal{D}}_y\vb{\hat{u}}-\left(\vb{\mathcal{D}}_y\left(\frac{1}{\R}+\vb{\nu}_t\right)\vb{\hat{u}^\dag}\right)^H\vb{\hat{u}}\right]{\rm{d}}z- \\
&\hskip-0.5cm\oint\left[\left(\vb{a}^2\left(\frac{1}{\R}+\vb{\nu}_t\right)\vb{\hat{u}^\dag}\right)^H\vb{\mathcal{D}}_y\vb{\hat{u}}-\left(\vb{\mathcal{D}}_y\vb{a}^2\left(\frac{1}{\R}+\vb{\nu}_t\right)\vb{\hat{u}^\dag}\right)^H\vb{\hat{u}}+2\left(\vb{ab}\left(\frac{1}{\R}+\vb{\nu}_t\right)\vb{\hat{u}^\dag}\right)^H\vb{\mathcal{D}}_z\vb{\hat{u}}\right]{\rm{d}}z- \\
&\hskip-0.5cm2{\rm{i}}\alpha^\ast\oint\left(\vb{a}\vb{\hat{u}^\dag}\right)^H\vb{\hat{u}}{\rm{d}}z-\oint\left(\vb{a}\vb{\nu}_{tx}\vb{\hat{u}^\dag}\right)^H\vb{\hat{u}}{\rm{d}}z-\oint\left(\vb{\nu}_{ty}\vb{\hat{u}^\dag}\right)^H\vb{\hat{u}}{\rm{d}}z-\oint\left(\vb{a}\vb{\nu}_{tx}\vb{\hat{u}^\dag}\right)^H\vb{\hat{u}}{\rm{d}}z- \\
&\hskip-0.5cm\oint\left(\vb{a}\vb{\nu}_{ty}\vb{\hat{u}^\dag}\right)^H\vb{\hat{v}}{\rm{d}}z-\oint\left(\vb{a}\vb{\nu}_{tz}\vb{\hat{u}^\dag}\right)^H\vb{\hat{w}}{\rm{d}}z+\oint\left(\vb{a}\vb{\hat{u}^\dag}\right)^H\vb{\hat{p}}{\rm{d}}z
\end{aligned}
\end{equation}
and on vertical boundaries is
\begin{equation} \label{eqn:BCUVert}
\begin{aligned}
&\hskip-0.5cm-\oint\left(\vb{b}\vb{\bar{u}}\vb{\hat{u}^\dag}\right)^H\vb{\hat{u}}{\rm{d}}y-\oint\left(\vb{\bar{w}}\vb{\hat{u}^\dag}\right)^H\vb{\hat{u}}{\rm{d}}y+\oint\left[\left(\left(\frac{1}{\R}+\vb{\nu}_t\right)\vb{\hat{u}^\dag}\right)^H\vb{\mathcal{D}}_z\vb{\hat{u}}-\left(\vb{\mathcal{D}}_z\left(\frac{1}{\R}+\vb{\nu}_t\right)\vb{\hat{u}^\dag}\right)^H\vb{\hat{u}}\right]{\rm{d}}y+ \\
&\hskip-0.5cm\oint\left[\left(\vb{b}^2\left(\frac{1}{\R}+\vb{\nu}_t\right)\vb{\hat{u}^\dag}\right)^H\vb{\mathcal{D}}_z\vb{\hat{u}}-\left(\vb{\mathcal{D}}_z\vb{b}^2\left(\frac{1}{\R}+\vb{\nu}_t\right)\vb{\hat{u}^\dag}\right)^H\vb{\hat{u}}-2\left(\vb{\mathcal{D}}_y\vb{ab}\left(\frac{1}{\R}+\vb{\nu}_t\right)\vb{\hat{u}^\dag}\right)^H\vb{\hat{u}}\right]{\rm{d}}y+ \\
&\hskip-0.5cm2{\rm{i}}\alpha^\ast\oint\left(\vb{b}\vb{\hat{u}^\dag}\right)^H\vb{\hat{u}}{\rm{d}}y+\oint\left(\vb{b}\vb{\nu}_{tx}\vb{\hat{u}^\dag}\right)^H\vb{\hat{u}}{\rm{d}}y+\oint\left(\vb{\nu}_{tz}\vb{\hat{u}^\dag}\right)^H\vb{\hat{u}}{\rm{d}}y+\oint\left(\vb{b}\vb{\nu}_{tx}\vb{\hat{u}^\dag}\right)^H\vb{\hat{u}}{\rm{d}}y+ \\
&\hskip-0.5cm\oint\left(\vb{b}\vb{\nu}_{ty}\vb{\hat{u}^\dag}\right)^H\vb{\hat{v}}{\rm{d}}y+\oint\left(\vb{b}\vb{\nu}_{tz}\vb{\hat{u}^\dag}\right)^H\vb{\hat{w}}{\rm{d}}y-\oint\left(\vb{b}\vb{\hat{u}^\dag}\right)^H\vb{\hat{p}}{\rm{d}}y
\end{aligned}
\end{equation}
Although these expressions are given in huge forms, they can be much simplified since the adjoint mode only needs to be computed in the near wake region. Therefore we can consider all direct and adjoint perturbations on the far-field boundaries to be zero, then only the boundary terms on $y/W=0$ and $z/W=0$ should be further considered. On $y/W=0$ we have the following conditions and equation \ref{eqn:BCUSide} can be simplified into
\begin{subequations} \label{eqn:BCU_Y0}
\begin{align}
&\vb{a}=\vb{\bar{v}}=\vb{\nu}_{ty}=\vb{\hat{u}}_y=0 \ {\rm{on}} \ y/W=0 \\
&\oint\left[\left(\frac{1}{\R}+\vb{\nu}_t\right)\vb{\mathcal{D}}_y\vb{\hat{u}^\dag}\right]^H\vb{\hat{u}}{\rm{d}}z
\end{align}
\end{subequations}
and on $z/W=0$ these conditions can be applied with equation \ref{eqn:BCUVert} can be simplified into
\begin{subequations} \label{eqn:BCU_Z0}
\begin{align}
&\vb{b}=\vb{\bar{w}}=\vb{\hat{u}}=0 \ {\rm{on}} \ z/W=0 \\
&\oint\left[\left(\frac{1}{\R}+\vb{\nu}_t\right)\vb{\hat{u}^\dag}\right]^H\vb{\mathcal{D}}_z\vb{\hat{u}}{\rm{d}}z
\end{align}
\end{subequations}
Therefore all adjoint boundary conditions appropriate to cancel the boundary terms are summarized as follows
\begin{subequations} \label{eqn:adjointBC}
\begin{align}
&\frac{\partial{\vb{\hat{u}^\dag}}}{\partial{y}}=\frac{\partial{\vb{\hat{w}^\dag}}}{\partial{y}}=\frac{\partial{\vb{\hat{p}^\dag}}}{\partial{y}}=0,\vb{\hat{v}^\dag}=0 \ {\rm{on}} \ y/W=0\\
&\vb{\hat{u}^\dag}=\vb{\hat{v}^\dag}=\vb{\hat{w}^\dag}=\vb{\hat{p}^\dag}=0 \ {\rm{on}} \ y/W=1.3\\
&\vb{\hat{u}^\dag}=\vb{\hat{v}^\dag}=\vb{\hat{w}^\dag}=0,\frac{\partial{\vb{\hat{p}^\dag}}}{\partial{z}}=0 \ {\rm{on}} \ z/W=0\\
&\vb{\hat{u}^\dag}=\vb{\hat{v}^\dag}=\vb{\hat{w}^\dag}=\vb{\hat{p}^\dag}=0 \ {\rm{on}} \ z/W=1.3
\end{align}
\end{subequations}\par}
\begin{figure}
  \centering
  \begin{subfigure}{0.85\textwidth}
  \subcaption{}
  \includegraphics[ width=1\textwidth]{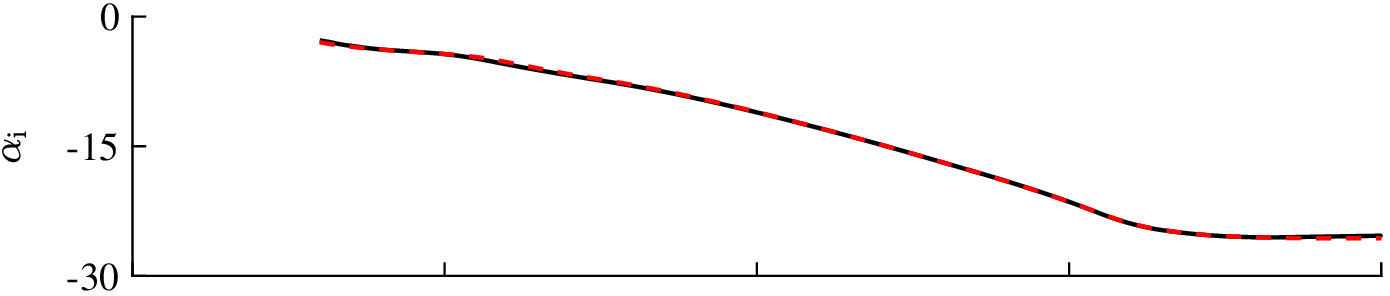}
  \end{subfigure}
  \begin{subfigure}{0.85\textwidth}
  \subcaption{}
  \includegraphics[width=1\textwidth]{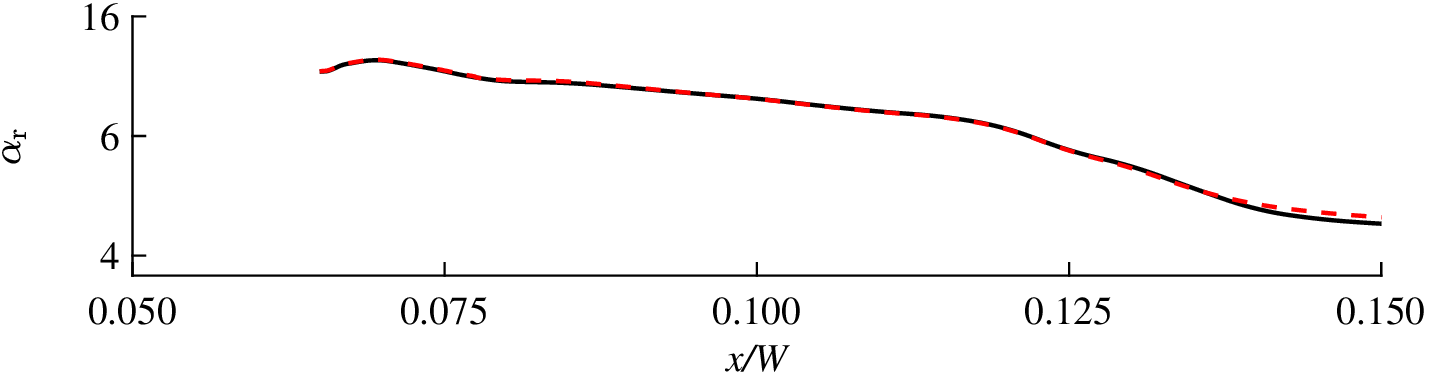}
  \end{subfigure}
  \caption{
  Local complex streamwise wavenumber of the adjoint mode based on direct LNS equation (black solid line), and its complex conjugation based on adjoint LNS equation (red dashed line). (\textit{a}) The real component; (\textit{b}) The imaginary component.
  }
  \label{fig:AdjointBranch}
\end{figure}
{To validate the adjoint operators and boundary conditions, we take the conjugation of the complex streamwise wavenumber of the adjoint mode computed based on the adjoint method and compare it with the results from the direct approach, as shown in figure \ref{fig:Wavemaker}(\textit{e,f}). The comparisons are shown in figure \ref{fig:AdjointBranch}. Good agreement can be observed between the two approaches, with discrepancies being observed only downstream of $x/W=0.135$. This is due to the fact that, downstream of this streamwise location, the adjoint eigenvalue is quite far from the real axis, which would lead to a convergence issue of the eigenvalue problem. At locations around the wavemaker regions, the two lines are almost identical, which confirms that the adjoint operators and boundary conditions presented in this paper can provide with accurate prediction on the adjoint mode and subsequently structural sensitivity.\par}

\bibliographystyle{jfm}
\bibliography{references}

\end{document}